\documentclass{IEEEtran}
\usepackage[utf8]{inputenc}
\usepackage{fullpage}
\usepackage{physics}
\usepackage{cite}
\usepackage{xcolor}
\usepackage{hyperref}
\usepackage{graphicx,subcaption}
\usepackage{amsmath}
\usepackage{mathtools}
\usepackage[T1]{fontenc}

\DeclareUnicodeCharacter{2212}{-}
\begin{document}
\title{Entanglement distribution in two-dimensional square grid network}

\author{\IEEEauthorblockN{Eneet Kaur and
Saikat Guha}\\
\IEEEauthorblockA{Wyant College of Optical Sciences, The University of Arizona, \\1630 E. University Blvd., Tucson, AZ, 85721, USA}}

\maketitle
\begin{abstract}
We study entanglement generation in a quantum network where repeater nodes can perform $n$-qubit Greenberger-Horne-Zeilinger(GHZ) swaps, i.e., projective measurements, to fuse $n$ imperfect-Fidelity entangled-state fragments. We show that the distance-independent entanglement distribution rate found previously for this protocol, assuming perfectly-entangled states at the link level, does not survive. This is true also in two modified protocols we study: one that incorporates $l \to 1$ link-level distillation, and another that spatially constrains the repeater nodes involved in the swaps. We obtain analytical formulas for a GHZ swap of multiple Werner states, which might be of independent interest. Whether the distance-independent entanglement rate might re-emerge with a spatio-temporally-optimized scheduling of GHZ swaps and multi-site block-distillation codes, remains open.
\end{abstract}

\section{Introduction}
A quantum network would be essential to provide quantum resources to spatially separated parties for various applications related to quantum technologies. One such resource is entanglement, which is essential for quantum key distribution \cite{Bennett84,Ekert1991}, quantum sensing \cite{Giovannetti2004}, and distributed quantum computing \cite{Van_Meter_2007,Cleve_1997}. However, entanglement distribution across an optical fiber has a fundamental restriction, as found in \cite{TGW14,Pirandola2017}, also known as the repeaterless bound. The entanglement distribution rates decay linearly in $\eta = \exp(-\alpha L)$, where $\alpha$ is a constant determined by the optical fiber and $L$ is the distance between nodes. The maximum attainable rate across an optical fiber is $− \log(1 − \eta)\approx  1.44\eta$, for $\eta \ll 1$, ebits (pure Bell states shared between two parties) per transmitted optical mode. To circumvent the repeaterless bound, we require quantum repeaters --- small quantum computers --- placed along the optical fibers. A collection of quantum repeaters forms a quantum network.

A well-studied model for a quantum network is that of a linear chain of quantum repeaters \cite{BriegalPRL}. A linear chain of the quantum network can be highly susceptible to failure in any repeater node. We, therefore, anticipate that a quantum network will have a more complex topology --- a star network \cite{Star-network}, a grid network \cite{Pant2019,Patil_2022}, or a ring network and a sphere network \cite{Schoute}, among others \cite{Samura2020}. Entanglement distribution protocols in complex networks are called routing protocols.  The underlying topology can be utilized differently, namely by exploiting various paths in the network \cite{Pant2019} using global and local link-level knowledge.  \cite{Schoute} introduced routing protocols for sphere and ring networks, assuming that each link generates a perfect, lossless entangled pair in every time slot and the repeater carries out a perfect Bell measurement. \cite{Chakraborty_2020} constructed an efficient linear-programming formulation, where the authors approached entanglement routing using a multi-commodity flow-based approach with perfect gate operations, probabilistic Bell measurements, and imperfect channels. \cite{Patil_2022} took a different approach and considered multipartite swapping in quantum networks, thereby creating robust intermediate entanglement states in the network. This problem assumed probabilistic entanglement generation, probabilistic swapping, and distribution of a perfect entangled state. Under the aforementioned set of assumptions, the authors of \cite{Patil_2022} were able to prove the distance-independent entanglement generation rate. The distance-independent regions were further studied in \cite{time-multiplexed} under space-time multiplexed (GHZ) measurements.  Distance independence for quantum communication capacity has also been studied in \cite{Quantao-distance-independence} and for imperfect yet pure states across links in \cite{Percolation1,Percolation2,Siomau_2016,Acn2007,Perseguers2008}. 

Entanglement swapping is a common ingredient in quantum repeaters and uses small-length entanglement links to create a longer-length entanglement link. However, one major drawback of swapping is that the created states tend to be noisier than the input states. Thus, if we take bipartite states of imperfect fidelities and swap them repeatedly, the resultant state would be separable. To this end, we must introduce distillation in various protocol stages. An analysis of the order of performing distillation and swapping in a linear chain was studied in \cite{BriegalPRL, Zoller1991}. Even when examined on a linear chain of quantum networks, this problem can be highly complicated. In \cite{routing-network}, the authors introduced routing and distillation protocols on grid networks.  

In this work, we analyze bipartite entanglement distribution in a square network with intermediate nodes allowed to perform GHZ measurements. A square network is characterized by $p,F$, where $p$ is the probability of heralding an entangled state on a link level, and $F$ is the fidelity of the heralded state. We show that the distance independence of rates found in \cite{Patil_2022} is highly fragile and vanishes when $F\neq1$. We then explore entanglement routing with GHZ swaps in the grid network and show that the optimal routing depends on the network parameters $p, F$. We add a distillation scheme to the routing algorithm and show how a naive distillation scheme can boost the entanglement distribution rates for specific parameters. 

\section{Overview}
\subsection{Routing protocols}
\begin{figure}
    \centering
    \includegraphics[width = .75\linewidth]{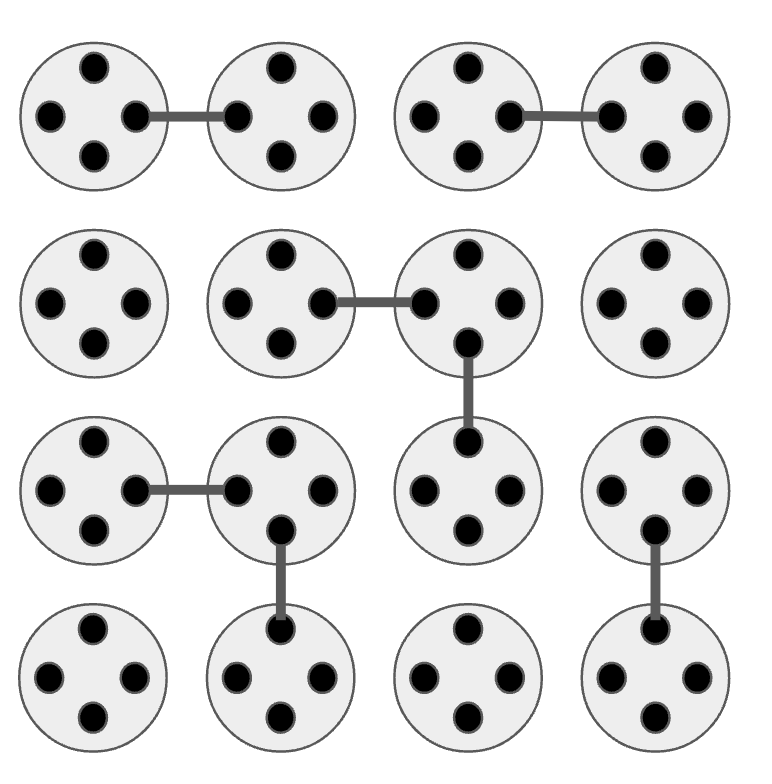}
    \caption{We represent a grid quantum network by a graph $G$, the grey circles represent a quantum repeater, and the black dots or the nodes represent the quantum memories. The edges between the memories represent a bipartite state.}
    \label{fig:example}
\end{figure}

We consider a two-dimensional square grid quantum network; see Fig.~\ref{fig:example}. Grey nodes represent a quantum repeater. Each quantum repeater is equipped with four quantum memories, capable of sharing a bipartite quantum state with quantum memories of the adjacent node with probability $p$. We divide the nodes present in the network into three categories: helper nodes, idle nodes, and consumer nodes. The helper nodes play an active role during the protocol --- by performing GHZ swaps, the idle nodes do not utilize their quantum memories during the protocol run, and the consumer nodes share the final entanglement. We designate a node as a helper or idle node depending on the location of the consumer
nodes and the parameters $p$ and $F$. A protocol begins with communicating, to all nodes, the location
of the consumer nodes. The probability of establishing an entangled state $p$ is known to all
nodes and is fixed. We decide which nodes to utilize based on the site of consumer nodes, link level probability $p$, and the fidelity $F$. 

In the first time step of the protocol, the quantum memories of the helper and consumer nodes attempt to establish entanglement with the quantum memories of the adjacent helper nodes. The entanglement distribution on each link succeeds with the probability $p$, and the resultant shared state is a Werner state $\rho$ of fidelity $F$, written as
\begin{multline}
    \rho = F \op{\Phi^+} + \frac{(1-F)}{3}\left(\op{\Phi^-}\right.\\\left.+\op{\Psi^+}+\op{\Psi^-}\right),
\end{multline}
where the vectors $\left\{\ket{\Phi^{\pm}},\ket{\Psi^{\pm}}\right\}$ form the Bell basis. We assume that the quantum memories on the top send one-half of the pair to the memories on the bottom, and the quantum memories on the left send one-half of the pair to nodes on the right. 
In the second time step, each helper or consumer node obtains information about the heralded entanglement 
of the $k$-local nodes. By $k$-local nodes, we mean that the Manhattan distance between the nodes is $k$. We call this as the $k$-hop communication. 
This is a departure from the protocol considered in \cite{Patil_2022}, where only the local link-level heralding information is available, i.e., a repeater knows the success-failure outcomes at each time slot of its own link generation attempts (across its neighboring edges). In the third step of the protocol, the helper nodes perform GHZ$(m)$ projections on $m$ quantum memories or local $X$ operations, where $m$ depends on the information available to the nodes and $2\leq m\leq 4$. The measurement results of the swap and the heralded memories are communicated to the consumer nodes. The consumer nodes can abandon the round or use the generated state depending on the classical information received. 

We define the GHZ basis for $m$-qubit systems with the basis vectors $\ket{\psi_{j,i_1,i_2\cdots i_{m-1}}}$ as
\begin{equation}
    \ket{\psi_{j,i_1,i_2\cdots i_{m-1}}} = \bigotimes_{\alpha=1}^{m-1}X_\alpha^{i_\alpha}Z_0^j\ket{\textrm{GHZ}(m)},
\end{equation}
where $i_{\alpha},j\in [0,1]$, 
and 
\begin{equation}
    \ket{\textrm{GHZ}(m)} = \frac{1}{\sqrt{2}}\left(\ket{00\cdots 0}+\ket{11\cdots 1}\right).
\end{equation}

All the GHZ measurements are implemented at the same time step --- allowed as all measurements are local and thus commute. At the end of the protocol, the consumer memory nodes may share an entangled state. The final state --- shared between the consumer nodes --- is a GHZ diagonal state with the coefficients determined by the time steps' successes and the outcome of the intermediate GHZ measurements. The consumer nodes will have to wait for the information from all the nodes to determine which quantum memories share entanglement and the exact form of the entangled state. The nodes can choose to use the entanglement without knowing the exact form, such as with QKD applications, where the corrections based on the measurement results of the helper nodes can be applied to the raw classical data. However, if the consumer nodes need to know the exact form of the state before utilizing it, they would have to wait for communication from all the intermediate nodes, which would introduce additional latency in the protocol. 

The protocol rate is the distillable entanglement of the final state per use of the quantum network. In this work, we use coherent information --- a lower bound on entanglement distillation of a state optimized over protocols with one-way LOCC (local operations and classical communication)\cite{Winter2003} to access the entanglement quality of the generated quantum state.

\subsection{Results}

The major contributions of this work are as follows:
\begin{itemize}
    \item In Section~\ref{sec:distance-independence}, we outline a method to obtain the final state of the protocol. We then provide numerical evidence that given a square grid network, with $m\geq 2$, imperfect link level Bell states, entanglement generation with the aforementioned protocol at a distance-independence rate is impossible. 
    \item  In Section~\ref{sec:region}, we consider modified routing protocols for entanglement generation in square grid networks and GHZ measurements on the helper nodes, with the modification being the region sizes. We show that the region of helper nodes to consider for entanglement generation is related to the link level probabilities and initial fidelities. We then consider rate envelopes for the achievable entanglement routing protocols using square grids with GHZ measurements. 
    \item In Appendix~\ref{appendix:werner-swap} and Appendix~\ref{appendix:diff-fidelities}, we give an analytical expression of the states obtained after we measure one half of $n$ Werner states with $\textrm{GHZ}(n)$ basis.
    \item In Appendix~\ref{appendix:linearity-proof}, we give a method of modeling the aforementioned protocol such that the maximal size of the intermediate entangled state created in the network consists of $N+c$ qubits, where $c \in [1,9]$ and depends on the position of Alice and Bob in the network. 
\end{itemize}
Our results highlight the need for carefully accounting for the link-level fidelities of the distributed states in a quantum network. We expect that the distance-independent quantum communication rates found in \cite{Quantao-distance-independence,Percolation1,Percolation2,Siomau_2016} will also be fragile to noise present in the network.

\section{Previous work}

A slight variation of the aforementioned protocol was introduced in \cite{Patil_2022} but with the initial fidelity $F=1$. We explicate the difference in Section~\ref{sec:distance-independence}. A version with Bell swap or $m=2$ has also been studied in \cite{Pant2019}, again with the initial fidelity $F=1$. In \cite{Patil_2022}, the authors studied the connectivity between two nodes of a two-dimensional square-grid graph as a function of link probabilities $p$ and fusion success probabilities $q$ and the distance between the consumers. They found that when $q = 1$, for specific topologies --- including the square grid network, the connectivity problem translates to a bond percolation problem. This implies that the network undergoes a phase transition for $p\geq p_c$, where $p_c$ is a threshold that
depends on the lattice geometry. That is, the probability that any two randomly chosen nodes are connected remains constant with the distance between the nodes. One can find a unique giant cluster in the network such that the probability of two nodes being a part of the cluster is distance-independent. 
For $q\leq 1$, the authors of \cite{Patil_2022} translated the existence of distance-independent connectivity to that of a site percolation problem and obtained the critical values of $q$ and $p$ above which the network undergoes a phase transition. When the fidelity of the link-level bipartite state in the above protocol is one, the distance-independent node connectivity translates to distance-independent entanglement generation. Authors of \cite{Gayne} studied the quantum entanglement of a switch serving $k$ users in a star topology. \cite{Lee2022} introduced a quantum router architecture for high-fidelity entanglement distribution in quantum networks. In \cite{Jiang2007,Kenneth-repeater}, the authors incorporate entanglement distillation to
obtain schemes for entanglement distribution in a quantum internet. This problem can be challenging, and the authors address them by introducing optimization schemes.

Apart from the works mentioned above, percolation-based phase transitions have also been studied for various quantum network topologies. In \cite{Samura2020}, the authors studied the statistical properties of random and photonic networks. They showed a continuous phase transition between a disconnected and a highly-connected phase characterized by the computation of critical exponents. In \cite{Quantao-distance-independence}, the authors analyze phase transitions for quantum communication capacities, assuming that it is possible to distill perfect Bell states. In \cite{time-multiplexed}, the authors showed distance-independent entanglement generation in a quantum network using space-time Multiplexed GHZ measurements.

\section{Protocols and results}\label{sec:distance-independence}
\subsection{Detailed Protocol}\label{sec:original}

\begin{figure}
    \centering
    \includegraphics[width = \linewidth]{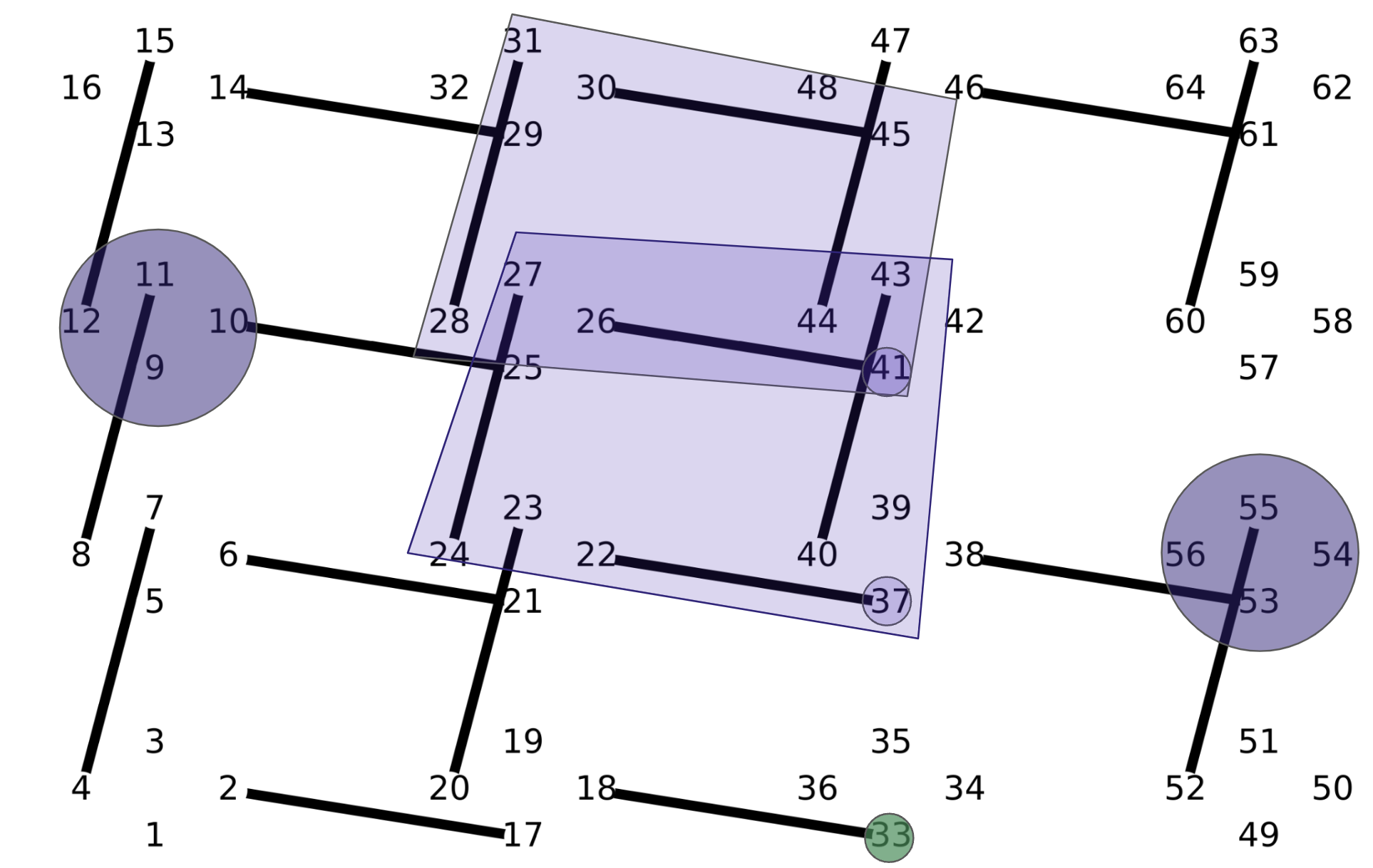}
    \caption{This figure represents the graph created after the entanglement distribution step. The numbers represent the vertices or the quantum memories. The edges represent an entangled state between the vertices. We see quantum memories numbered $(41)$ and $(37)$ lie at the bottom right corner of the square formed by the heralded entanglement of helper nodes. In the protocol, we perform an $X$ measurement on the node. Quantum memory $(33)$ is the only single memory in the node to share a heralded entangled state with an adjacent memory. Thus in the protocol, we perform $X$ measurement on this memory qubit.}
    \label{fig:example-cases}
\end{figure}

We represent the state of the network with a graph. 
We begin with a graph $G(V,\emptyset)$ with $V$ representing the set of vertices modeling the quantum memories. In the first time step, the quantum memories attempt entanglement generation with the adjacent nodes which succeed with probability $p$. We model the state of the network after the first time step as a graph $G(V,E)$, where $v \in  V$ represents the quantum memory, and $e\in E$ represents an edge created with probability $p$, see figure~\ref{fig:example-cases} for an example. 

We assume that the quantum memories on the top send one-half of an entangled pair to the memories on the bottom, and the quantum memories on the left send one-half of an entangled pair to nodes on the right. In the second time step, the nodes send information about the heralded entanglement on their quantum memories to the $k$-neighboring nodes. The choice of $k$ depends on the grid size and $p$. The need for $k$-hop communication is outlined in \ref{subsection: k-level-importance}. The helper nodes perform the following actions in the $(k+2)$ time step based on the information received:

\begin{itemize}
    \item If a quantum memory lies at the bottom right corner of a $2k+2$-edged polygon formed by the heralded entanglement of the quantum memories, then the helper node performs an $X$ measurement on the memory. This information is available to the node by the classical communication step, where $k$ is the level of communication. 
    \item If the leftover memories, after the $X$ measurement in a node, have more than one heralded entanglement, the node performs a GHZ measurement on the memories.
    \item If the leftover memories in a node have only one heralded entanglement, the node performs $X$ measurement. 
    \item The measurement results are communicated to the consumer nodes. If there exists a $l$-edged polygon with $l> 2k+2$, and such that Alice's or Bob's memories are not parts of the polygon, the round is abandoned.
\end{itemize}
For example cases of the actions performed by the nodes, see Figure~\ref{fig:example-cases}.

\subsection{Modeling of protocol}\label{sec:model}

\begin{figure*}\label{fig:entanglement_progress}
\begin{subfigure}{0.45\textwidth}
   \includegraphics[width=\linewidth]{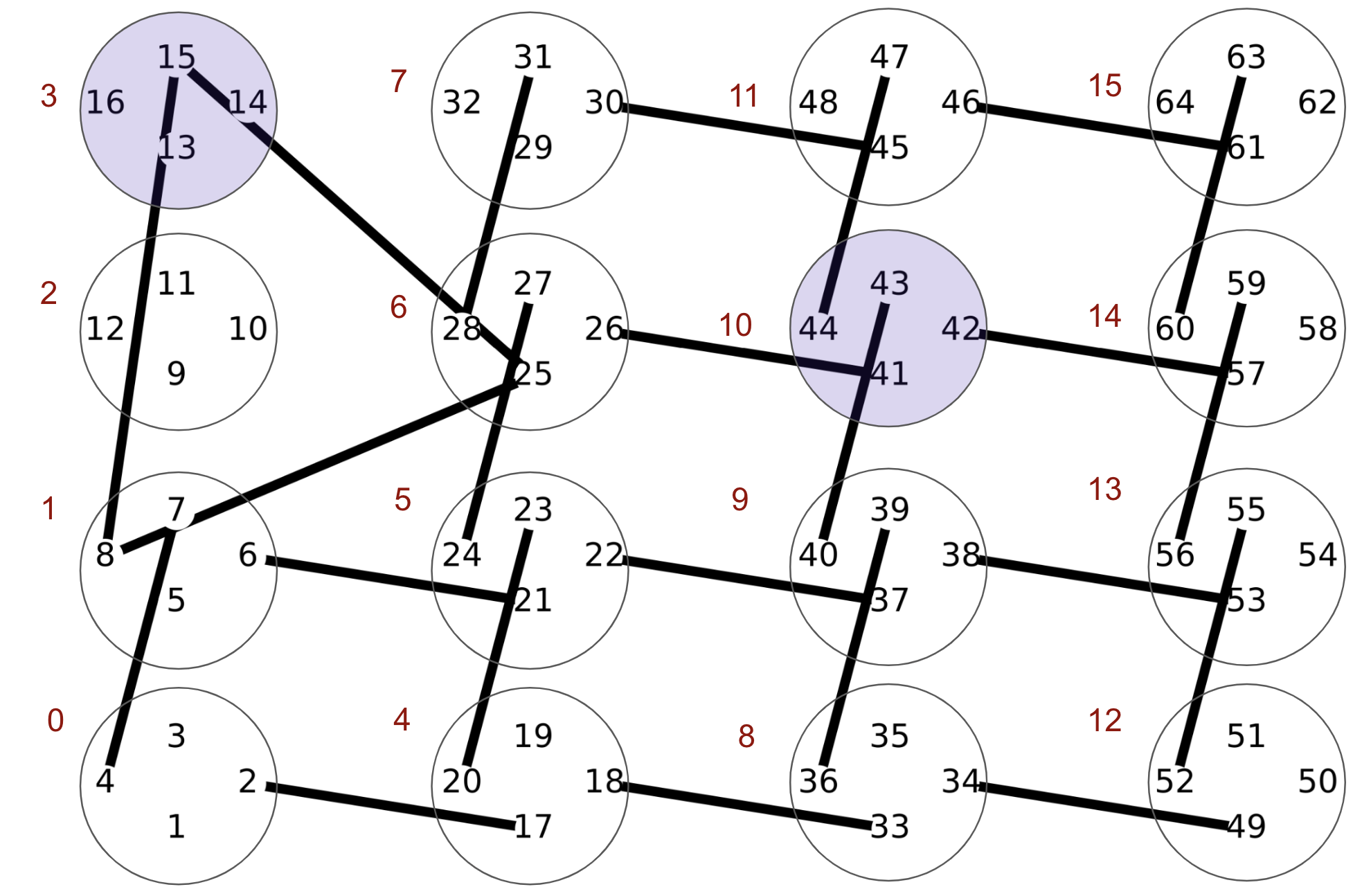}
   \caption{$G_1:$ Graph representation after first swap} \label{fig:x_a}
\end{subfigure}
\hspace*{\fill}
\begin{subfigure}{0.45\textwidth}
   \includegraphics[width=\linewidth]{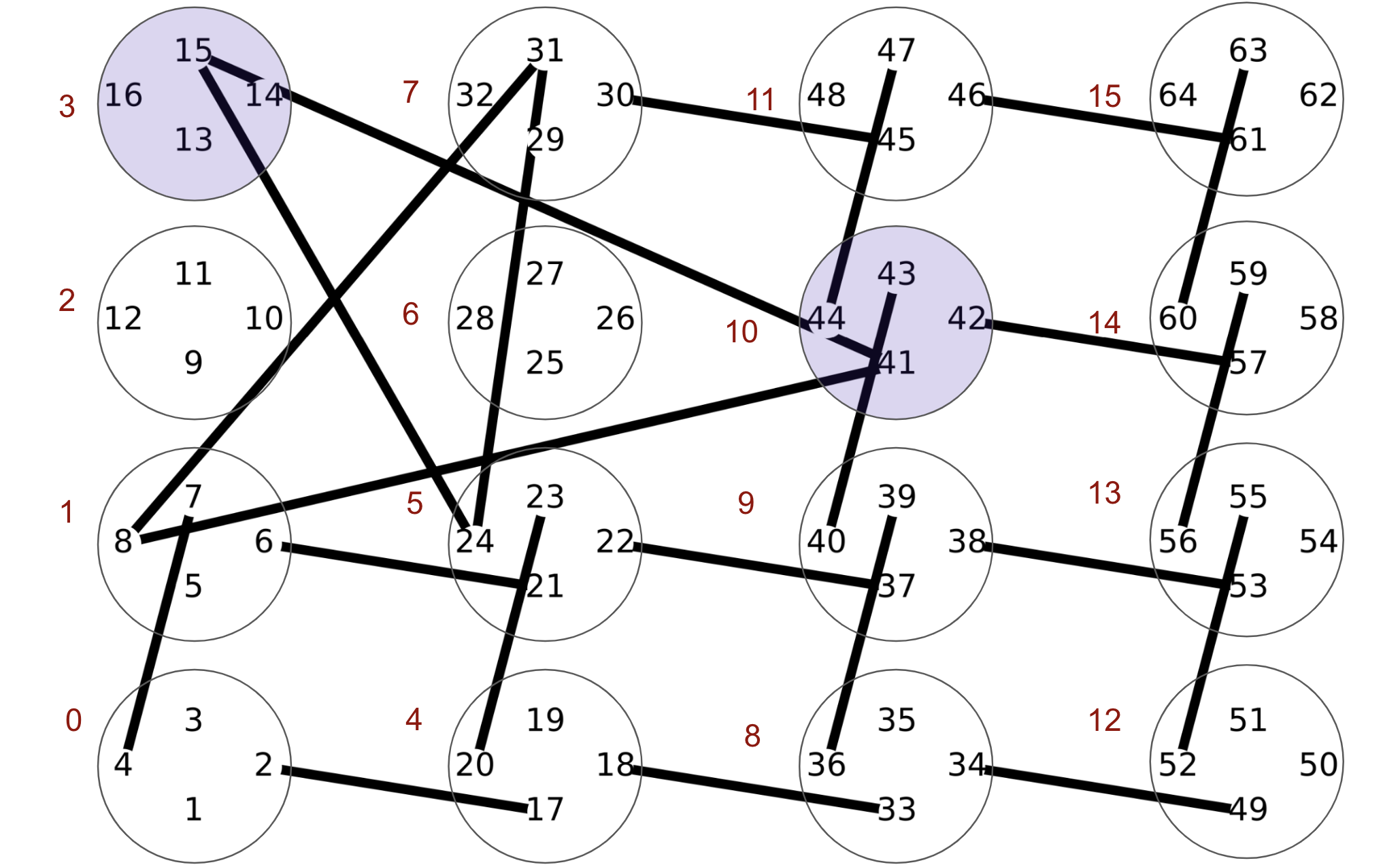}
   \caption{$G_2$: Graph representation after second swap} \label{fig:x_b}
\end{subfigure}

\bigskip
\begin{subfigure}{0.45\textwidth}
   \includegraphics[width=\linewidth]{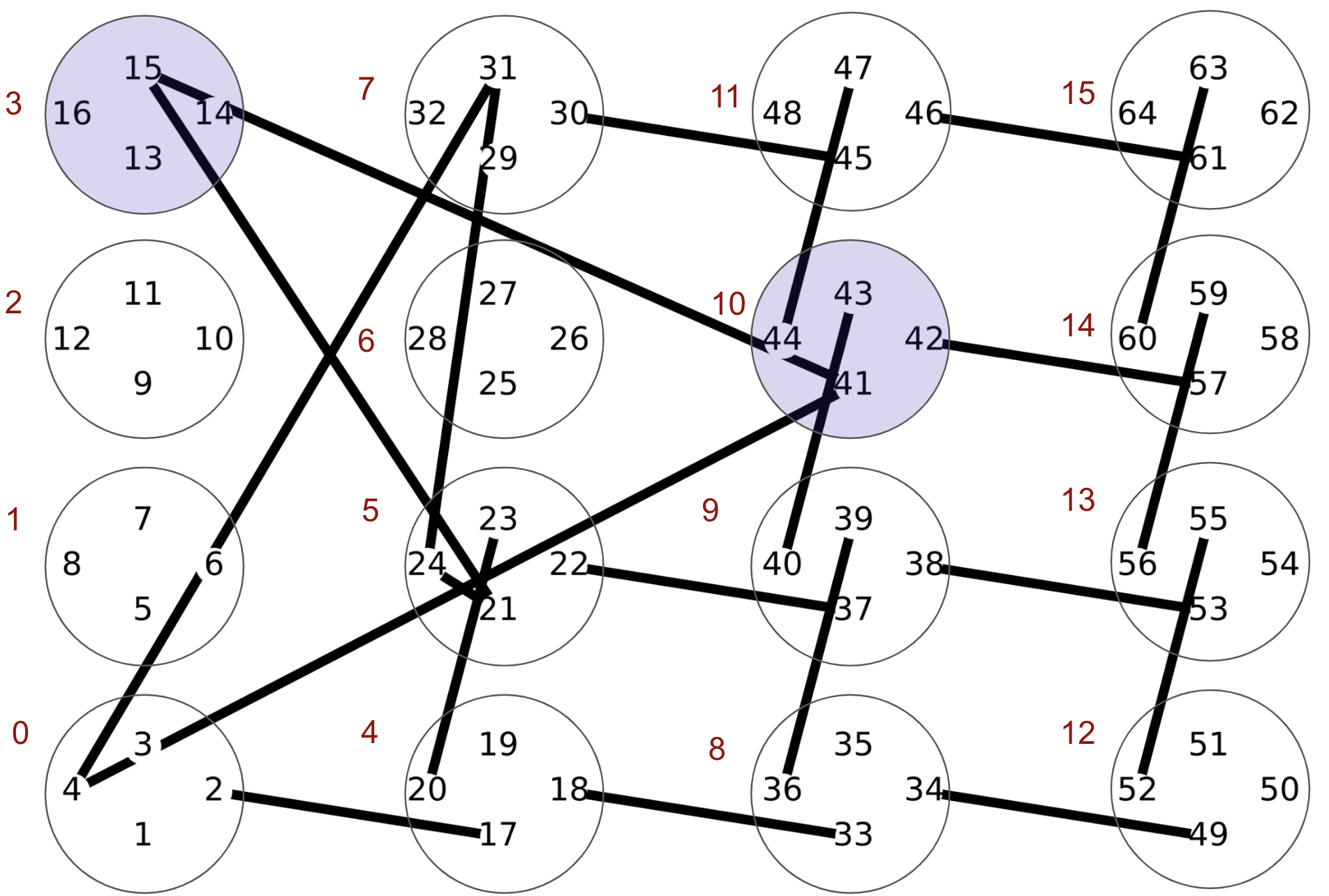}
   \caption{$G_3$: Graph representation after third swap} \label{fig:x_c}
\end{subfigure}
\hspace*{\fill}
\begin{subfigure}{0.45\textwidth}
   \includegraphics[width=\linewidth]{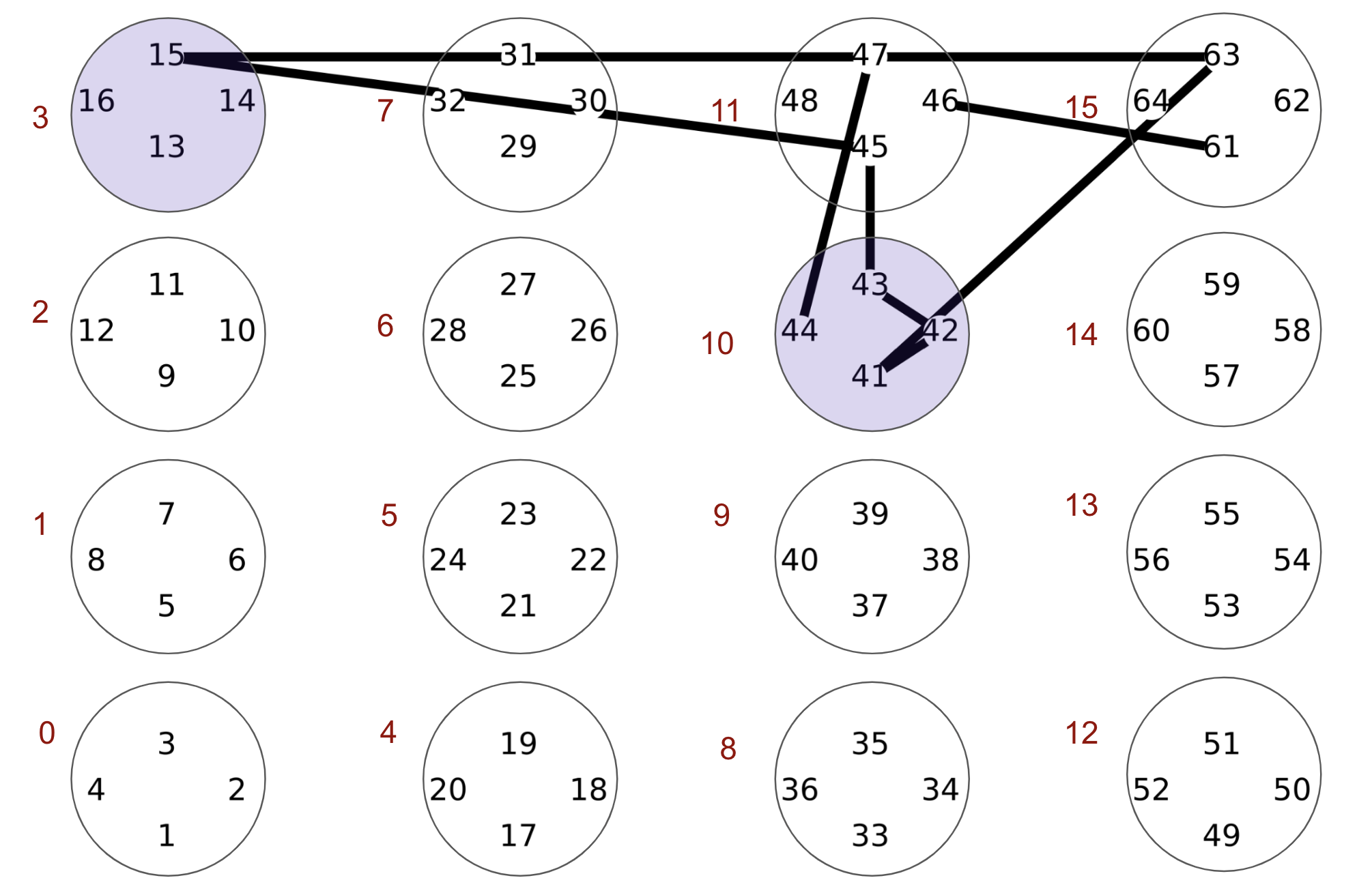}
   \caption{$G_{11}$} \label{fig:x_d}
\end{subfigure}
\bigskip
\begin{subfigure}{0.45\textwidth}
   \includegraphics[width=\linewidth]{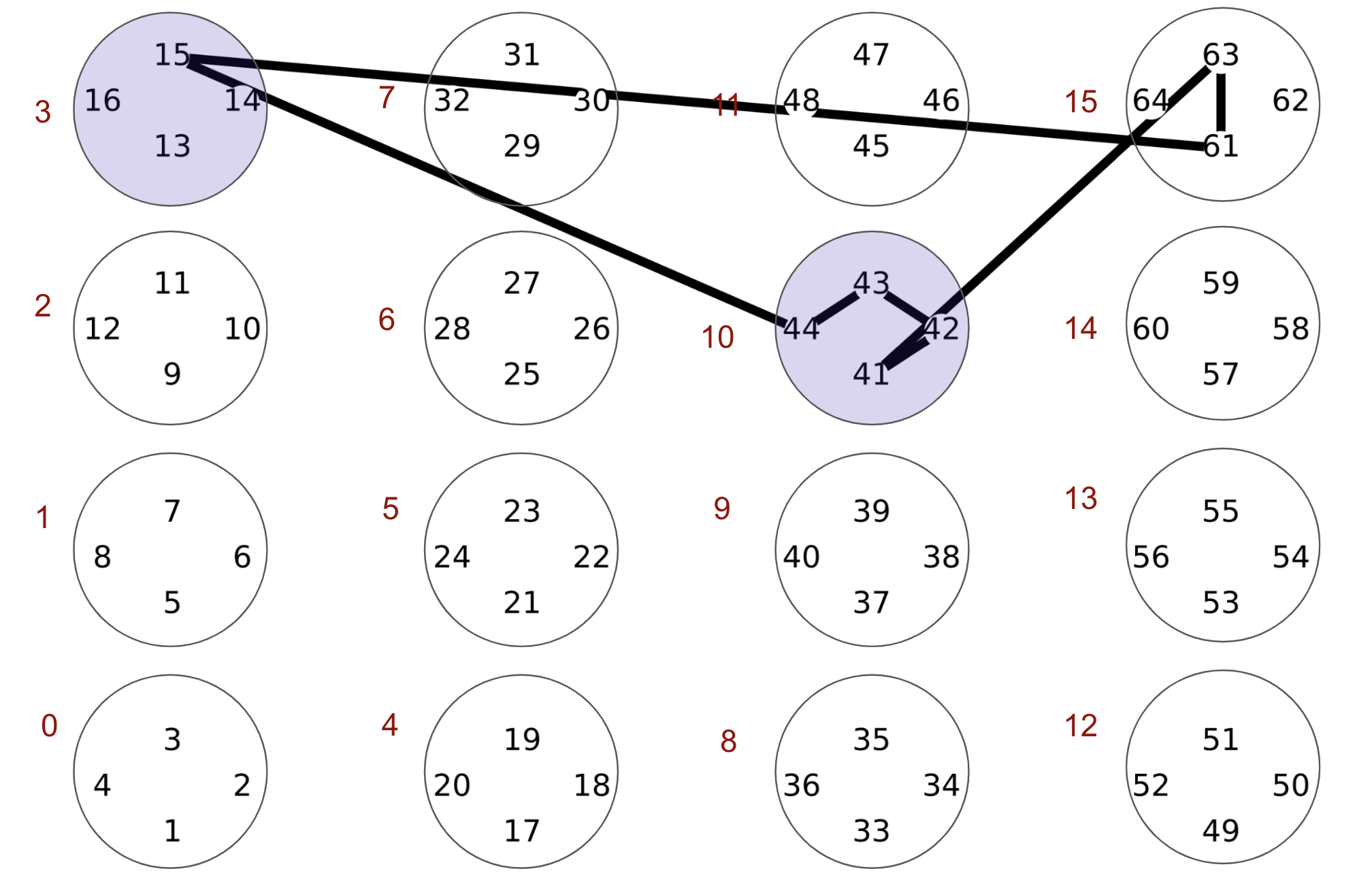}
   \caption{$G_{12}$} \label{fig:x_e}
\end{subfigure}
\hspace*{\fill}
\begin{subfigure}{0.45\textwidth}
   \includegraphics[width=\linewidth]{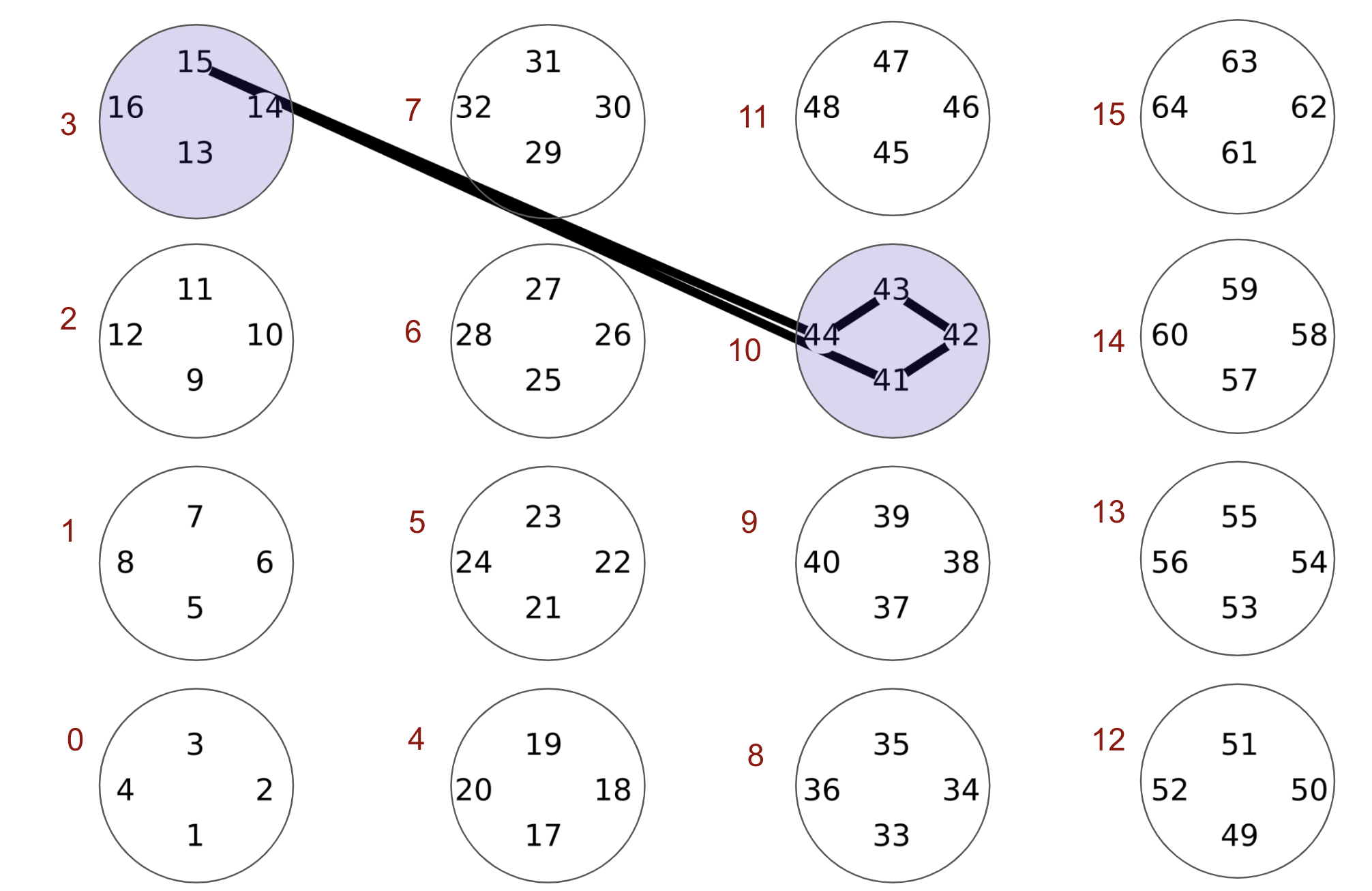}
   \caption{$G_{13}$} \label{fig:x_f}
\end{subfigure}

\hspace*{\fill}

\caption{In this figure, we describe the sequence of graphs $G_i$ created during the swapping process, with $i \in [0,n^2-3]$.}
\end{figure*}

While the $\textrm{GHZ}$ and $X$ measurements are commutative and can be performed simultaneously, we assume a sequential swapping for simulation purposes. With each sequential swap, a new graph $G'(V,E')$ is created. The graph's vertex set remains unchanged, while the edge set $E'$ changes. The sequence of the swaps does not change the final graph and the final shared state, however, it does change the intermediate graphs' states.

We give an example of the graph progression in Fig~\ref{fig:x_a}-\ref{fig:x_f}. We label the nodes by the red numbers, and the memories within are numbered in black. The sequence is as follows:
\begin{itemize}
    \item For our strategy, we start with any consumer node, in this example denoted by node~$3$. For the first swap, choose any neighboring node of $3$, sharing a heralded quantum state with quantum memories of node~$3$. In this example, we choose node~$2$. We model the state of the network after the first swap by $G_1(V,E_1)$, see figure~\ref{fig:x_a}. We call the state shared by the connected memories $(15, 25, 8)$ as the main state $\rho^{\star}$. 
    \item For the second swap, we choose any helper node with a quantum memory in the main state $\rho^{\star}$, in this example denoted by node~$6$. The state of the network after the second swap is given by $G_2(V,E_2)$, see figure~\ref{fig:x_b}. The main state is given by the connected component $(15,41,31,24,8)$.
    \item For the third swap, we chose, we chose node 1. For the fourth swap, we chose node 5. We see that for node 5, two of the memories (24,21) belong to the main state. The node performs an $X$ measurement on memory 21. This information was available to the node due to the presence of a four-edged polygon involving the edge (6,21) in the graph. 
    \item We continue this process till the memories left in the main state is that of the consumer nodes to obtain the final state of the network represented by $G_{13}$, see figure~\ref{fig:x_f}. 
\end{itemize}

In Appendix~\ref{appendix:linearity-proof}, we introduce a slightly different modeling of the protocol such that the maximum size of the entangled state during the swapping sequence scales linearly in the grid size $N$, which might be of independent interest.


With each swap, we keep track of the density operators of the entangled states in the network. With the simulation strategy described above, we need to keep track of the density operator of only the main state. We start with the first swap and keep track of the memory nodes and the density operator of the resultant state. For the first swap, we swap $m$ Werner states via a $m$-GHZ projection. We provide the details of the resultant state in Appendix~\ref{appendix:werner-swap}, which might be of independent interest. We observe that the resultant state is a GHZ diagonal state.

Next, we choose a swapping node. If the quantum memory lies at the bottom right corner of a polygon, we perform an $X$ operator on the memory. We give the expression for the resultant state in Appendix~\ref{appendix-X-swap}. This step ensures that we never encounter a situation with two memories of the same node belonging to the state $\rho^{\star}$.

We next perform the swap between the incident bipartite states and the main state. The expression for this part has been implemented in the numerical simulation.  We then obtain the resultant state. This process is repeated till the main state is shared only between the consumer nodes.

\subsection{Importance of $k$-hop communication}\label{subsection: k-level-importance}

In the above modeling of the protocol, we observe that if a quantum memory lies at the bottom right corner of a polygon, we perform the $X$ measurement on the memory node. The information of whether a quantum memory belongs to a $2k+2$-edged polygon requires classical information from the $k^{\textrm{th}}$ neighbor. 

If we had chosen to include this spurious memory in the GHZ swap, then the measurement results of the GHZ swap are no longer Pauli equivalent. That is, we would have to keep track of different measurement results from \textit{each} swap that occurs when two or more memories of the nodes belong to the main state. Note that this is not an artifact of the modeling of the protocol --- rather is an intrinsic property of the protocol introduced in \cite{Patil_2022}. To avoid keeping track of multiple states arising in the network, we allow for $k$-hop communication. For details on the impact of $k$-hop communication, see Appendix~\ref{appen:k-level}. The fraction of the attempts aborted after $k$-hop communication depends on the grid size and the link probability $p$.

\subsection{Numerical results}
\begin{figure*}
\begin{subfigure}{0.5\textwidth}
   \includegraphics[width=\linewidth]{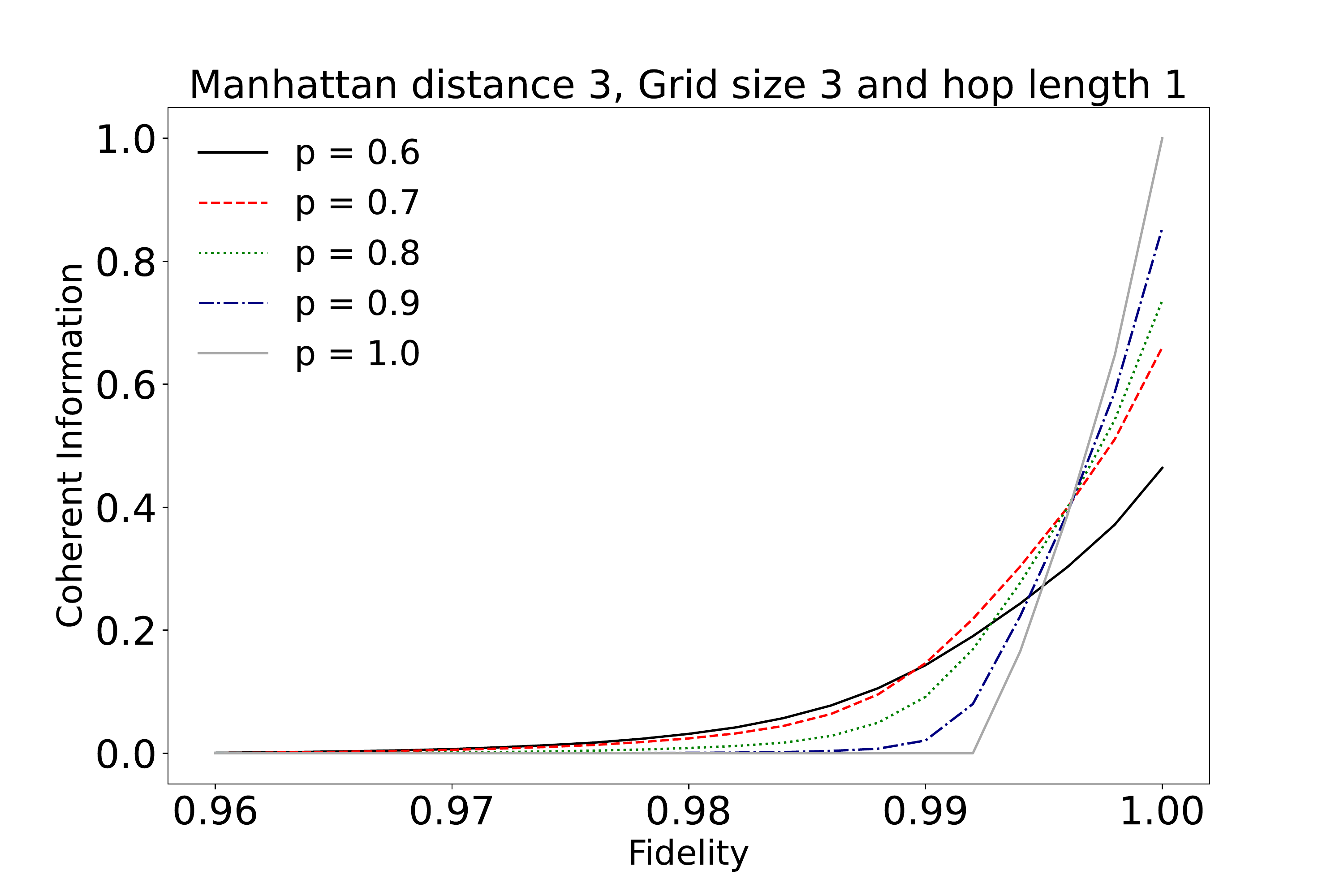}
   \caption{Consumer nodes = (2,3)} \label{fig:man-distance-link_a}
\end{subfigure}
\hspace*{\fill}
\begin{subfigure}{0.5\textwidth}
   \includegraphics[width=\linewidth]{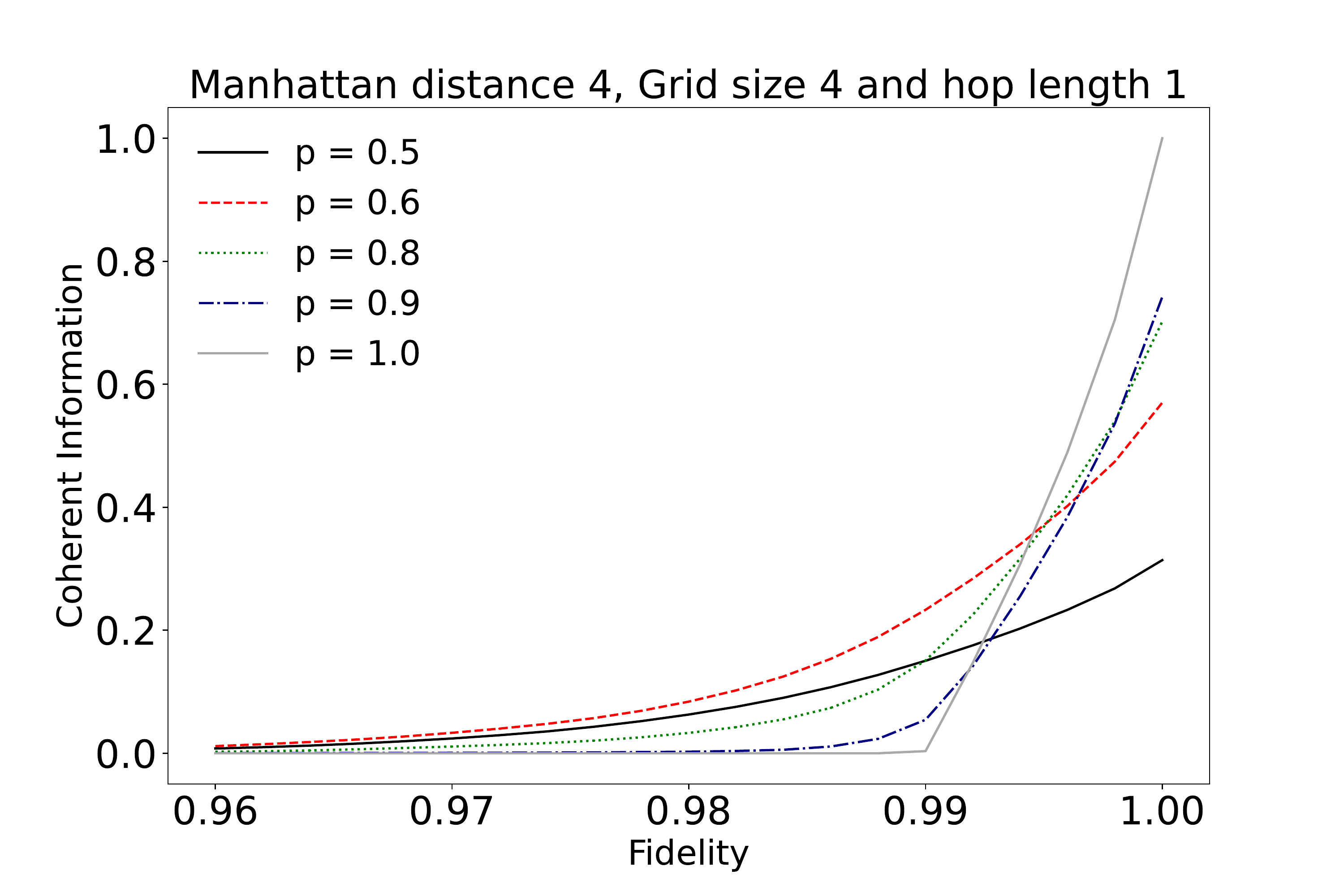}
   \caption{ Consumer nodes = (2,8)} \label{fig:man-distance-link_b}
\end{subfigure}
\hspace*{\fill}
\begin{subfigure}{0.48\textwidth}
   \includegraphics[width=\linewidth]{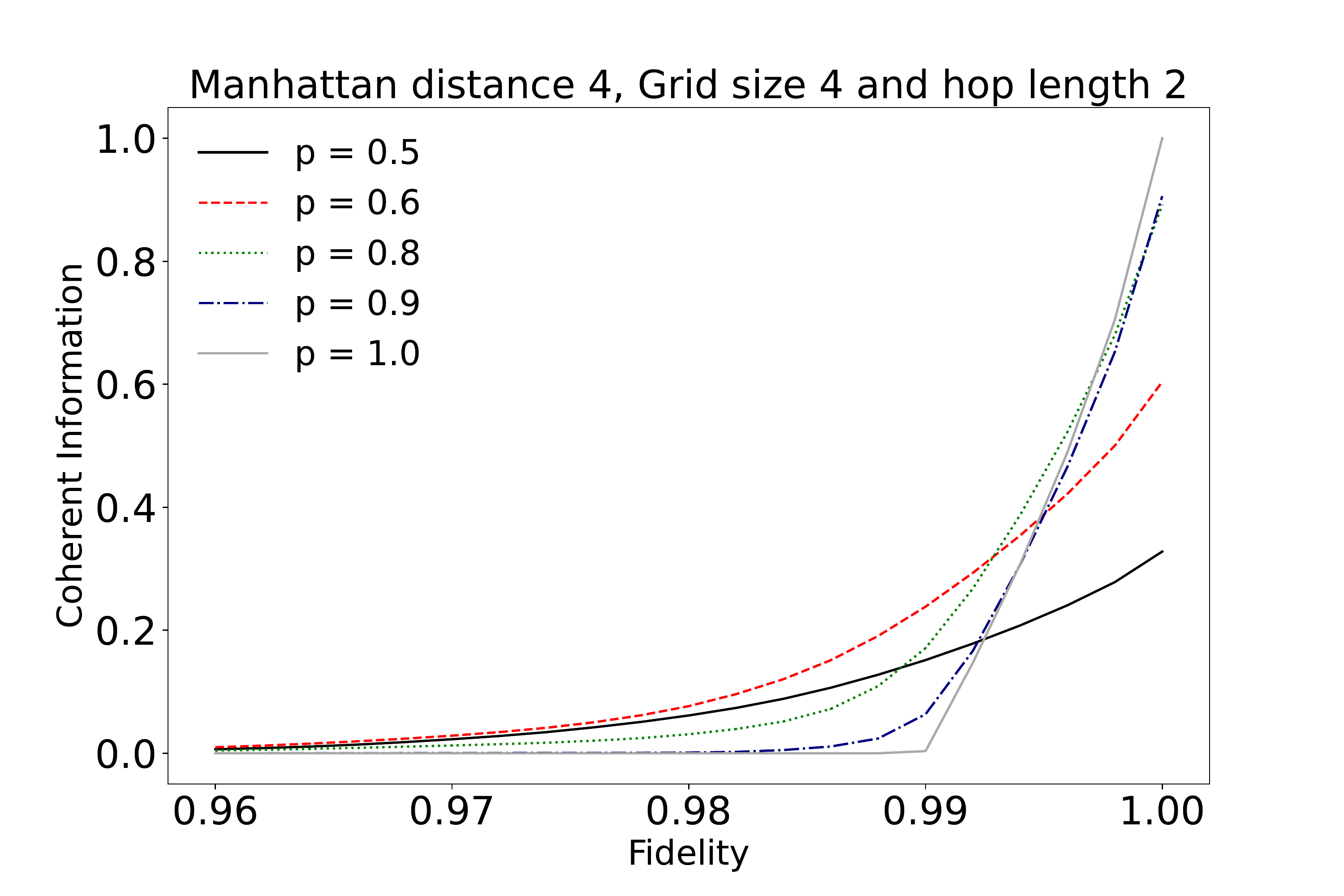}
   \caption{Consumer nodes = (2,8)} \label{fig:man-distance-link_c}
\end{subfigure}
\hspace*{\fill}
\begin{subfigure}{0.48\textwidth}
   \includegraphics[width=\linewidth]{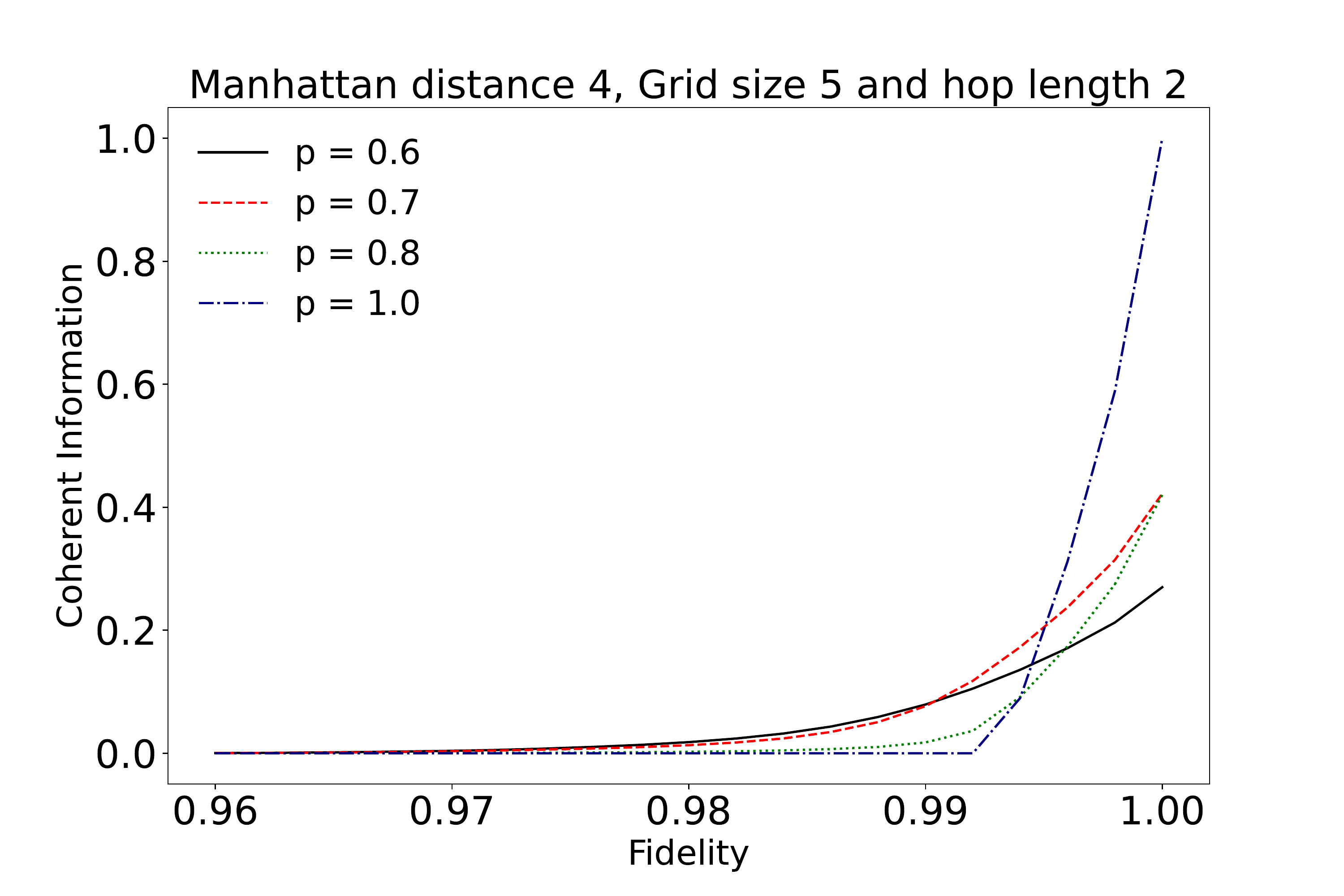}
   \caption{ Consumer nodes = (2,15)} \label{fig:man-distance-link_d}
\end{subfigure}
\hspace*{\fill}
\caption{In this figure, we fix the consumer nodes and the grid size. We plot the coherent information of the state generated from the protocol given above. We observe that the optimal link probability varies depending on the initial link fidelity.}
\label{fig:man-distance-link}
\end{figure*}

In Figure~\ref{fig:man-distance-link}, we choose networks with varying grid sizes, consumer node locations, link probabilities, and $k$. After following the protocol elucidated in this work, we plot the coherent information-- a lower bound on distillable entanglement-- of the state shared between Alice and Bob. We see in the figures~\ref{fig:man-distance-link_a}-\ref{fig:man-distance-link_d}, three competing phenomena --- connectivity of the two consumer nodes, the number of states swapped to connect the nodes, and the number of aborted attempts. Fixing the grid size, as $p$ increases, the probability that the two nodes are connected increases. However, with the increase in $p$, the number of states that are swapped during the protocol also increases. As we observed in Appendix~\ref{appen:k-level}, the fraction of attempts aborted first increases with $p$ and then starts to decrease with $p$. An increase in the number of noisy swapped states implies that more noise is pushed into the final state. We thus observe that, for low fidelities, the coherent information is higher for low link probability. Note that this is not a one-to-one relation, we still need to balance the lack of connectivity that arises from low link probabilities. For example, link probability $p=0.6$ reduces the number of states in the network. For lower fidelities, we favor a low number of states in the network, so that the output state has less noise. However, when the fidelities are higher, the numerics favor a higher number of states in the network, implying the optimal link probability is higher. This is because the final state resulting from a larger number of swaps of high initial fidelity is not very noisy. 

\begin{figure*}[hbt!]\label{fig:fixed-manhattan}
\begin{subfigure}{0.5\textwidth}
   \includegraphics[width=\linewidth]{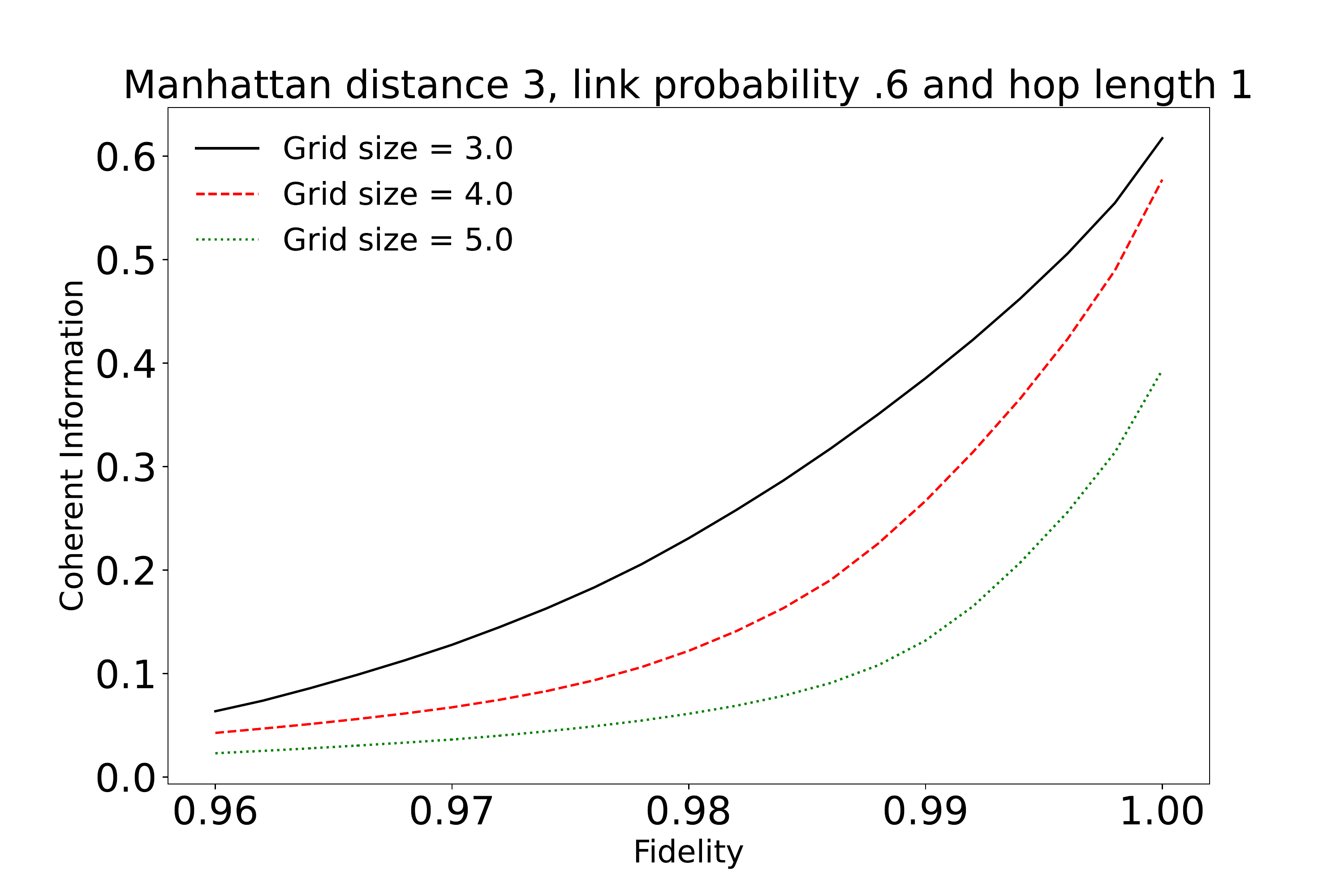}
   \caption{Consumer nodes (2,3) for Grid size 3} \label{fig:pob_a}
\end{subfigure}
\hspace*{\fill}
\begin{subfigure}{0.5\textwidth}
   \includegraphics[width=\linewidth]{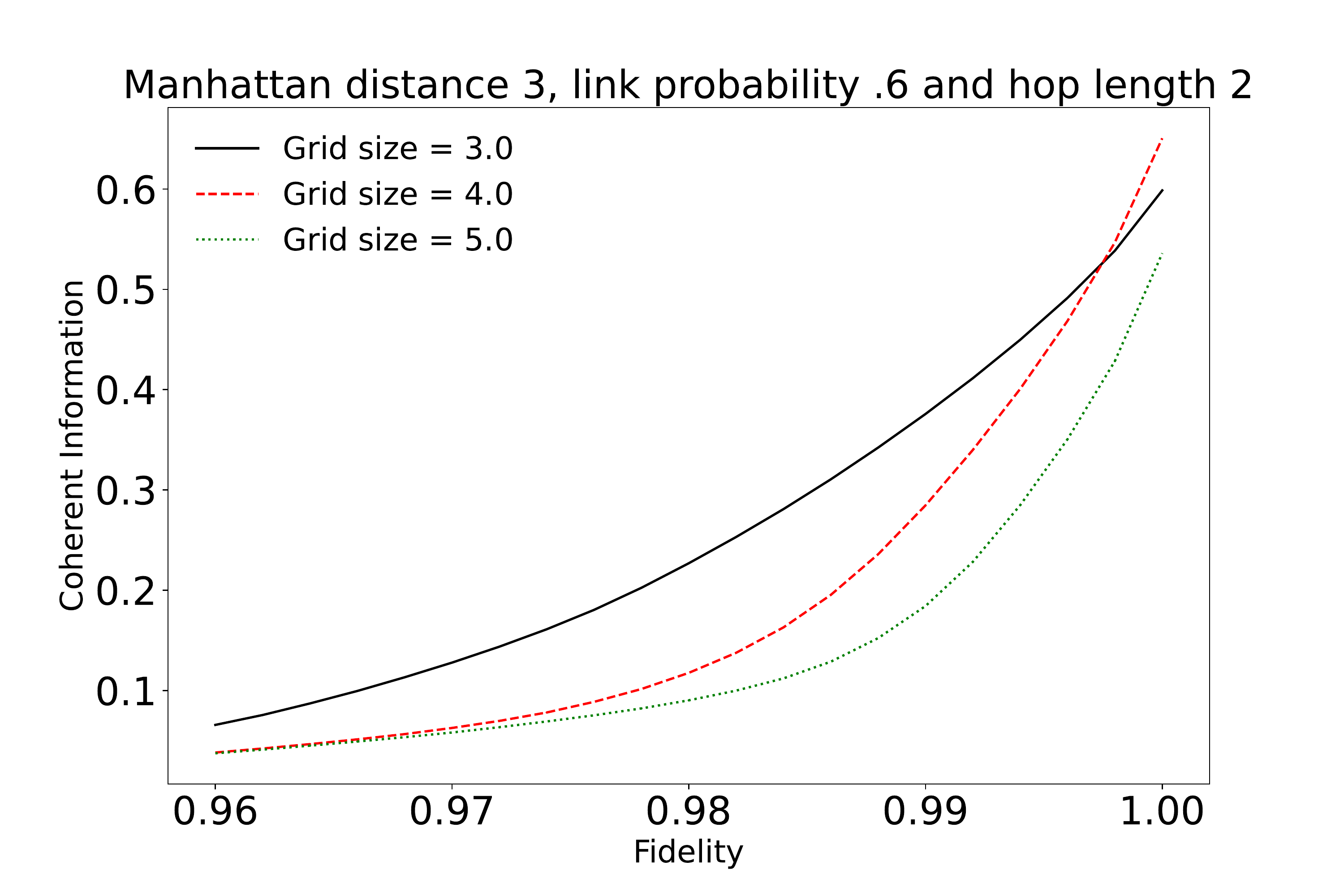}
   \caption{Consumer nodes (2,3) for Grid size 3} \label{fig:pob_b}
\end{subfigure}
\hspace*{\fill}
\begin{subfigure}{0.45\textwidth}
   \includegraphics[width=\linewidth]{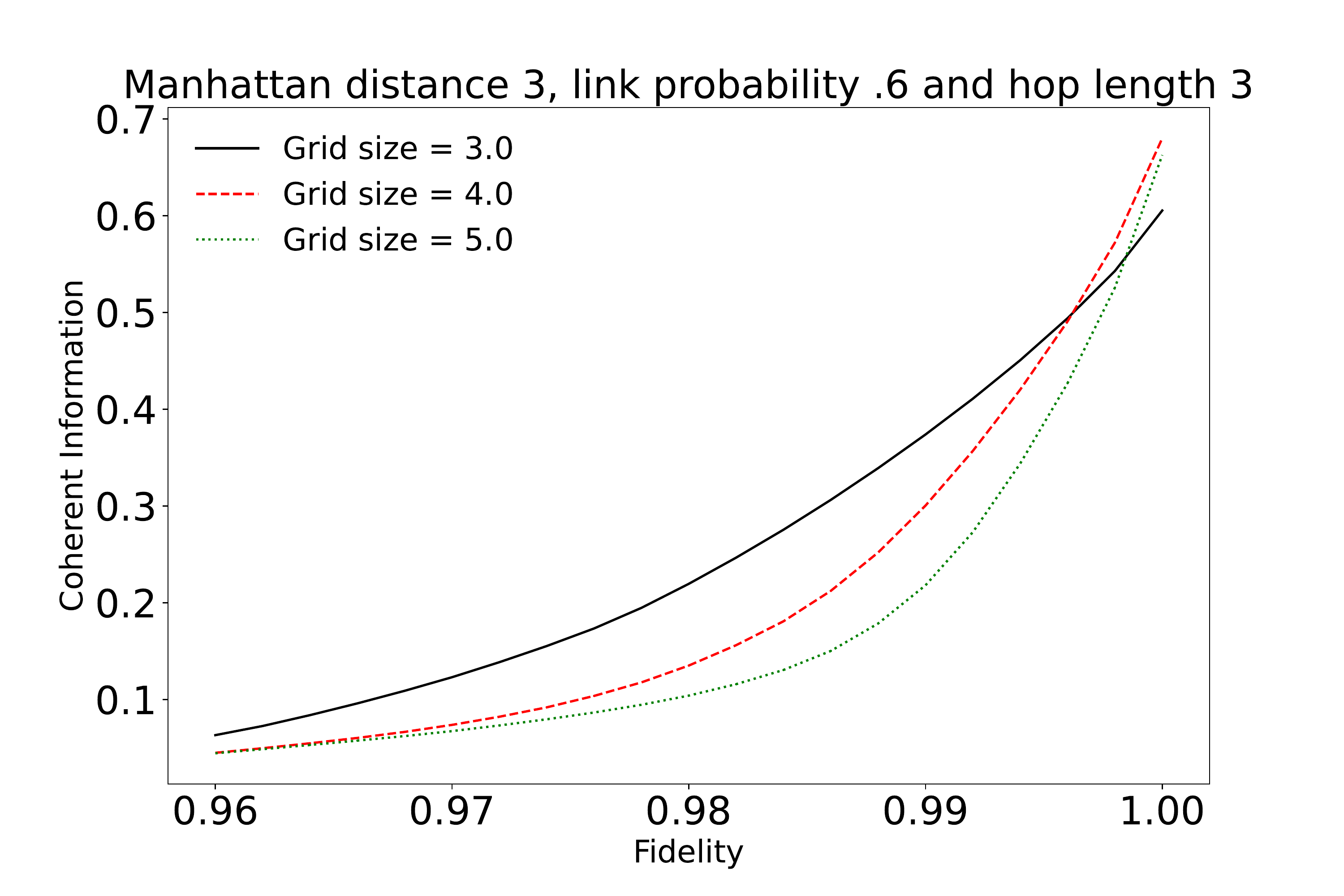}
   \caption{Consumer nodes (2,6) for Grid size 3} \label{fig:prob_c}
\end{subfigure}
\hspace*{\fill}
\begin{subfigure}{0.45\textwidth}
   \includegraphics[width=\linewidth]{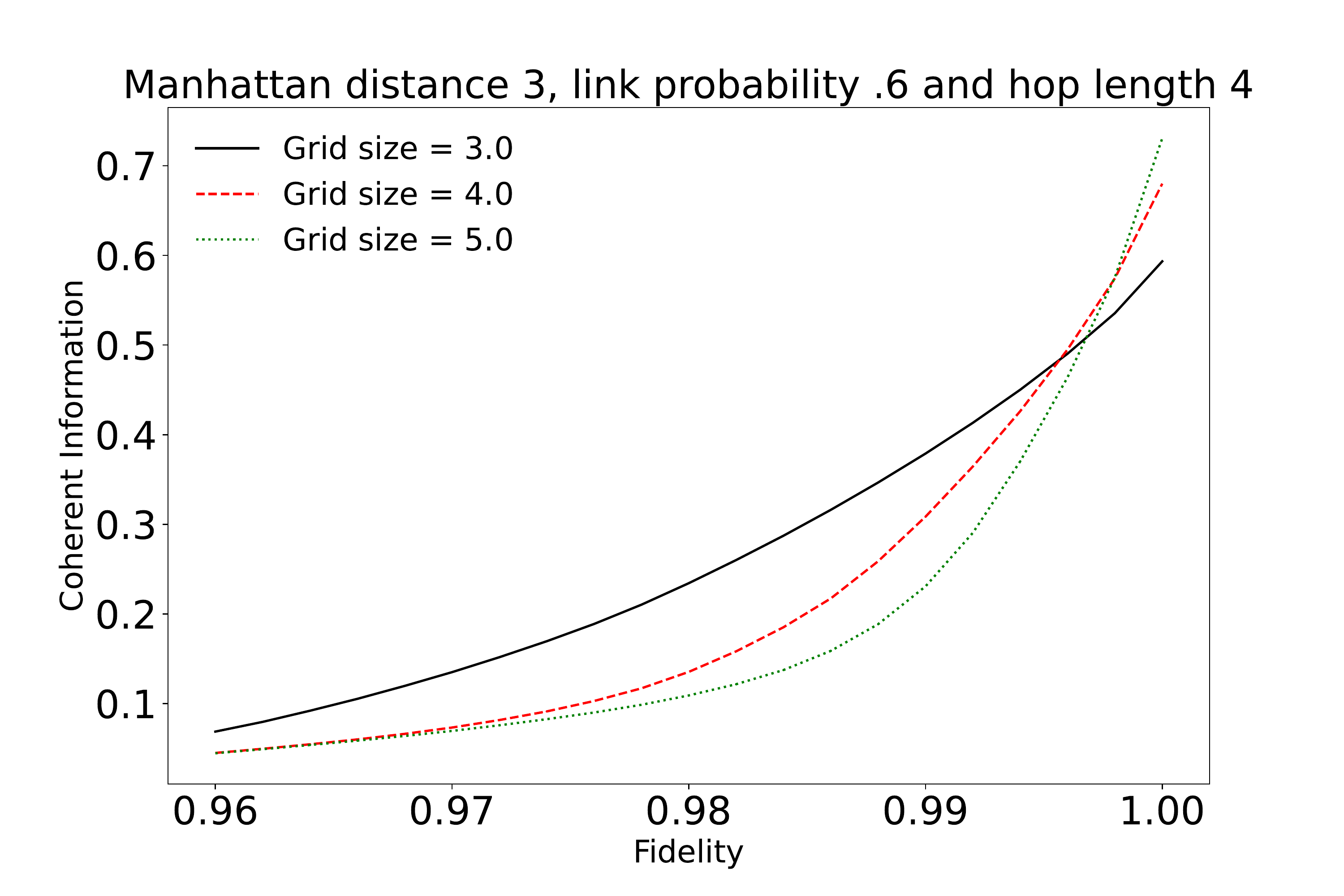}
   \caption{Consumer nodes (2,6) for Grid size 3} \label{fig:prob_d}
\end{subfigure}
\hspace*{\fill}
\caption{In these figures, we fix the consumer node locations and the link probability. We vary the grid size and the hop length. The consumer node locations are given with respect to grid size 3. }
\end{figure*}

In Figures~\ref{fig:pob_a}-\ref{fig:prob_d}, we fix the Alice and Bob nodes, thereby fixing the Manhattan distance, link probability, $k$-hop communication, and modifying the grid size. We see that for low fidelities, low grid sizes are favorable, irrespective of $k$. While for the higher initial fidelities, bigger grid sizes are favorable. This can be explained by larger grid sizes providing higher connectivity. We also note the impact of $k$ on coherent information. For higher grid sizes, the fraction of aborted attempts is higher when $k$ is low. Thus, we see that for $k=1$, even for high fidelities, the rate is higher for grid size 3. This changes as $k$ increases. 

We explore this region's choice with respect to initial fidelities further in the next section. We observe that the distance-independent entanglement generation rate observed in \cite{Patil_2022} is extremely fragile with respect to initial fidelities. Increasing the grid size, which is crucial for the percolation or phase transition of connectivity in the graph, can be detrimental to the coherent information of the final state for non-perfect fidelities. 

\subsection{ Optimized region selection}\label{sec:region}
\begin{figure}\label{fig:regions_1}
\begin{centering}
\begin{subfigure}{0.25\textwidth}
   \includegraphics[width=\linewidth]{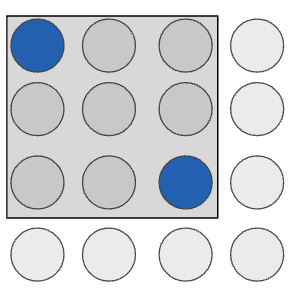}
   \caption{$R_0$ for grid size four} \label{fig:region_1_grid_four}
\end{subfigure}
\hspace*{\fill}
\begin{subfigure}{0.25\textwidth}
   \includegraphics[width=\linewidth]{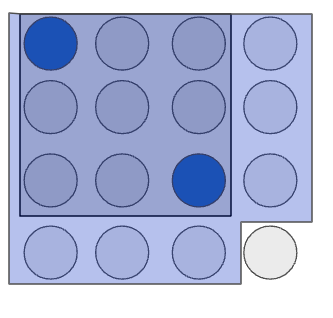}
   \caption{$R_1$ for grid size four} \label{fig:region_2_grid_four}
\end{subfigure}
\hspace*{\fill}
\caption{Example of region selection for grid size 4. The blue nodes denote the position of Alice and Bob. }
\end{centering}
\end{figure}

In this section, we fix the node location of Alice and Bob, and we optimize over various network sizes and region selections in the network. Given an initial graph $G$ of grid size $N$, we divide $G$ as follows. First, consider the set of all nodes in the set of all shortest paths from Alice and Bob. These nodes constitute region $R_0$. We then consider the set of all nodes in the set of the first and the second shortest paths. This constitutes the region $R_1$. We will continue this process till we have covered the whole network. The division based on regions allows for selecting the nodes based on their importance in the protocol. We call the nodes not used during the protocol idle nodes. For an example of region selection, see Figure~\ref{fig:region_1_grid_four} and \ref{fig:region_2_grid_four}.

\begin{figure}\label{fig:regions}
\begin{subfigure}{0.5\textwidth}
   \includegraphics[width=\linewidth]{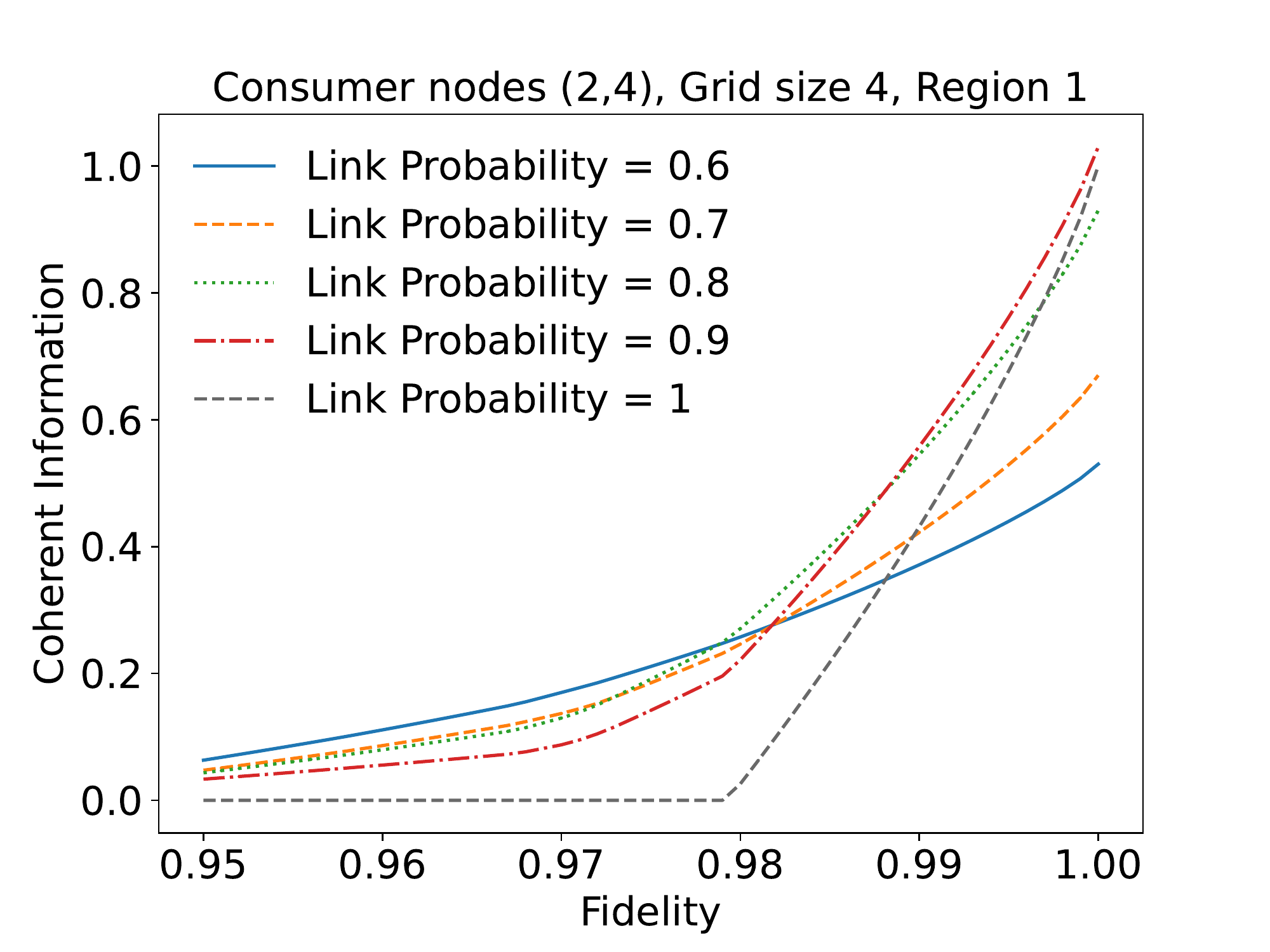}
   \caption{Region one for grid size four} \label{fig:region_a}
\end{subfigure}
\hspace*{\fill}
\begin{subfigure}{0.5\textwidth}
   \includegraphics[width=\linewidth]{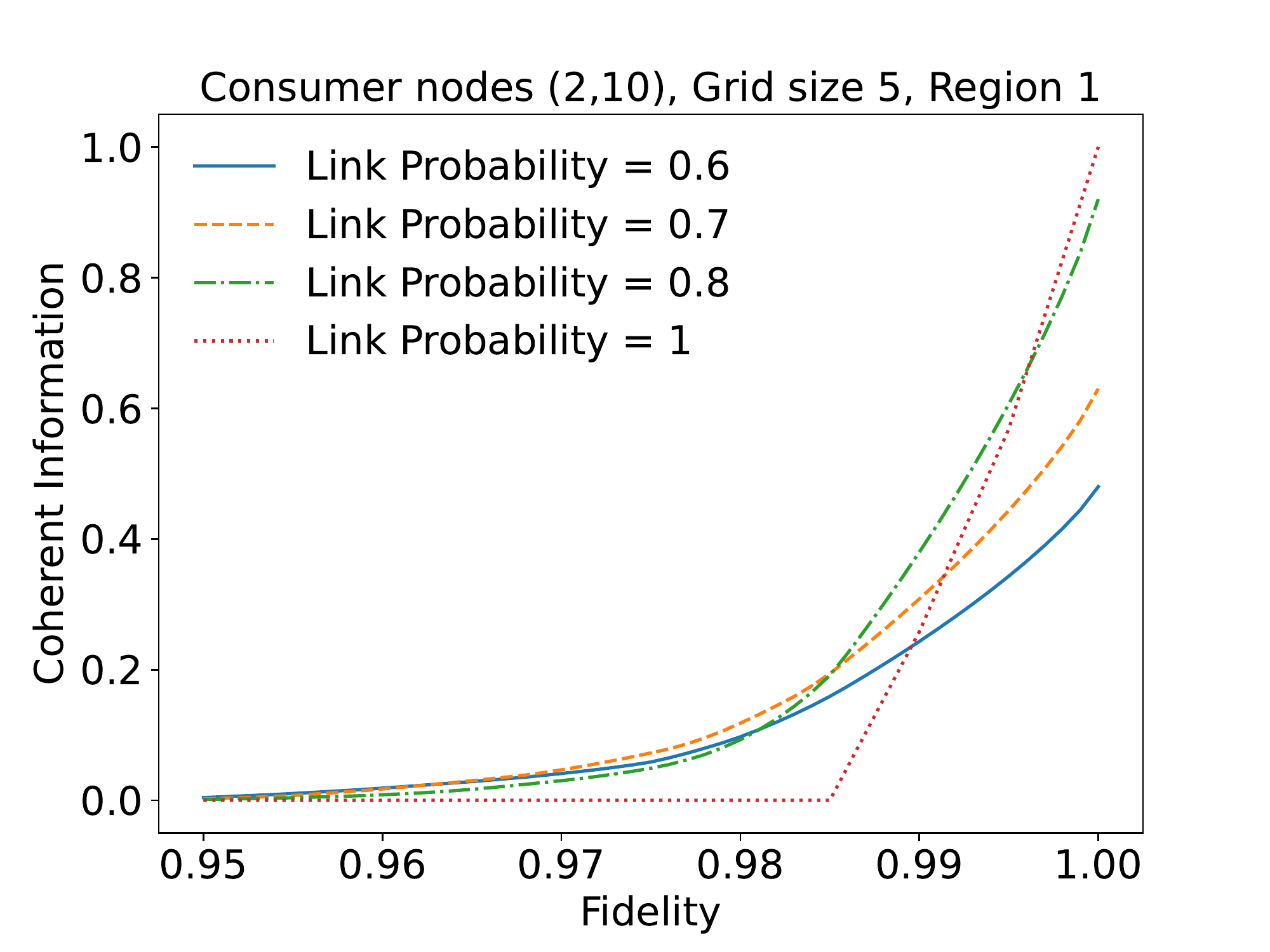}
   \caption{Region one for grid size five} \label{fig:region_b}
\end{subfigure}
\hspace*{\fill}
\caption{We fix the regions of the network based on the consumer nodes. We plot the coherent information over various link probabilities. }
\end{figure}

In Figures~\ref{fig:region_a}-\ref{fig:region_b}, we fix the region considered based on the location of the consumer nodes. We allow for global communication, so as to depict the effect of region selection.  The region's optimal link probability depends on the initial fidelity. As the fidelity increases, the optimal link probability tends to increase. We note here that the choice of the region can be further optimized, i.e., we might subdivide region one into several subregions depending on the number of shortest paths considered. We do not perform a fine-grained optimization on the choice of regions in this work. 

\begin{figure*}

\begin{subfigure}{0.5\textwidth}
   \includegraphics[width=\linewidth]{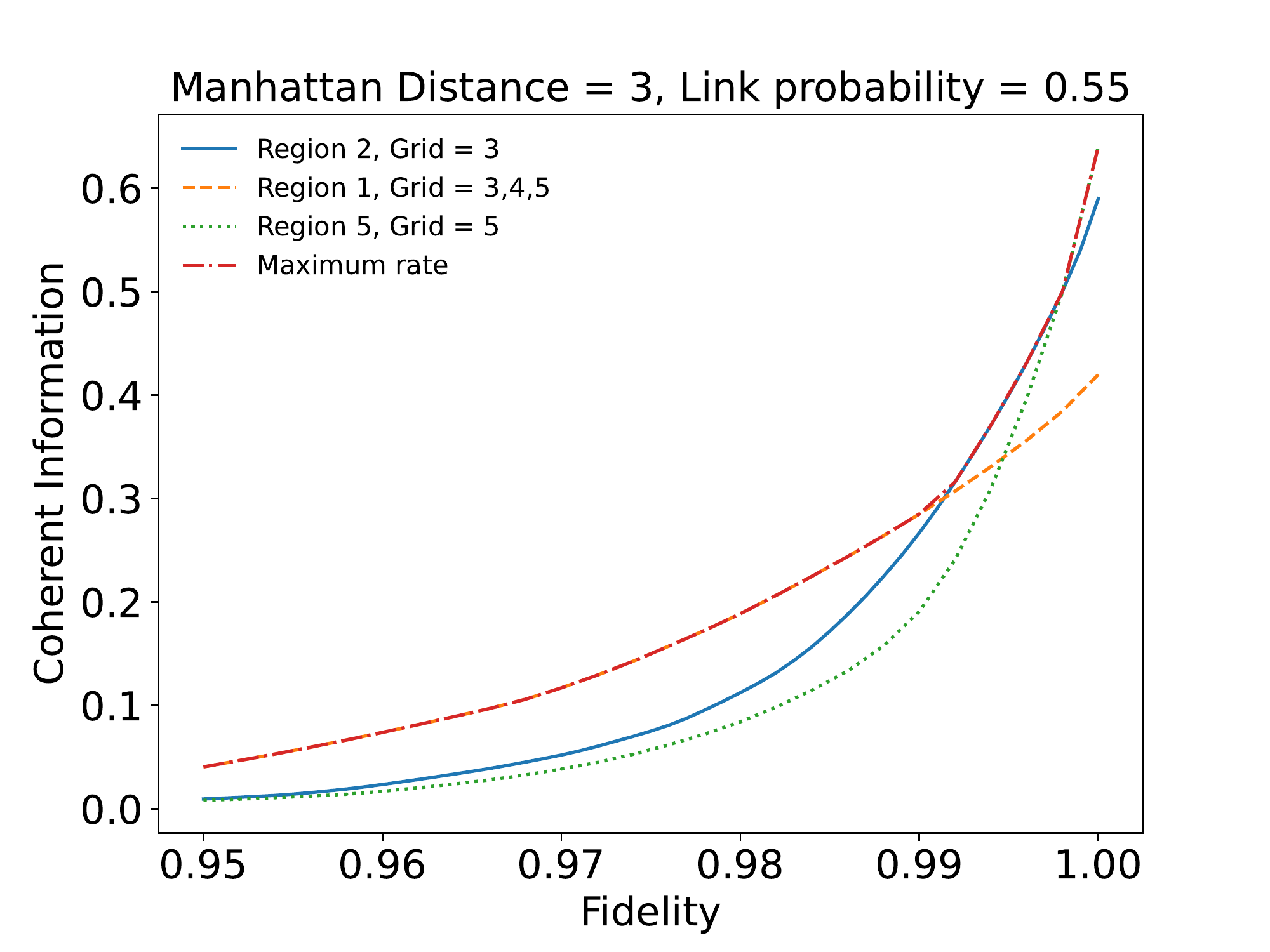}
   \caption{Rate envelopes for Manhattan distance three and link probability 0.55.} \label{fig:region1_c}
\end{subfigure}
\hspace*{\fill}
\begin{subfigure}{0.5\textwidth}
   \includegraphics[width=\linewidth]{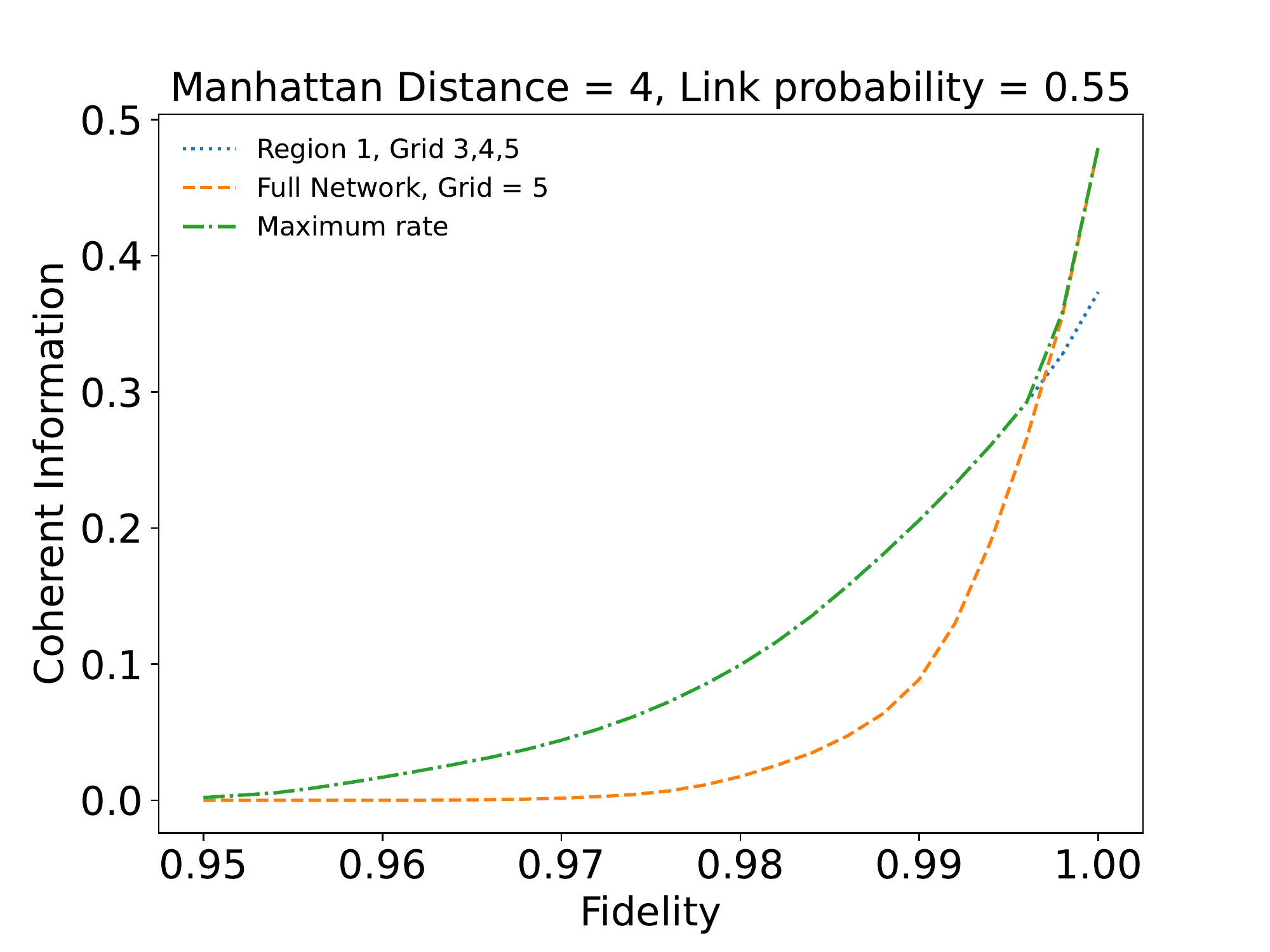}
   \caption{Rate envelopes for Manhattan distance four and link probability 0.55.} \label{fig:region1_d}
\end{subfigure}
\hspace*{\fill}
\begin{subfigure}{0.45\textwidth}
   \includegraphics[width=\linewidth]{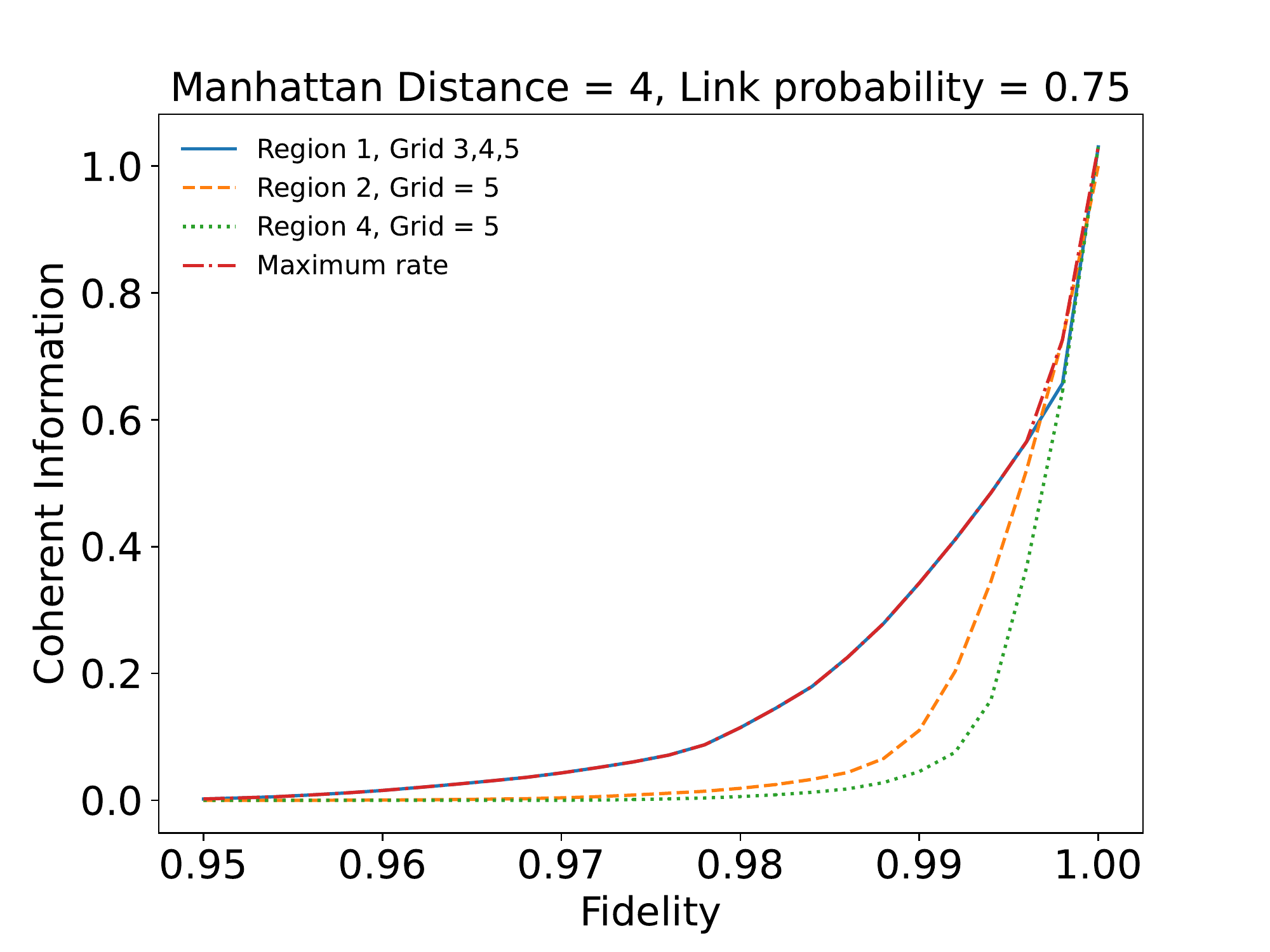}
   \caption{Rate envelopes for Manhattan distance four and link probability 0.75.} \label{fig:region1_e}
\end{subfigure}
\hspace*{\fill}
\begin{subfigure}{0.45\textwidth}
   \includegraphics[width=\linewidth]{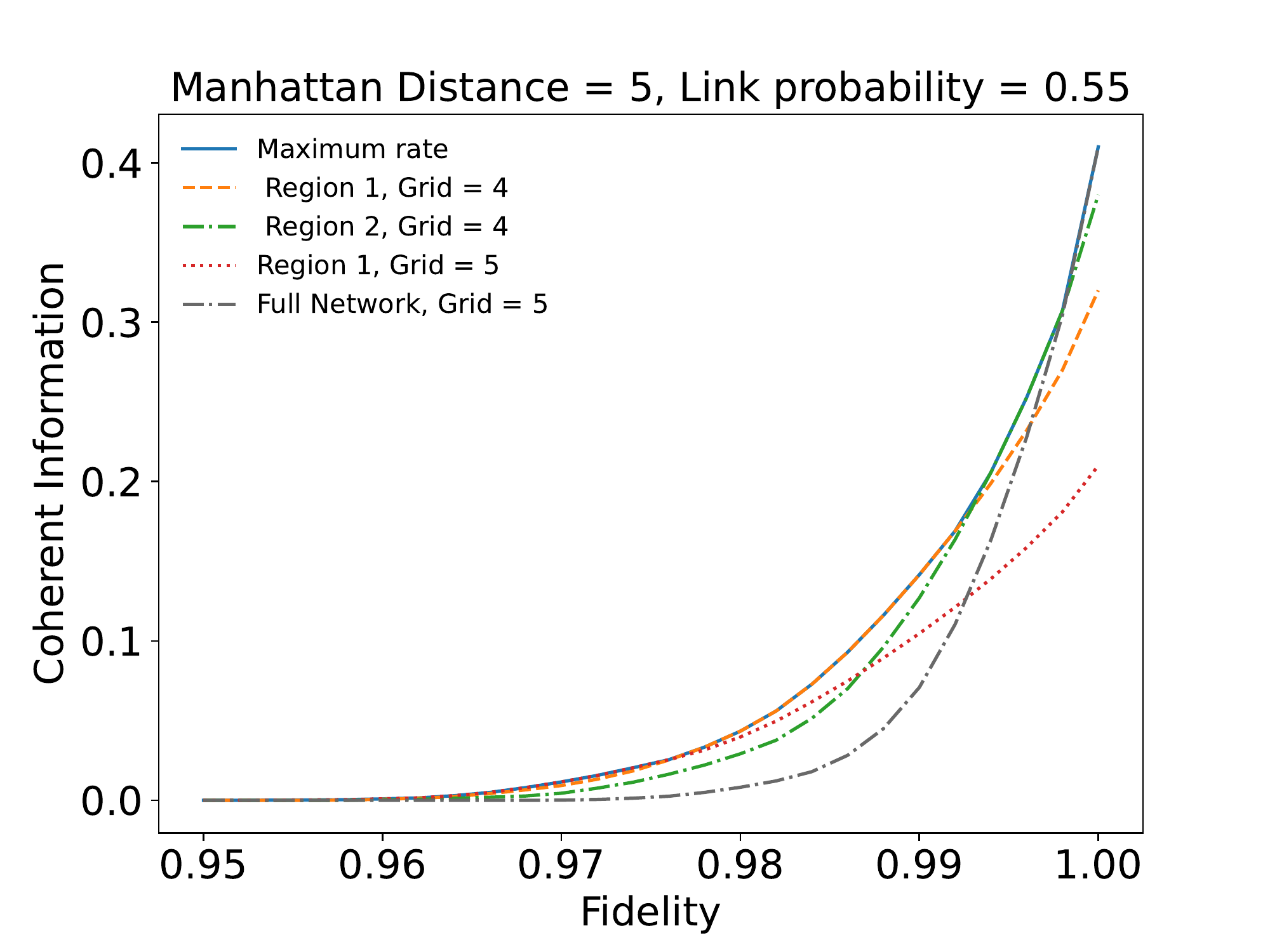}
   \caption{Rate envelopes for Manhattan distance five and link probability 0.55.} \label{fig:region1_f}
\end{subfigure}
\caption{In this figure, we plot the rate envelopes of the protocol introduced in this work for different Manhattan distances and link probabilities.}
\label{fig:region-size}

\end{figure*}

In Figure~\ref{fig:region-size}, we fix the link probabilities, the consumer node location, and plot rate envelopes for the protocol mentioned above. We observe that the optimal region is a function of the initial fidelity and the link probability. If the link probability is high and initial fidelity is low, it is better to consider smaller regions. 

\begin{figure}
    \centering
    \includegraphics[width = \linewidth]{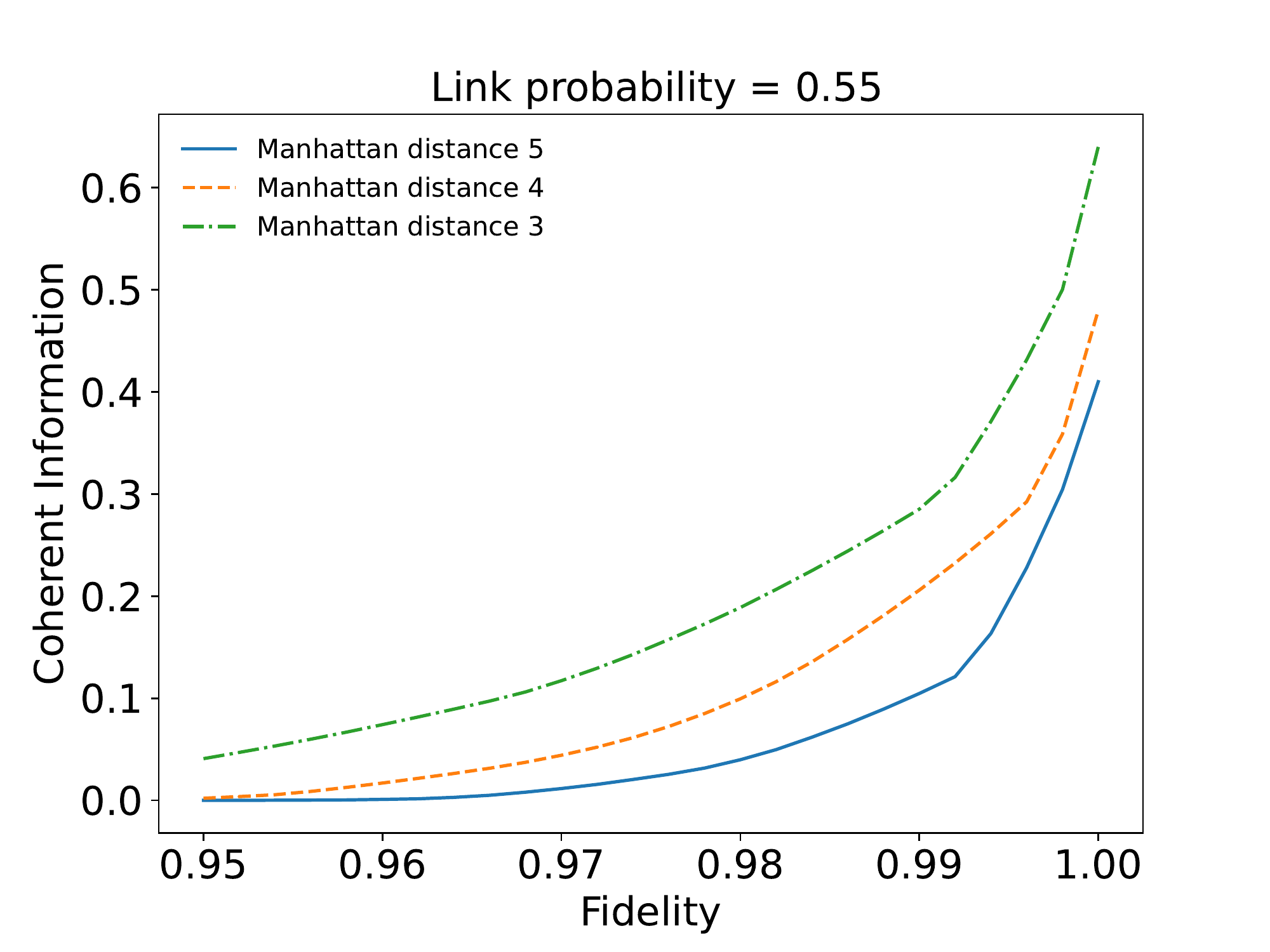}
    \caption{In this figure, we fix the Manhattan distance and optimize over various protocols given above. We see that rate decays with the distance.}
    \label{fig:dist-depen}
\end{figure}

In Figure~\ref{fig:dist-depen}, we fix the Manhattan distance and link probability and plot the achievable distillation rates for the protocol described above. 

\section{Link level distillation in grid networks}\label{sec:distillation-grid-network}

This section explores a slightly different version of the aforementioned protocol. In this protocol, we allow for link-level distillation. Entanglement distillation allows for the conversion of $n$ number of low-fidelity Bell pairs into $l$ number of high-fidelity Bell pairs, where $l<n$ via local operations with probability $p$. Since the initial idea in \cite{Benett1996,Bennett1996-2,Deutsch1996}, other works such as \cite{Hans-thesis,Krastanov2019} have developed and worked on new distillation schemes. This work considers the distillation scheme introduced in \cite{Benett1996}, recalled in Appendix~\ref{appen-distillation}. Optimizing for different distillation schemes would be an interesting direction to explore.

\begin{figure}
\begin{subfigure}{0.4\textwidth}
   \includegraphics[width=\linewidth]{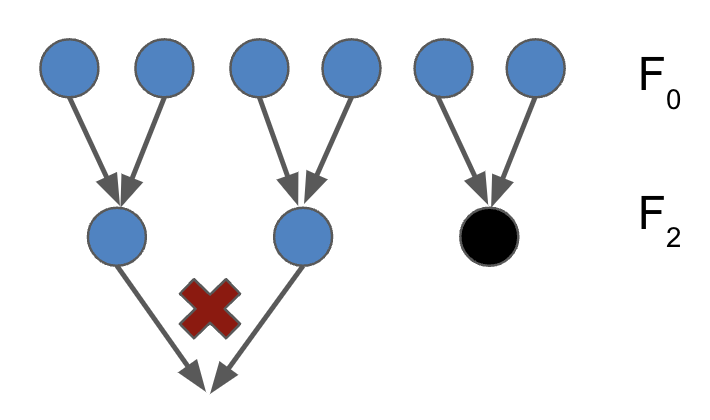}
   \caption{We assume that six entangled states are heralded. We perform $2\rightarrow 1$ entanglement purification on these entangled states in the first step. Let us assume that all the attempts are successful. In the second part, we use the outputs of two of the successes and perform another $2\rightarrow 1$ distillation. If this attempt of distillation fails, we use the leftover pair with fidelity $F_2$. } 
\end{subfigure}
\begin{subfigure}{0.4\textwidth}
   \includegraphics[width=\linewidth]{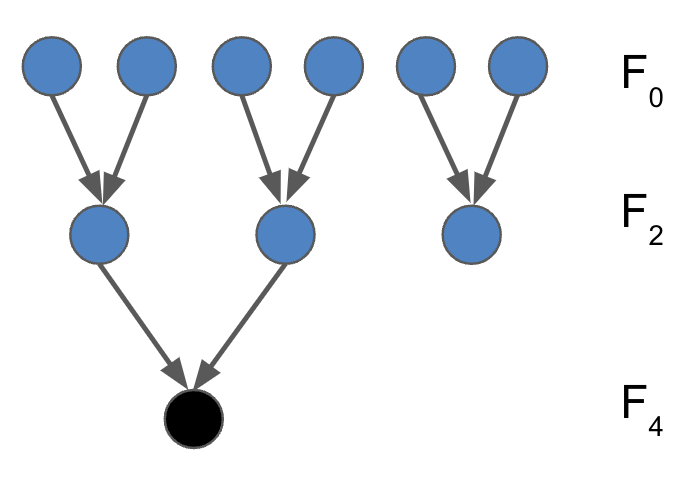}
   \caption{We assume that six entangled states are heralded. Let us assume that all the distillation attempts are successful. In the second part, we use the outputs of two of the successes and perform another $2\rightarrow 1$ distillation. If this attempt at distillation succeeds, we use the newly generated pair with fidelity $F_4$. } 
\end{subfigure}
\begin{subfigure}{0.4\textwidth}
   \includegraphics[width=\linewidth]{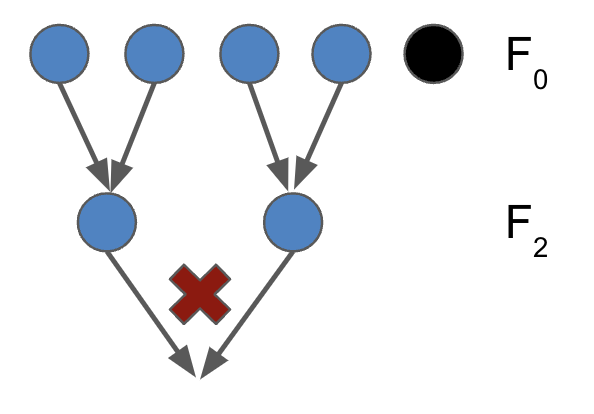}
   \caption{We assume that five entangled states are heralded. Let us assume that all the distillation attempts are successful. In the second part, we use the outputs of two successes and perform another $2\rightarrow 1$ distillation. If this attempt at distillation fails, we use the leftover node with fidelity $F_0$. } 
\end{subfigure}
\caption{Representative situations of distillation protocols}
\label{fig:purification-example}
\end{figure}

In this protocol, we wait for $t$ time steps for the entanglement distribution part of the protocol. Each memory gets $t$ trials and attempts to generate entangled pairs between the neighboring nodes. The entanglement distribution is assumed to be probabilistic with link probability $p$. Let us suppose that two neighboring memories share $l_1$ pairs. Then, the memories repeatedly perform $2\rightarrow 1$ distillation on states with the same fidelity. If a particular distillation step fails, the nodes could still share a low-fidelity pair. The protocol makes use of the better fidelity state. A representative situation in a protocol is given in Figure~\ref{fig:purification-example}. 

In the next time step, the nodes perform entanglement swapping. We can still model the modified protocol by a graph $G(V,E)$, where $V$ is the set of vertices and $E$ is the edges. However, due to the distillation procedure, the link probability changes to $p_1$, and we have different fidelities for the link level states existing in the network. For example, when $t=2$, the nodes can have fidelity $F$ if only one pair was heralded or $F_2$ if two pairs were heralded and the $2\rightarrow 1$ distillation succeeded.  For further details, see Appendix~\ref{appen-distillation}.

\begin{figure*}

\begin{subfigure}{0.5\textwidth}
   \includegraphics[width=\linewidth]{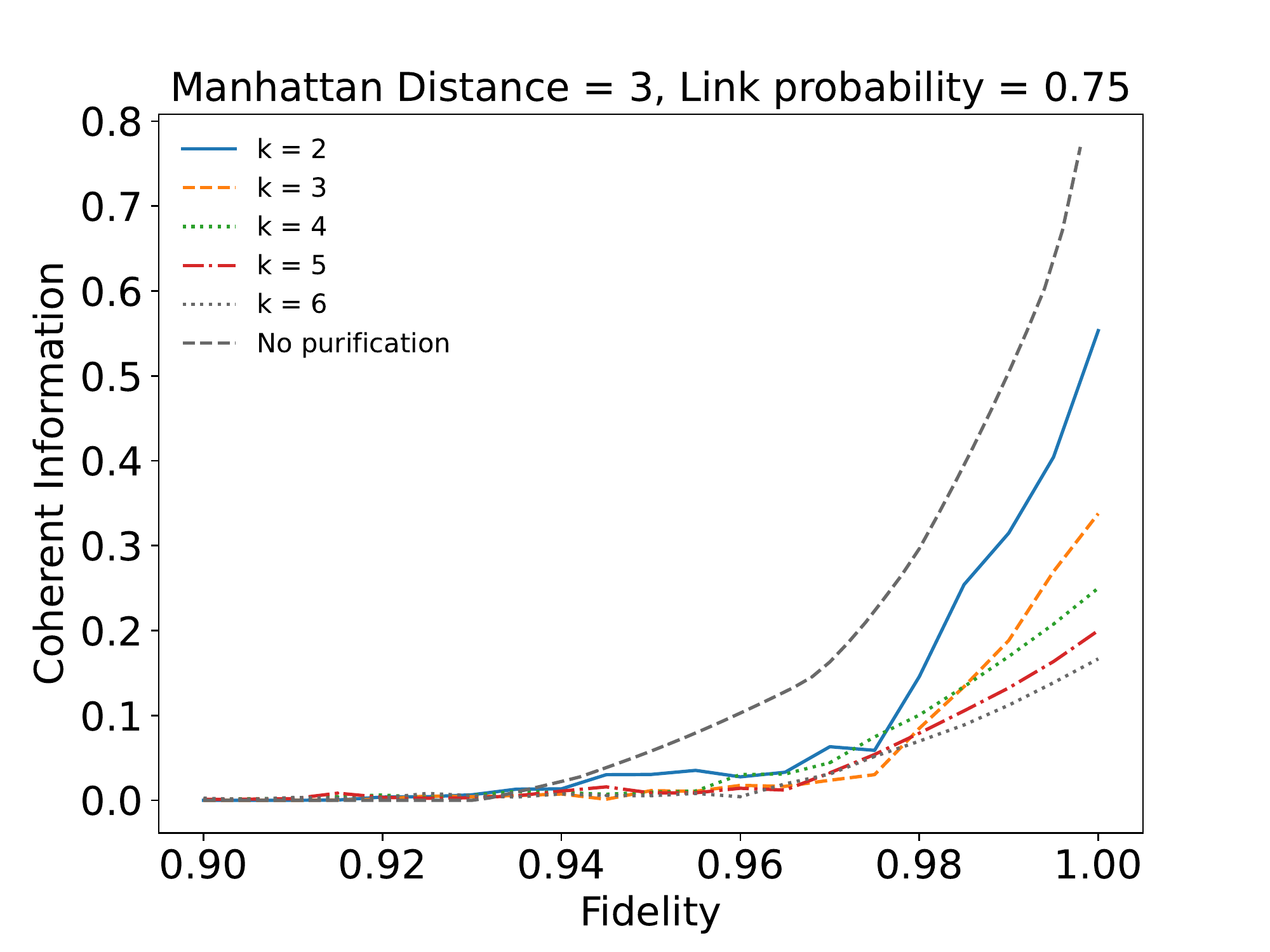}
   \caption{Grid size three, Manhattan distance three, Link prob = 0.75} \label{fig:region_c}
\end{subfigure}
\hspace*{\fill}
\begin{subfigure}{0.5\textwidth}
   \includegraphics[width=\linewidth]{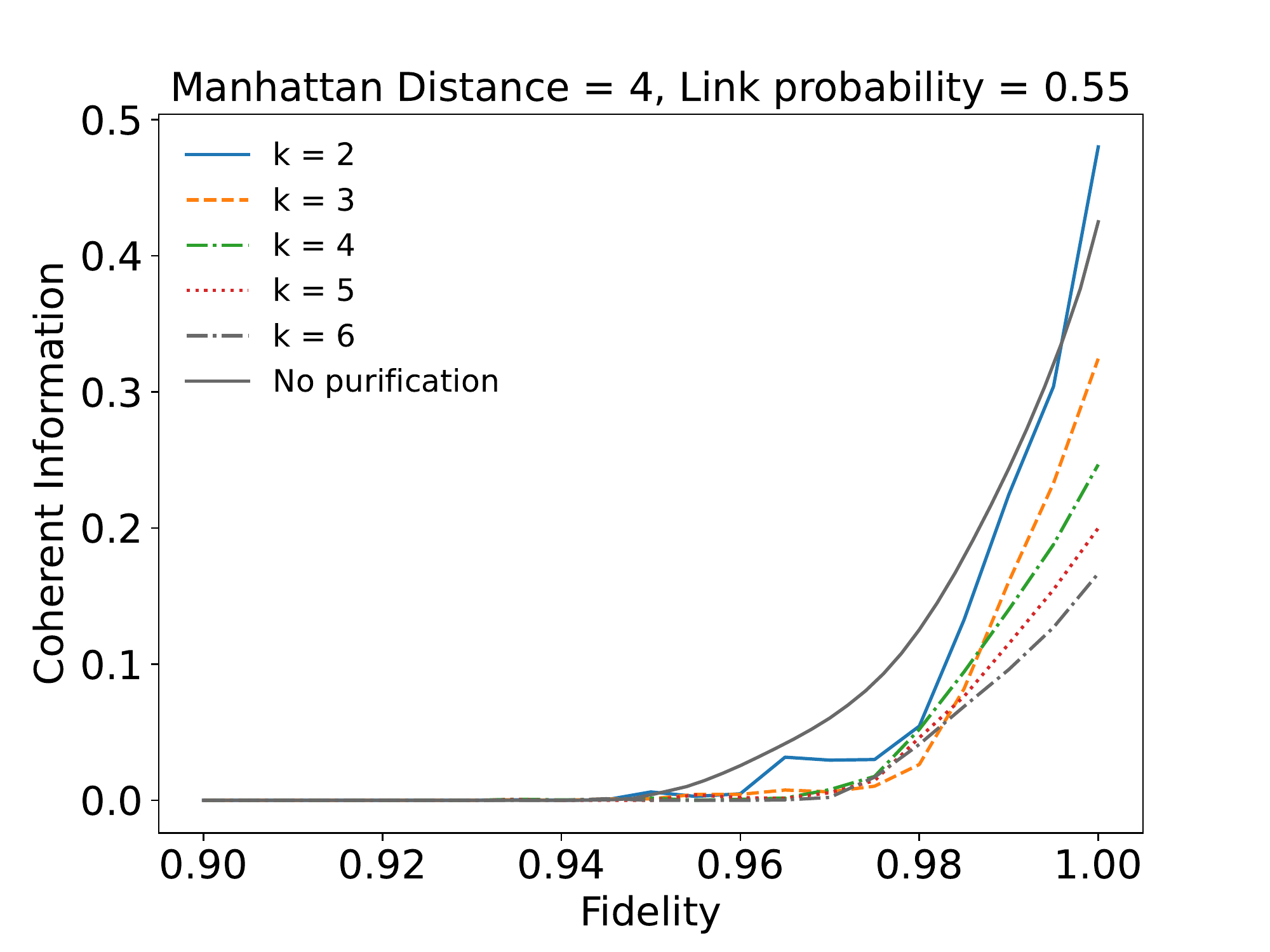}
   \caption{Grid size three, Manhattan distance four, Link prob = 0.55} \label{fig:region_d1}
\end{subfigure}
\hspace*{\fill}
\begin{subfigure}{0.45\textwidth}
   \includegraphics[width=\linewidth]{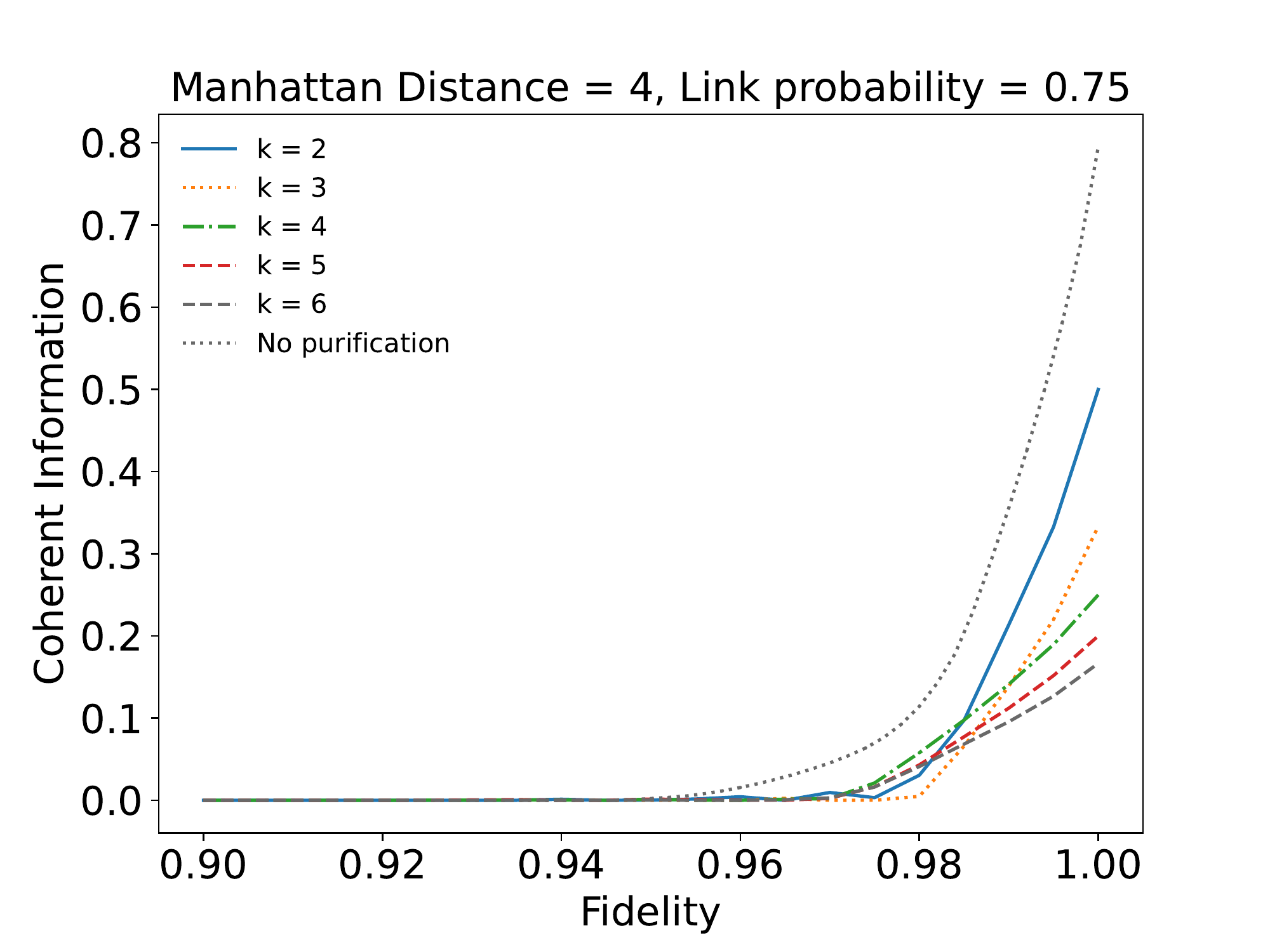}
   \caption{Grid size three, Manhattan distance four, Link prob = 0.75} \label{fig:region_e}
\end{subfigure}
\hspace*{\fill}
\begin{subfigure}{0.45\textwidth}
   \includegraphics[width=\linewidth]{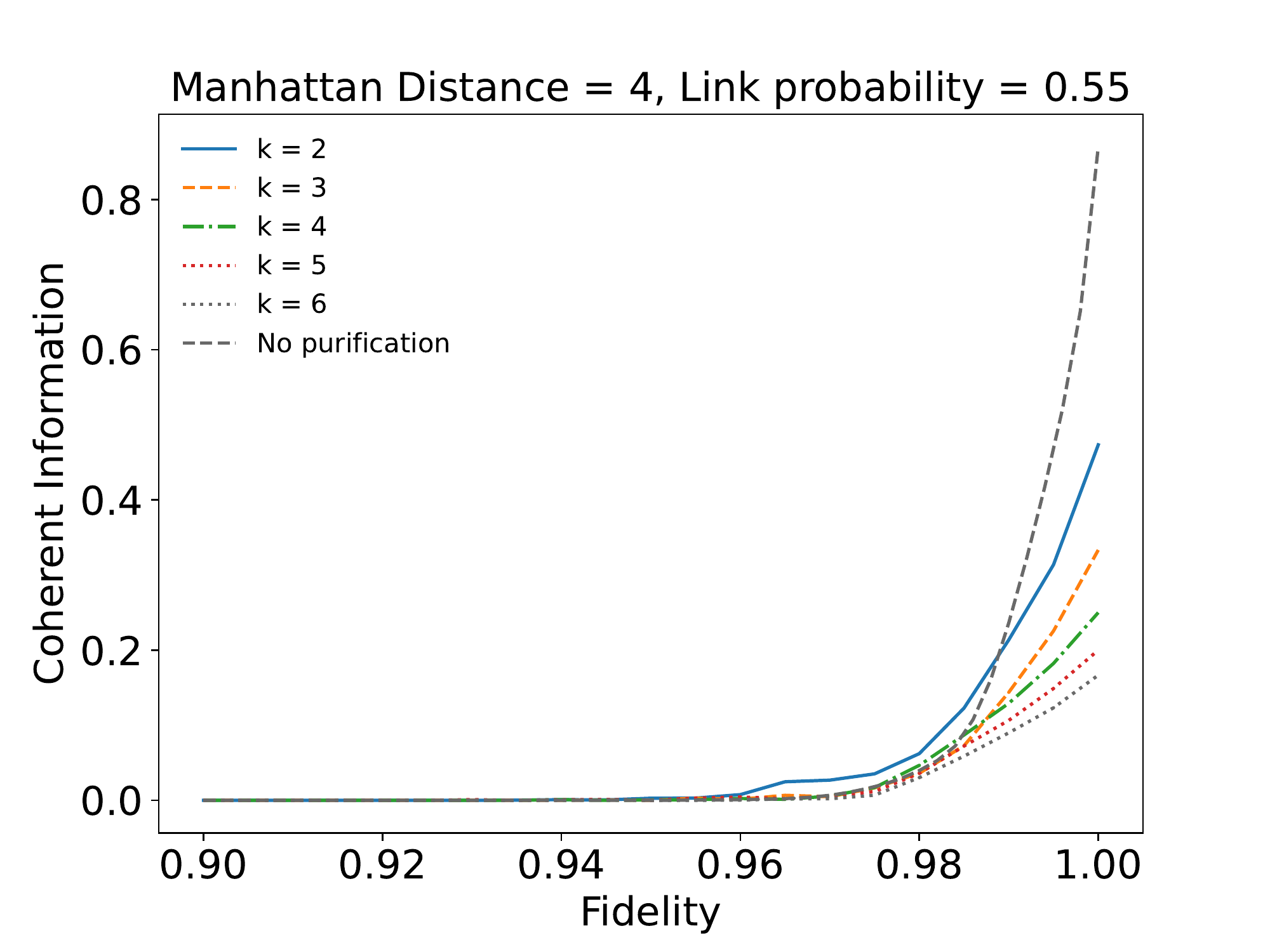}
   \caption{Grid size four, Manhattan distance five, Link prob = 0.75} \label{fig:region_f}
\end{subfigure}

\caption{We plot the coherent information for the modified protocol elucidated in Section~\ref{sec:distillation-grid-network}. }
\label{fig:distillation-rates}
\end{figure*}

For numerical simulation, we use the virtual protocol introduced above. We keep track of the main state $\rho^\star$, and keep track of the fidelities of the bipartite states used for swapping. We allow for global communication to depict the effect of link level distillation. For the first step, we swap $m$ Werner state of different fidelity via $m$-GHZ measurements. We give the form for the resultant state in Appendix~\ref{appendix:diff-fidelities}. 
The entanglement distillation rate is given as $I(A;B)/t$, where $I(A;B)$ is the coherent information, and $t$ is the number of time steps for the entanglement distribution. We find that the improvements in the rates obtained from link-level distillation circuits, in general, are minute and appear only for very low fidelities, as depicted in Figure~\ref{fig:distillation-rates}. When link probabilities are low, waiting for $t=2$ time steps can be helpful.

\section{Conclusion}
In this work, we have analyzed entanglement distribution in grid networks in the presence of noise. Our work highlights the need for carefully considering noise for entanglement generation rates. We briefly also explored the addition of link-level entanglement distillation to the protocol. One interesting future direction would be to move beyond the BBPSSW distillation scheme and explore other distillation schemes such as \cite{Krastanov2019}. A further extension would be to incorporate multi-hop entanglement distillation into this protocol. While our work rules out distance independence in entanglement generation rates with link-level non-perfect entanglement distillation, it is still to be determined if distance independence in entanglement generation rates can be recovered with multi-hop entanglement distillation.  

We have explored the routing protocol for entanglement generation in grid networks, considering the initial fidelity. We observe that the protocols should carefully account for the number of swaps used. The number of swaps or repeaters used in the protocol increases the connectivity of the two parties. However, many swaps between Werner states decrease the final states' quality. Thus, the protocol for entanglement generation must carefully balance the two opposing factors. We explore the interplay carefully in this work. We also give an analytical expression for swapping $n$ Werner states with a $\textrm{GHZ}(n)$ swap. We also give a method of modeling the protocol that restricts the size of the intermediate entangled states to $n+c$, where $c\in [1,9]$. 
\newpage

\section{Acknowledgements}

E.K. would like to acknowledge funding support from the U.S. Army Research Office MURI program, contract
number W911NF2110325. S.K. acknowledge National Science Foundation (NSF) Engineering Research Center
for Quantum Networks (CQN), grant number 1941583. The numerical codes and data are available in \cite{num_grid_network}.

\bibliographystyle{IEEEtran}
\bibliography{IEEEabrv,main}{}
\onecolumn
\appendices
\section{Swapping of $n$ Werner states with an $n$-GHZ swap}\label{appendix:werner-swap}
Consider the Werner State given as
\begin{equation}
\rho_{AB} = p \Phi_{AB} + (1-p)\pi_{A}\otimes\pi_B,
\end{equation}
where $\Phi_{AB}$ is the maximally entangled state. Consider an $n$-GHZ state:
\begin{equation}
    \ket{\textrm{GHZ}(n)} = \frac{\ket{0}^{\otimes n}+\ket{1}^{\otimes n}}{\sqrt{2}}
\end{equation}
In this section, we obtain the form of the unnormalized state resulting from the measurement: 
\begin{equation}
 \bra{\textrm{GHZ}(n)}_{B_1\cdots B_n}\bigotimes_{i=1}^n\rho_{A_iB_i}\ket{\textrm{GHZ}(n)}_{B_1\cdots B_n}\label{eq:GHZ-measurement}
\end{equation}
We define the GHZ basis for $n$-qubit systems with the basis vectors $\ket{\psi_{j,i_2,i_3\cdots i_{n}}}$ as
\begin{equation}
    \ket{\psi_{j,i_2,i_3\cdots i_{n}}} = \bigotimes_{\alpha=2}^{n}X_\alpha^{i_\alpha}Z_1^j\ket{\textrm{GHZ}(n)},
\end{equation}
where $i_{\alpha},j\in [0,1]$. The notation $P_{k}$ implies that the Pauli operator $P$ acts on the $k^{\textrm{th}}$ qubit. We alternatively observe that the basis vectors with the action of $m$ $X$ operators on the GHZ state can be written as a vector with the action of $n-m$ $X$ operators on $GHZ(n)$. This implies that all basis vectors with $m\geq \lceil n/2\rceil$ $X$ operators can be converted into vectors with $m\leq \lfloor n/2\rfloor$ $X$ operators. 
In this notation, the basis for a $3$-qubit system is given as:
\begin{align}
    \ket{\psi_{0,0,0}} &= \ket{\textrm{GHZ}(3)}\quad\quad \ket{\psi_{1,0,0}} = Z_1\ket{\textrm{GHZ}(3)}\\\ket{\psi_{0,1,0}} &= X_2\ket{\textrm{GHZ}(3)}\quad\quad \ket{\psi_{1,1,0}} = X_2Z_1\ket{\textrm{GHZ}(3)}\\\ket{\psi_{0,0,1}} &= X_3\ket{\textrm{GHZ}(3)}\quad\quad \ket{\psi_{1,0,1}} = X_3Z_1\ket{\textrm{GHZ}(3)}\\
    \ket{\psi_{0,1,1}} &= X_2X_3\ket{\textrm{GHZ}(3)}= X_1\ket{\textrm{GHZ}(3)}\\ \ket{\psi_{1,1,1}} &= X_2X_3Z_1\ket{\textrm{GHZ}(3)} = -X_1Z_1\ket{\textrm{GHZ}(3)}
\end{align}
We first observe that the state in Equation~\ref{eq:GHZ-measurement} is diagonal in the GHZ basis. The diagonal coefficients corresponding to the basis states are given as
\begin{align}
    \ketbra{\textrm{GHZ}}: \frac{1}{2^n}\left(p^n\ +\ \sum_{i=1}^{n-2}\frac{^nC_i }{2^{i+1}}p^{n-i}(1-p)^i\  + \ \frac{n}{2^n}(1-p)^{n-1}p+\frac{1}{2^n}(1-p)^n\right)\\
    Z_1\ketbra{\textrm{GHZ}}Z_1: \frac{1}{2^n}\left( \sum_{i=1}^{n-2}\frac{^nC_i }{2^{i+1}}p^{n-i}(1-p)^i\  + \ \frac{n}{2^n}(1-p)^{n-1}p+\frac{1}{2^n}(1-p)^n\right)
\end{align}
We now group the basis states according to the number of $X$ operators acting on the state $\ket{\textrm{GHZ}}$. The maximum number of $X$ operators is given by $\lfloor n/2\rfloor$. Let us define a set of basis vectors $S_k^n$ with $k \in \left[1,\lfloor n/2 \rfloor\right]$ such that 
\begin{equation}
    S_k^n = \left\{\ket{\psi_{j,i_2,i_3,\cdots i_{n}}}:  \sum_{\alpha=2}^ni_{\alpha} = k \bigg| \sum_{\alpha =2}^n i_{\alpha} = n-k\right\}
\end{equation}

For the basis states $\ket{\psi} \in S_k^n$, the diagonal coefficient is given as
\begin{equation}\label{eq:setSk}
    \frac{1}{2^n}\left(\sum_{i=k}^{n-2} \frac{^{n-k}C_{i-k}}{2^{i+1}}p^{n-i}(1-p)^i + \ \frac{n}{2^n}(1-p)^{n-1}p+\frac{1}{2^n}(1-p)^n\right)
\end{equation}
\subsection{GHZ(3) swap}
Now, as a test case, let us consider that a node performs a 3-qubit GHZ swap on three pairs of Werner state to obtain an unnormalized state $\sigma_{A_1A_2A_3}$. 
\begin{equation}
 \sigma_{A_1A_2A_3}= \bra{\textrm{GHZ}}_{B_1\cdots B_3}\bigotimes_{i=1}^3\rho_{A_iB_i}\ket{\textrm{GHZ}_{B_1\cdots B_3} }
\end{equation}
We can write the state $\sigma_{A_1A_2A_3}$ as 
\begin{multline}
   \sigma_{A_1A_2A_3}= p^3 \bra{\textrm{GHZ}}_{B_1B_2B_3}\Phi_{A_1B_1}\otimes\Phi_{A_2B_2}\otimes\Phi_{A_3B_3}\ket{\textrm{GHZ}}_{B_1B_2B_3}+\\
    (1-p)p^2\bra{\textrm{GHZ}}_{B_1B_2B_3}(\left(\Phi_{A_1B_1}\otimes\Phi_{A_2B_2}\otimes \pi_{A_3}\otimes \pi_{B_3}\right)+\left(\Phi_{A_1B_1}\otimes\Phi_{A_3B_3}\otimes \pi_{A_2}\otimes \pi_{B_2}\right)\\+\left(\Phi_{A_3B_3}\otimes\Phi_{A_2B_2}\otimes \pi_{A_1}\otimes \pi_{B_1}\right))\ket{\textrm{GHZ}}_{B_1B_2B_3}\\+
    (1-p)^2p\bra{\textrm{GHZ}}_{B_1B_2B_3} \left(\Phi_{A_1B_1}\otimes \pi_{A_2}\otimes \pi_{B_2}\otimes \pi_{A_3}\otimes \pi_{B_3}+\Phi_{A_2B_2}\otimes \pi_{A_1}\otimes \pi_{B_1}\otimes \pi_{A_3}\otimes \pi_{B_3}+\right.\\\left.\Phi_{A_3B_3}\otimes \pi_{A_2}\otimes \pi_{B_2}\otimes \pi_{A_1}\otimes \pi_{B_1}\right)\ket{\textrm{GHZ}}_{B_1B_2B_3}+\\
    (1-p)^3 \bra{\textrm{GHZ}}_{B_1B_2B_3}\pi_{A_1}\otimes \pi_{B_1}\otimes\pi_{A_2}\otimes \pi_{B_2}\otimes\pi_{A_3}\otimes \pi_{B_3}\ket{\textrm{GHZ}}_{B_1B_2B_3}.
\end{multline}
Then,
\begin{multline}
   \sigma_{A_1A_2A_3}= \frac{p^3}{8} \ketbra{\textrm{GHZ}}_{A_1A_2A_3} \\+ \frac{(1-p)p^2}{16}\left(\left(\ketbra{00}_{A_1A_2}+\ketbra{11}_{A_1A_2}\right)\otimes \pi_{A_3}+\left(\ketbra{00}_{A_3A_2}+\ketbra{11}_{A_3A_2}\right)\otimes \pi_{A_1}+\right.\\\left.\left(\ketbra{00}_{A_1A_3}+\ketbra{11}_{A_1A_3}\right)\otimes \pi_{A_2}\right)+\\
    \frac{3(1-p)^2p+(1-p)^3}{8}\pi_{A_1}\otimes\pi_{A_2}\otimes\pi_{A_3}\label{eq:GHZ-3}
\end{multline}
We see that the state $(\ketbra{00}+\ketbra{11})_{A_1A_2}\otimes \pi_{A_3}$ can be stated as 
\begin{equation}
    \frac{\ketbra{\textrm{GHZ}}+Z_1\ketbra{\textrm{GHZ}}Z_1}{2}+\frac{X_3\ketbra{\textrm{GHZ}}X_3+X_3Z_1\ketbra{\textrm{GHZ}}Z_1X_3}{2}
\end{equation}
We can also expand the other terms in Equation~\ref{eq:GHZ-3} in the GHZ basis to obtain the following coefficients: 
\begin{align}
    \op{\textrm{GHZ}} : \frac{p^3}{8}+\frac{(1-p)p^2}{8}\frac{3}{4}+\frac{3(1-p)^2p+(1-p)^3}{8}\frac{1}{8}\\
    Z_1\op{\textrm{GHZ}}Z_1:\frac{(1-p)p^2}{8}\frac{3}{4}+\frac{3(1-p)^2p+(1-p)^3}{8}\frac{1}{8}\\
    \mathrm{ else}:\frac{(1-p)p^2}{8}\frac{1}{4}+\frac{3(1-p)^2p+(1-p)^3}{8}\frac{1}{8}.
\end{align}
\subsection{GHZ(4) swap}
We next consider the case with a GHZ(4) swap. We need to analyze the terms of the form: 
\begin{itemize}
    \item The first scenario corresponds to the swapping operator acting on four Bell states. The term given below has a coefficient $p^4$.
    \begin{equation}
    \bra{\textrm{GHZ}}_{B_1B_2B_3B_4}\bigotimes_{i=1}^4\Phi_{A_iB_i}\ket{\textrm{GHZ}}_{B_1B_2B_3B_4} = \ketbra{\textrm{GHZ}}_{A_1A_2A_3A_4}\frac{1}{2^4}
    \end{equation}
    \item The second scenario corresponds to swapping on three Bell states and maximally mixed states. This term has a coefficient $p^3(1-p)$. There are four possible permutations. 
    \begin{multline}
        \bra{\textrm{GHZ}}_{B_1B_2B_3B_4}\bigotimes_{i=1}^3\Phi_{A_iB_i}\otimes\pi_{A_4}\otimes\pi_{B_4}\ket{\textrm{GHZ}}_{B_1B_2B_3B_4} \\=\frac{1}{2^4}\frac{1}{4}\left([0000]_{A_1A_2A_3A_4}+[1111]_{A_1A_2A_3A_4}+[1110]_{A_1A_2A_3A_4}+[0001]_{A_1A_2A_3A_4}\right),
    \end{multline}
    where $[ijkl]:= \op{ijkl}$.
    Now, it is simple to see that the above operators can be written as 
    \begin{equation}
        \frac{1}{2^4}\frac{1}{4}\left(\ketbra{\textrm{GHZ}}+Z_1\ketbra{\textrm{GHZ}}Z_1+X_4\ketbra{\textrm{GHZ}}X_4+X_4Z_1\ketbra{\textrm{GHZ}}Z_1X_4\right).
    \end{equation}
\item The third scenario consists of swapping on two Bell states and two maximally mixed states. The term has a coefficient $p^2(1-p)^2$. There are $^4C_2$ terms of the form given below: 
\begin{multline}
    \bra{\textrm{GHZ}}_{B_1B_2B_3B_4}\bigotimes_{i=1}^2\Phi_{A_iB_i}\otimes\pi_{A_3}\otimes\pi_{B_3}\otimes\pi_{A_4}\otimes\pi_{B_4}\ket{\textrm{GHZ}}_{B_1B_2B_3B_4} \\\frac{1}{2^4}\frac{1}{8} [0000]_{A_1A_2A_3A_4}+[1111]_{A_1A_2A_3A_4}+[0001]_{A_1A_2A_3A_4}+[1110]_{A_1A_2A_3A_4}\\+ [0010]_{A_1A_2A_3A_4}+[1101]_{A_1A_2A_3A_4}+[0011]_{A_1A_2A_3A_4}+[1100]_{A_1A_2A_3A_4}
\end{multline}
Now, we can write the above operator as 
\begin{multline}
    \frac{1}{2^4}\frac{1}{8}\left(\ketbra{\textrm{GHZ}}+Z_1\ketbra{\textrm{GHZ}}Z_1+X_3\ketbra{\textrm{GHZ}}X_3\right.\\+X_3Z_1\ketbra{\textrm{GHZ}}Z_1X_3\left.+X_4\ketbra{\textrm{GHZ}}X_4+X_4Z_1\ketbra{\textrm{GHZ}}Z_1X_4\right.\\+X_3X_4\ketbra{\textrm{GHZ}}X_3X_4\left.+X_4X_3Z_1\ketbra{\textrm{GHZ}}Z_1X_3X_4\right)
\end{multline}

\item The next contribution comes from the swapping operator acting on one Bell state and three maximally mixed states. The coefficient for the term is $p(1-p)^3$. There are $^4C_3$ terms of this form. We obtain:
\begin{multline}
 \bra{\textrm{GHZ}}_{B_1B_2B_3B_4}\bigotimes\Phi_{A_1B_1}\otimes \pi_{A_2}\otimes\pi_{B_2}\otimes\pi_{A_3}\otimes\pi_{B_3}\otimes\pi_{A_4}\otimes\pi_{B_4}  \ket{\textrm{GHZ}}_{B_1B_2B_3B_4} \\= \frac{1}{2^4}\frac{1}{2^4}\sum_j\ketbra{\psi^j}_{A_1A_2A_3A_4},
\end{multline}
where $j$ sums over the GHZ Basis states. 
\item The last contribution comes from a swap over all maximally mixed states. The coefficient for the term is $(1-p)^4$ We will obtain: 
\begin{equation}
    \frac{1}{2^4}\frac{1}{2^4}\sum_j\ketbra{\psi^j}_{A_1A_2A_3A_4},
\end{equation}
where $j$ sums over the GHZ basis states. 
\end{itemize}

\subsection{GHZ(n) swap}
We can similarly write down the swapping of $n$ Werner states by $\ket{\textrm{GHZ}(n)}$. We can write all contributions from swapping on $n-m$ Bell states and $m$ maximally mixed states, where $m\in[0,n]$. We write down the cases for $m=0$, $m=1$, $m=2$, $m=n-1$ and $m=n$ below. 
\begin{itemize}
    \item $\mathbf{m=0}$: The first scenario corresponds to the swapping operator acting on $n$ Bell states. The term given below has a coefficient $p^n$.
    \begin{equation}
    \bra{\textrm{GHZ}}_{B_1\cdots B_n}\bigotimes_{i=1}^n\Phi_{A_iB_i}\ket{\textrm{GHZ}}_{B_1\cdots B_n} = \ketbra{\textrm{GHZ}}_{A_1\cdots A_n}\frac{1}{2^n}
    \end{equation}
    \item $\mathbf{m=1}$: The second scenario corresponds to swapping on $n-1$ Bell states and a single maximally mixed state. The term has a coefficient $p^{n-1}(1-p)$. There are $n$ permutations possible for the scenario given below: 
    \begin{multline}
        \bra{\textrm{GHZ}}_{B_1\cdots B_n}\bigotimes_{i=1}^{n-1}\Phi_{A_iB_i}\otimes\pi_{A_n}\otimes\pi_{B_n}\ket{\textrm{GHZ}}_{B_1\cdots B_n} \\=\frac{1}{2^{n}}\frac{1}{4}\left([0\cdots 00]_{A_1\cdots A_n}+[1\cdots11]_{A_1\cdots A_n}+[1\cdots10]_{A_1\cdots A_n}+[0\cdots01]_{A_1\cdots A_n}\right)\label{eq:m_1}
    \end{multline}
    Now, it is simple to see that the above operator can be written as 
    \begin{multline}
        \frac{1}{2^n}\frac{1}{4}\left(\ketbra{\textrm{GHZ}}+Z_1\ketbra{\textrm{GHZ}}Z_1\right.\\\left.+X_n\ketbra{\textrm{GHZ}}X_n+X_nZ_1\ketbra{\textrm{GHZ}}Z_1X_n\right)
    \end{multline}
The position of the $X$ operator will be the same as that of the maximally mixed state $\pi$ in \eqref{eq:m_1}. 
\item $\mathbf{m=2}$: The third scenario consists of swapping on $n-2$ Bell states and two maximally mixed states. The term has a coefficient $p^{(n-2)}(1-p)^2$. There are $^nC_2$ terms of the form given below: 
\begin{multline}
    \bra{\textrm{GHZ}}_{B_1\cdots B_n}\bigotimes_{i=1}^{n-2}\Phi_{A_iB_i}\otimes\pi_{A_{n-1}}\otimes\pi_{B_{n-1}}\otimes\pi_{A_n}\otimes\pi_{B_n}\ket{\textrm{GHZ}}_{B_1\cdots B_n} \\\frac{1}{2^{n}}\frac{1}{8} [0\cdots000]_{A_1\cdots A_n}+[1\cdots111]_{A_1\cdots A_n}+[0\cdots001]_{A_1\cdots A_n}+[1\cdots101]_{A_1\cdots A_n}\\+ [0\cdots010]_{A_1\cdots A_n}+[1\cdots110]_{A_1\cdots A_n}+[0\cdots011]_{A_1\cdots A_n}+[1\cdots100]_{A_1\cdots A_n}
\end{multline}
We can write the above operator as 
\begin{multline}
    \frac{1}{2^n}\frac{1}{8}\left(\ketbra{\textrm{GHZ}}+Z_1\ketbra{\textrm{GHZ}}Z_1+X_{n-1}\ketbra{\textrm{GHZ}}X_{n-1}\right.\\\left.+X_{n-1}Z_1\ketbra{\textrm{GHZ}}Z_1X_{n-1}+X_n\ketbra{\textrm{GHZ}}X_n+X_nZ_1\ketbra{\textrm{GHZ}}Z_1X_n\right.\\\left.+X_nX_{n-1}\ketbra{\textrm{GHZ}}X_{{n-1}}X_n+X_nX_{{n-1}}Z_1\ketbra{\textrm{GHZ}}Z_1X_{n-1}X_n\right)
\end{multline}

We see that each term contributes to coefficients of two operators in $S^n_1$ and one operator in $S^n_2$.
We can also observe that each operator in $S^n_1$ gets contributions from $(n-1)$ such terms, and each operator in $S^n_2$ gets contributions from a single term. 

\item For $\mathbf{m\leq n-2}$, the swapping operator acts on $n-m$ Bell states and $m$ maximally mixed states with the coefficient $p^{n-m}(1-p)^m$. There are $^nC_{m}$ terms of the form 
\begin{multline}
    \bra{\textrm{GHZ}}_{B_1\cdots B_n}\bigotimes_{i=1}^{n-m}\Phi_{i}\otimes \bigotimes_{j=n-m}^{n}\pi_{A_{j}}\otimes \pi_{B_{j}}\ket{\textrm{GHZ}}_{B_1\cdots B_n} = \frac{1}{2^n}\frac{1}{2^{m+1}}\sum_{l=0}^m\left([0..0l']+[1..1\bar{l}']\right),
\end{multline}
where $l'$ is a binary representation of $l$ and $\bar{l'}$ bitwise flip of $l'$ . We can express this state as 
\begin{multline}
    \frac{1}{2^n}\frac{1}{2^{m+1}}\left(\op{\textrm{GHZ}}+Z_1\op{\textrm{GHZ}}Z_1 +\sum_{i=n-m}^nX_i\op{\textrm{GHZ}}X_i +\right.\\\left. \sum_{i=n-m}^nX_iZ_1\op{\textrm{GHZ}}Z_1X_i + \right.\\\left.\sum_{i,j \in  P_2} \left(X_iX_j\op{\textrm{GHZ}}X_iX_j+X_iX_jZ_1\op{\textrm{GHZ}}Z_1X_iX_j\right) \right.\\ \left.\cdots+ X_{n-m}\ldots X_n\left(\op{\textrm{GHZ}}X_{n-m}\ldots X_n+X_{n-m}\ldots X_nZ_1\op{\textrm{GHZ}}Z_1X_{n-m}\ldots X_n\right) \right),
\end{multline}
where $P_2$ represents all unique pairs in the list $\left[n-m,\ldots, n\right]$. We see now that the contributions from these terms to 
\begin{itemize}
    \item operators in $S^n_1$: $p^{n-m}(1-p)^m\frac{1}{2^n2^{m+1}}\ ^{n-1}C_{m-1}$
    \item operators in $S^n_2$: $p^{n-m}(1-p)^m\frac{1}{2^n2^{m+1}}\ ^{n-2}C_{m-2}$.
    \item operators in $S^n_j$: $p^{n-m}(1-p)^m \frac{1}{2^n 2^{m+1}}\ ^{n-j}C_{m-j}$.
\end{itemize}
\item $\mathbf{m=n-1}$: The next contribution comes from the swapping operator acting on one Bell state and $n-1$ maximally mixed state. The coefficient for the term is $p(1-p)^{n-1}$. There are $^nC_{n-1}$ terms of this form. We obtain:
\begin{equation}
 \bra{\textrm{GHZ}}_{B_1\cdots B_n}\Phi_{A_1B_1}\otimes \bigotimes_{i=2}^{n}\pi_{A_i}\otimes\pi_{B_i} \ket{\textrm{GHZ}}_{B_1\cdots B_n} \\= \frac{1}{2^n}\frac{1}{2^n}\sum_j\ketbra{\psi^j}_{A_1\cdots A_n},
\end{equation}
where $j$ sums over the GHZ Basis states. 
\item $\mathbf{m=n}$: The last contribution comes from a swap over all maximally mixed states. The coefficient for the term is $(1-p)^n$. We obtain: 
\begin{equation}
    \frac{1}{2^n}\frac{1}{2^n}\sum_j\ketbra{\psi^j}_{A_1\cdots A_n},
\end{equation}
where $j$ sums over the GHZ Basis states. 
\end{itemize}

We observe the following: 
\begin{itemize}
    \item Let the number of maximally mixed states involved in the swap be $m$, where $m\leq n$. Then this term will contribute to the diagonal coefficients of the operators in the set $\left\{S_{i}^n\right\}_{i=0}^{m}$. (See Equation~\ref{eq:setSk}).
    \item Each coefficient will have the contribution $\frac{1}{2^{2n}}np(1-p)^{(n-1)}+\frac{1}{2^{2n}}(1-p)^n$ from the terms corresponding to $m=n$ and $m=n-1$.
\end{itemize}
Combining the coefficients, we obtain \eqref{eq:setSk}.

We next consider the scenario wherein the measurement result is $\bigotimes_{\alpha=2}^{n}X_\alpha^{i_\alpha}Z_1^j\ket{\textrm{GHZ}(n)}$, where $i_{\alpha},j \in [0,1]$. 
\begin{equation}
 \bra{\textrm{GHZ}(n)}_{B_1\cdots B_n}\bigotimes_{\alpha=2}^{n}X_{B_\alpha}^{i_\alpha}Z_{B_1}^j\bigotimes_{i=1}^n\rho_{A_iB_i}\bigotimes_{\alpha=2}^{n}X_{B_\alpha}^{i_\alpha}Z_{B_1}^j\ket{\textrm{GHZ}(n)}_{B_1\cdots B_n}
\end{equation}
By using the property $M_B\ket{\Phi}_{AB} = M_A\ket{\Phi}_{AB}$, where $M$ is some Pauli operator, we obtain that the resultant state is equivalent to the state obtained when the measurement result is $\ket{GHZ(n)}$ up to local Paulis on Alice's side. We can also conclude that the probability of obtaining the result $\bigotimes_{\alpha=2}^{n}X_\alpha^{i_\alpha}Z_1^j\ket{\textrm{GHZ}(n)}$ is $1/2^n$.

\section{Swapping of Werner states with different fidelity}\label{appendix:diff-fidelities}

We note that the factor $^nC_i$ in Equation~\ref{eq:setSk} appears due to the equal fidelities of the initial Werner states. Since we no longer have the same fidelities, this factor will change.  We again begin with analyzing the scenarios for swapping with $m$ maximally mixed state in the terms. The $n-m$ maximally entangled states are situated on the first $n-m$ systems. That is, we consider the case:
\begin{equation}
    \bra{\textrm{GHZ}}_{B_1\cdots B_n}\bigotimes_{i=1}^{n-m}\Phi_{A_iB_i} \otimes \bigotimes_{j={n-m+1}}^{n}\pi_{A_j}\otimes\pi_{B_j}\ket{\textrm{GHZ}}_{B_1\cdots B_n}
\end{equation}
\begin{itemize}
    \item For $m=0$, we obtain the state $\frac{1}{2^n}\op{\textrm{GHZ}}$ with the coefficient $\Pi_{i=1}^np_i$, where $p_i$ corresponds to the coefficient of the $i^{\mathrm{th}}$ Werner state.
    \item For $m=1$, we obtain the state
    \begin{align}
        \frac{1}{2^n}\frac{1}{4}\left(\ketbra{\textrm{GHZ}}+Z_1\ketbra{\textrm{GHZ}}Z_1\right.\\\left.+X_n\ketbra{\textrm{GHZ}}X_n+X_nZ_1\ketbra{\textrm{GHZ}}Z_1X_n\right)
    \end{align}
    with the coefficient $\Pi_{i=1}^{n-1} p_i(1-p_n)$. We thus see that terms of this type will contribute to coefficients of $\op{\textrm{GHZ}}$ and of $X_{i}\op{\textrm{GHZ}}X_i$. 
    \item For $m=2$, we obtain the state
    \begin{multline}
    \frac{1}{2^n}\frac{1}{8}\left(\ketbra{\textrm{GHZ}}+Z_1\ketbra{\textrm{GHZ}}Z_1\right.\\+\left.X_{n-1}\ketbra{\textrm{GHZ}}X_{n-1}+X_{n-1}Z_1\ketbra{\textrm{GHZ}}Z_1X_{n-1}\right.\\\left.+X_n\ketbra{\textrm{GHZ}}X_n+X_nZ_1\ketbra{\textrm{GHZ}}Z_1X_n+X_nX_{n-1}\ketbra{\textrm{GHZ}}X_{{n-1}}X_n\right.\\\left.+X_nX_{{n-1}}Z_1\ketbra{\textrm{GHZ}}Z_1X_{n-1}X_n\right),
\end{multline}
with the coefficient $\Pi_{i=1}^{n-2}p_i(1-p_n)(1-p_{n-1})$. 

We see that the terms of this form will contribute to the $\op{\textrm{GHZ}}$, $Z\op{\textrm{GHZ}}Z$, $X_{n-1}\op{\textrm{GHZ}}X_{n-1}$, $X_n\op{\textrm{GHZ}}X_n$, and $X_{n}X_{n-1}\op{\textrm{GHZ}}X_{n-1}X_n$.

We can track the terms above to obtain the following coefficients for the basis $\op{\textrm{GHZ}}$: 
\begin{multline}
    \frac{1}{2^n}\Pi_{i=1}^n p_i + \frac{1}{2^n}\sum_{i=1}^{n-2} \frac{1}{2^{i+1}}\frac{1}{i!}\sum_{j_1\in\left[1,n\right]}\sum_{j_2\in\left[1,n\right]\backslash j_1}\cdots\sum_{j_i\in\left[1,n\right]\backslash\Pi_{k= 1}^{i-1}j_k}\Pi_{l=j_1}^{j_{i}}(1-p_l) \ \Pi_{k\in [j_1,j_n]\backslash\Pi_{k=1}^{i}j_k}p_k\\ +\frac{1}{2^n}\Pi_{i=1}^n (1-p_i) + \frac{1}{2^n}\sum_{i=1}^np_i\Pi_{j\in[1,n]\backslash i}(1-p_j)\label{eq:GHZ-coeffi-different1}
\end{multline}

The coefficients for $Z_1\op{\textrm{GHZ}}Z_1$ are: 
\begin{multline}
 \frac{1}{2^n}\sum_{i=1}^{n-2} \frac{1}{2^{i+1}}\frac{1}{i!}\sum_{j_1\in\left[1,n\right]}\sum_{j_2\in\left[1,n\right]\backslash j_1}\cdots\sum_{j_i\in\left[1,n\right]\backslash\Pi_{k= 1}^{i-1}j_k}\Pi_{l=j_1}^{j_{i}}(1-p_l) \ \Pi_{k\in [j_1,j_n]\backslash\Pi_{k=1}^{i}j_k}p_k\\ +\frac{1}{2^n}\Pi_{i=1}^n (1-p_i)^n + \frac{1}{2^n}\sum_{i=1}^np_i\Pi_{j\in[1,n]\backslash i}(1-p_j)\label{eq:GHZ-coeffi-different}
\end{multline}

The coefficients for $X_{\alpha_k}\cdots X_{\alpha_2}X_{\alpha_1}\op{\textrm{GHZ}}X_{\alpha_1}X_{\alpha_2}\cdots X_{\alpha_k}$ and $X_{\alpha_k}\cdots X_{\alpha_2}X_{\alpha_1}Z_1\op{\textrm{GHZ}}Z_1X_{\alpha_1}X_{\alpha_2}\cdots X_{\alpha_k}$ are: 
\begin{multline}
 \frac{1}{2^n}\sum_{i=k}^{n-2} \frac{1}{2^{i+1}}\frac{1}{(i-k)!}\sum_{j_1\in\left[1,n\right]}\sum_{j_2\in\left[1,n\right]\backslash j_1}\cdots\sum_{j_i\in\left[1,n\right]\backslash\Pi_{k= 1}^{i-1}j_k}\Pi_{s=1}^k\delta_{j_s = \alpha_s} \ \Pi_{l=j_1}^{j_{i}}(1-p_l) \ \Pi_{k\in[1,n]\backslash\Pi_{k=1}^{i}j_k}p_k\\ +\frac{1}{2^n}\Pi_{i=1}^n (1-p_i)^n + \frac{1}{2^n}\sum_{i=1}^np_i\Pi_{j\in[1,n]\backslash i}(1-p_j)\label{eq:GHZ-X-coeffi-different}
\end{multline}
We see that the equations derived above reduce to Eq~\ref{eq:setSk}
when the Werner states have equal fidelities. 
\end{itemize}

\section{Measuring a single node in X basis}\label{appendix-X-swap}
Consider a mixed state in the $\textrm{GHZ}(n)$ diagonal basis 
\begin{equation}
    \rho_{}=\sum_{i_1i_2\cdots i_n}p_{i_1i_2i_3\cdots i_n}\op{\psi^{i_1i_2\cdots i_n}},
\end{equation}
where $\ket{\psi^{i_1i_2\cdots i_n}} = Z_1^{i_1}X_2^{i_2}\cdots X_n^{i_n}\ket{\Phi}$, and $\ket{\Phi}$ is a GHZ state. 

The final state after performing an $X$ measurement on the node is a $\textrm{GHZ}$ diagonal state characterized by the coefficients $\left\{q_{j_1j_2j_3\cdots j_{n-1}}\right\}_{j_1j_2j_3\cdots j_{n-1}}$. 

Consider that the node $m\neq 1$ of $\rho$ is measured in the $X$ basis. We obtain 
\begin{align}
 \bra{+}_m\rho\ket{+}_m &= \sum_{i_1i_2\cdots i_n}p_{i_1i_2i_3\cdots i_n}\ \braket{+}{\psi^{i_1i_2\cdots i_n}}\braket{\psi^{i_1i_2\cdots i_n}}{+}
\end{align}
We obtain
\begin{align}
    \bra{+}_mZ_1^{i_1}X_2^{i_2}\cdots X_n^{i_n}\ket{\Phi(n)} &= Z_1^{i_1}X_2^{i_2}\cdots X_{m-1}^{i_{m-1}}X_{m+1}^{i_{m+1}}\cdots X_n^{i_n}\bra{+}X_{m}^{i_m}\ket{\Phi(n)}\\
    &=\frac{1}{\sqrt{2}}Z_1^{i_1}X_2^{i_2}\cdots X_{m-1}^{i_{m-1}}X_{m}^{i_{m+1}}\cdots X_{n-1}^{i_{n}}\ket{\Phi(n-1)}
\end{align}
We thus obtain:
\begin{align}
    \bra{+}\rho\ket{+} &=\sum_{j_1j_2\cdots j_{n-1}}q_{j_1j_2\cdots j_{n-1}}\op{\psi^{j_1j_2\cdots j_{n-1}}} \\
    &= \frac{1}{2}\sum_{i_1i_2\cdots i_n}p_{i_1i_2i_3\cdots i_n} Z_1^{i_1}X_2^{i_2}\cdots X_{m-1}^{i_{m-1}}X_{m}^{i_{m+1}}\cdots X_{n-1}^{i_{n}}\ket{\Phi(n-1)}\nonumber\\&\quad\bra{\Phi(n-1)}Z_1^{i_1}X_2^{i_2}\cdots X_{m-1}^{i_{m-1}}X_{m}^{i_{m+1}}\cdots X_{n-1}^{i_n}
\end{align}
We obtain 
\begin{equation}
q_{j_1j_2\cdots j_{n-1}} = \frac{1}{2}\sum_{\mathcal{C}}p_{i_1i_2\cdots i_n},
\end{equation}
where 
\begin{equation}
    \mathcal{C} = \left\{i_1i_2i_3i_4\cdots i_n| j_1 = i_1, i_2 = j_2, i_3 = j_3, \cdots i_{m-1}= j_{m-1}, i_{m+1} = j_{m},i_{n}=j_{n-1}\right\}
\end{equation}
Let the $X$ measurement be on $m=1$. We obtain
\begin{align}
    \bra{+}Z_1^{i_1}X_2^{i_2}\cdots X_n^{i_n}\ket{\Phi(n)} &= X_2^{i_2}\cdots X_n^{i_n}\bra{+}Z_1^{i_1}\ket{\Phi(n)}\\
    &=\frac{(-1)^{\delta_{i_1,1}\delta_{i_2,1}}}{\sqrt{2}}X_2^{i_3\oplus i_2}\cdots X_{n-1}^{i_n\oplus i_2}Z_1^{i_1}\ket{\Phi(n-1)}
\end{align}
We thus obtain:
\begin{align}
    \bra{+}\rho\ket{+} &=\sum_{j_1j_2\cdots j_{n-1}}q_{j_1j_2\cdots j_{n-1}}\op{\psi^{j_1j_2\cdots j_{n-1}}} \\
    &= \frac{1}{2}\sum_{i_1i_2\cdots i_n}p_{i_1i_2i_3\cdots i_n} X_2^{i_3\oplus i_2}\cdots X_{n-1}^{i_n\oplus i_2}Z_1^{i_1}\ket{\Phi(n-1)}\nonumber\\&\quad\bra{\Phi(n-1)}X_2^{i_3\oplus i_2}\cdots X_{n-1}^{i_n\oplus i_2}Z_1^{i_1}
\end{align}
We obtain 
\begin{equation}
q_{j_1j_2\cdots j_{n-1}} = \sum_{\mathcal{C}}p_{i_1i_2\cdots i_n},
\end{equation}
where 
\begin{equation}
    \mathcal{C} = \left\{i_1i_2i_3i_4\cdots i_n| j_1 = i_1, j_2 = i_2 \oplus i_3, j_3 = i_2 \oplus i_4 \cdots j_{n-1} = i_n \oplus i_2 \right\}
\end{equation}
The measurement result corresponding to $\ket{-}$ is related to a local Pauli on the measurement result corresponding to $\ket{+}$.

\section{Proof for linearity}\label{appendix:linearity-proof}
In this section, we give a method of modeling the protocol given in Section~\ref{sec:original}, such that
the maximal size of the intermediate entangled state created in the network is $n+c$, where $c$ is a constant dependent on the position of Alice and Bob.

We call a node $m$-edge if it initially shares entanglement with quantum memories of $m$-nodes. In our set up $m\in[2,3,4]$. 

Modeling: We number the nodes as in Figure~\ref{fig:x_a}. We swap along $\left\{0,1,2,\cdots n^2-1, n^2\right\}$, skipping over the consumer nodes. As an example, in Figure~\ref{fig:x_a}, we would follow the pattern : $0\rightarrow 1\rightarrow 2\rightarrow 4\rightarrow 5 \rightarrow 6 \rightarrow 7 \rightarrow 8 \rightarrow 9 \rightarrow  11 \rightarrow 12 \rightarrow 13 \rightarrow 14 \rightarrow 15$. This is slightly different from the modeling of the protocol considered in Section~\ref{sec:model}

Let us consider the link probability $p=1$. We encounter the following types of node situations. 
\begin{itemize}
    \item Degree-three nodes with two memories sharing a bipartite entangled state with the adjacent node and one memory a part of an $n$-partite entangled state. After the 2-GHZ swap and an $X$ measurement, the resultant state has $n+4-3 = n+1$ nodes. 
    \item Degree-three nodes with one memory having bipartite entangled states and two memories being a part of the $n$-partite entangled state. The resultant state after the 3-GHZ swap has $n+2-3 = n-1$ nodes. 
    \item Degree-two nodes with one memory as a bipartite state and one as a part of $n$-partite entangled state. The resultant state after the swap has $n+2-2=n$ nodes.
    \item A degree-four node with two memories sharing bipartite states with adjacent memories and two edges as a part of an $n$-partite state. The resultant state is a $n$-partite state. 
\end{itemize}
\begin{figure*}
\begin{center}
\includegraphics[
width=1.72in
]{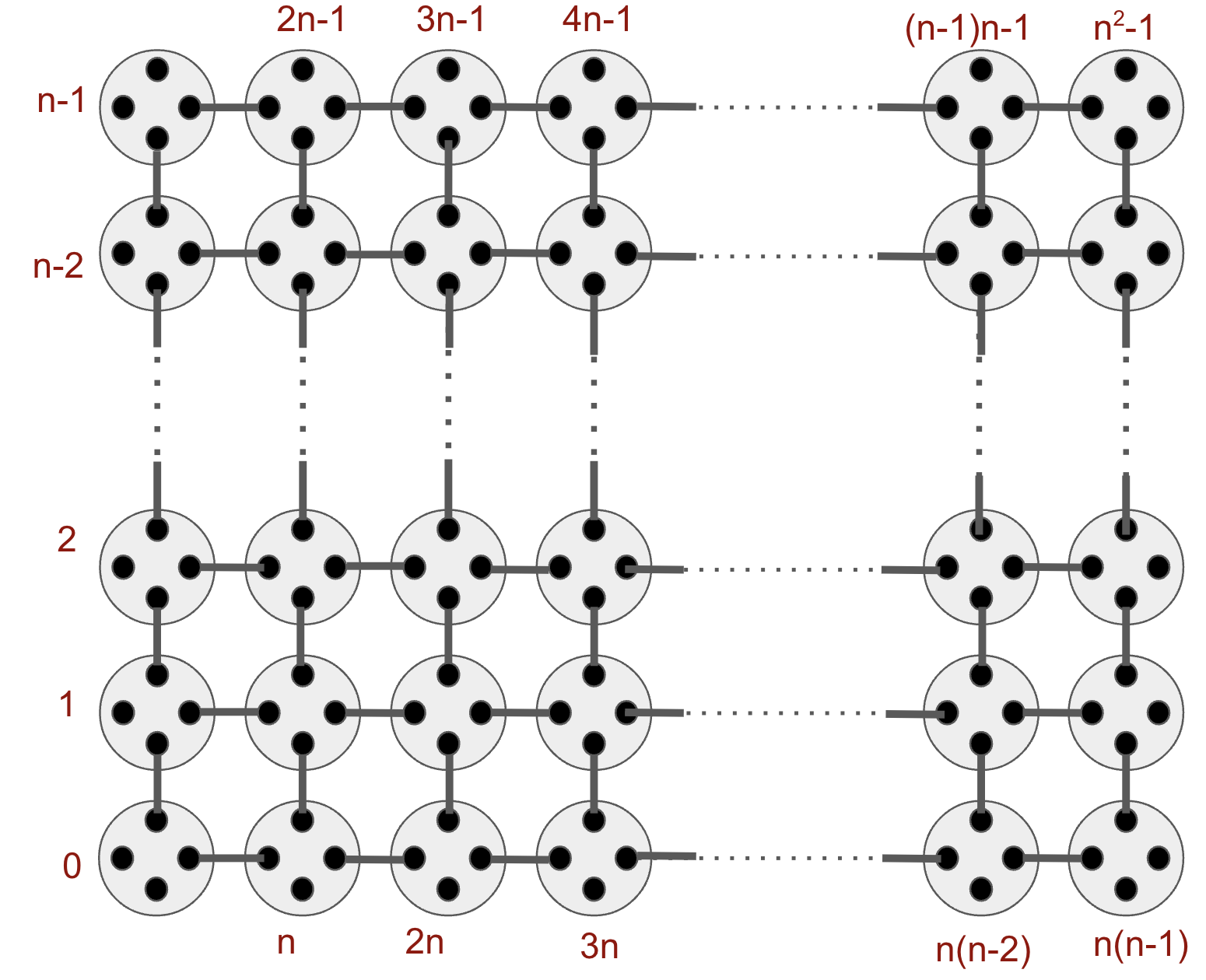}\hfil
\includegraphics[
width=1.72in
]{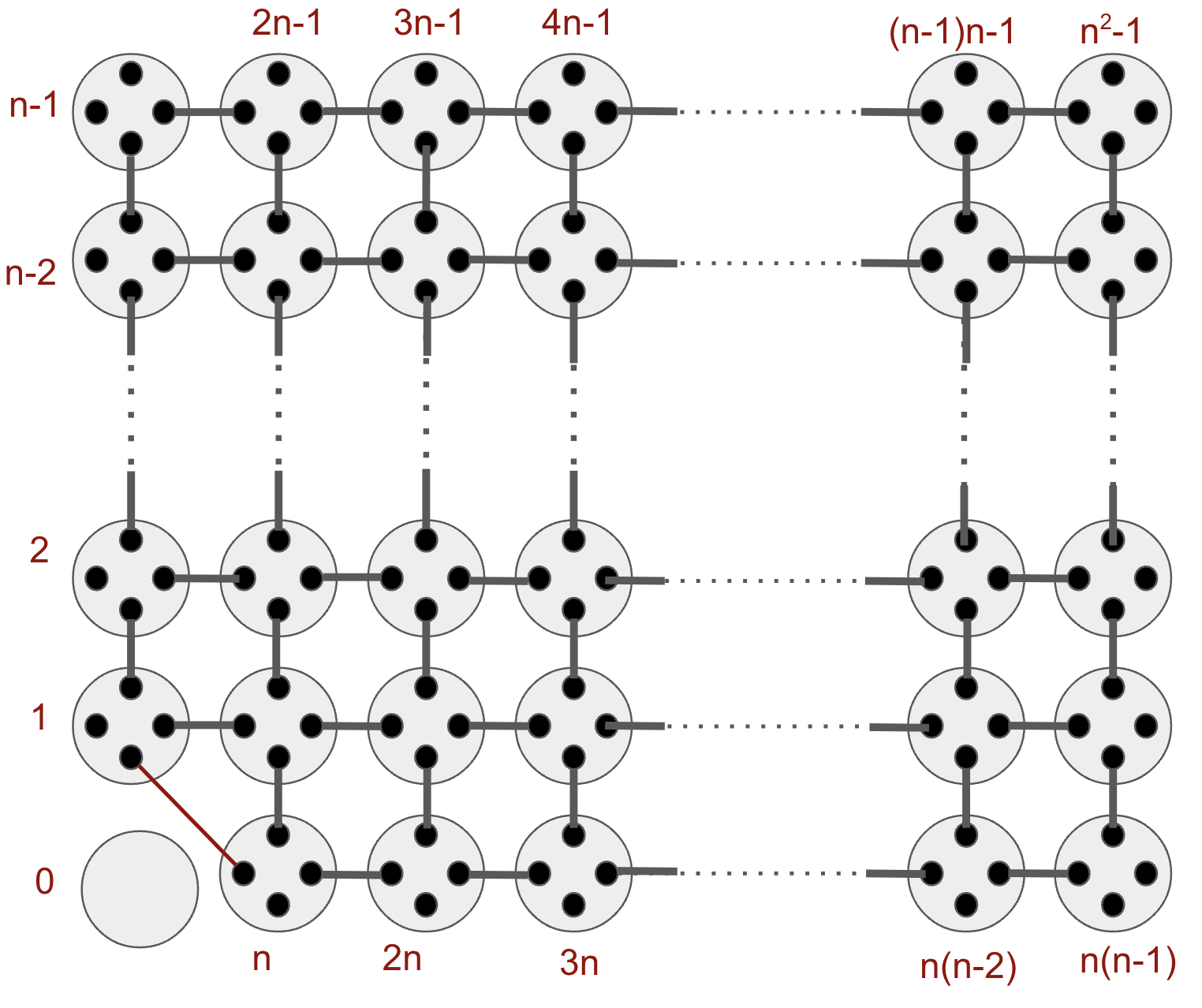}\hfil
\includegraphics[
width=1.72in
]{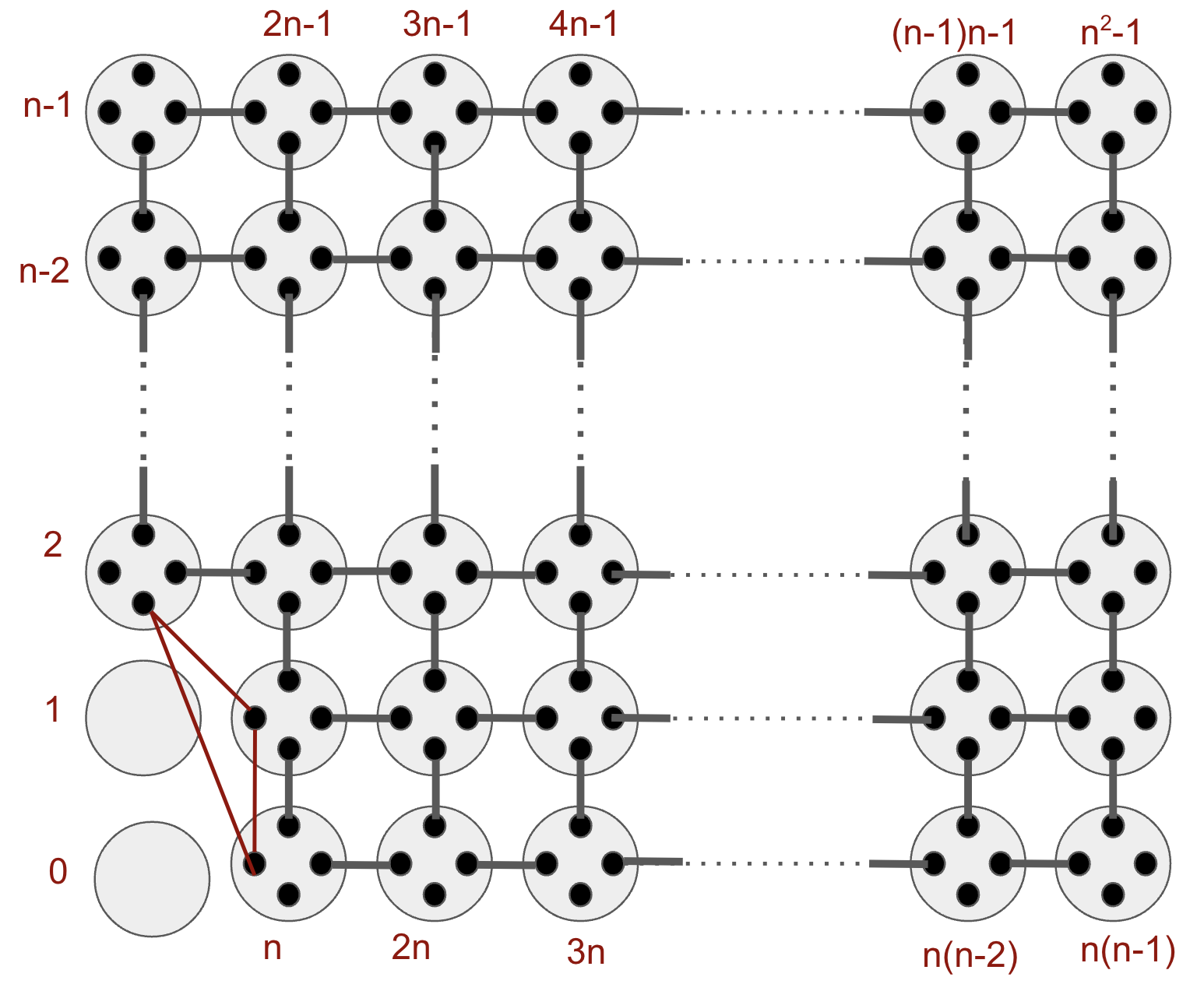}

\par\medskip\medskip
\includegraphics[
width=1.72in
]{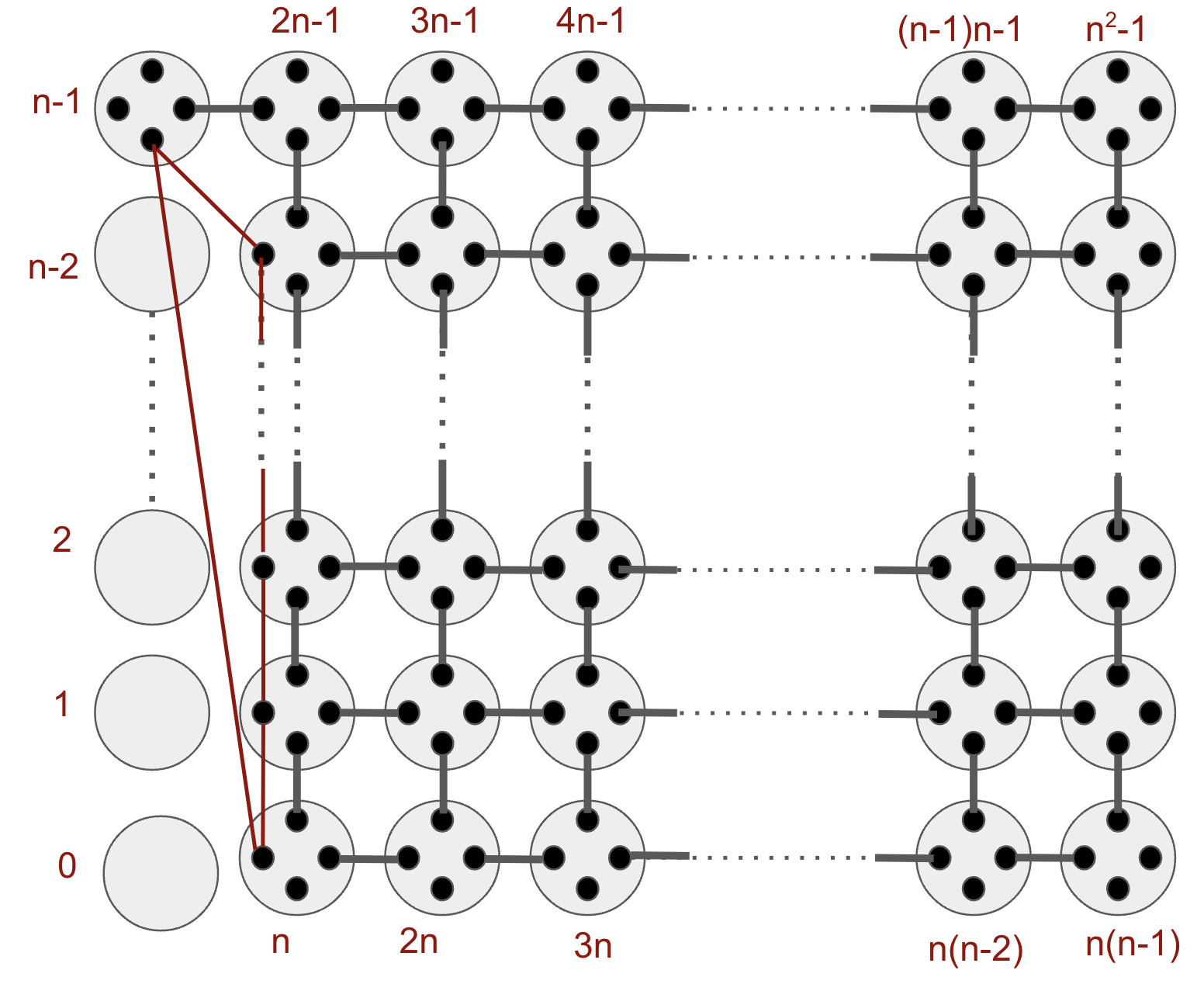}\hfil
\includegraphics[
width=1.72in
]{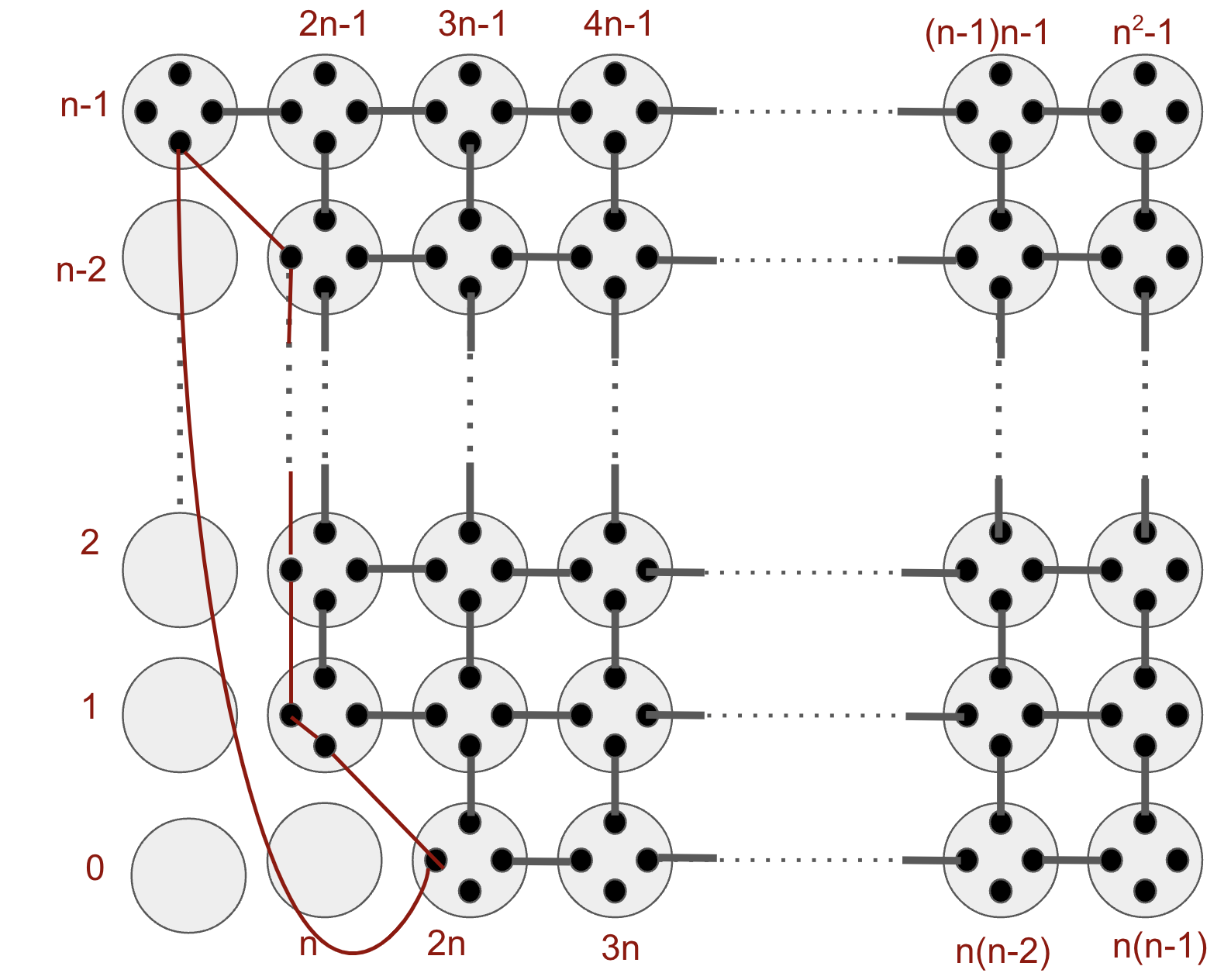}\hfil
\includegraphics[
width=1.72in
]{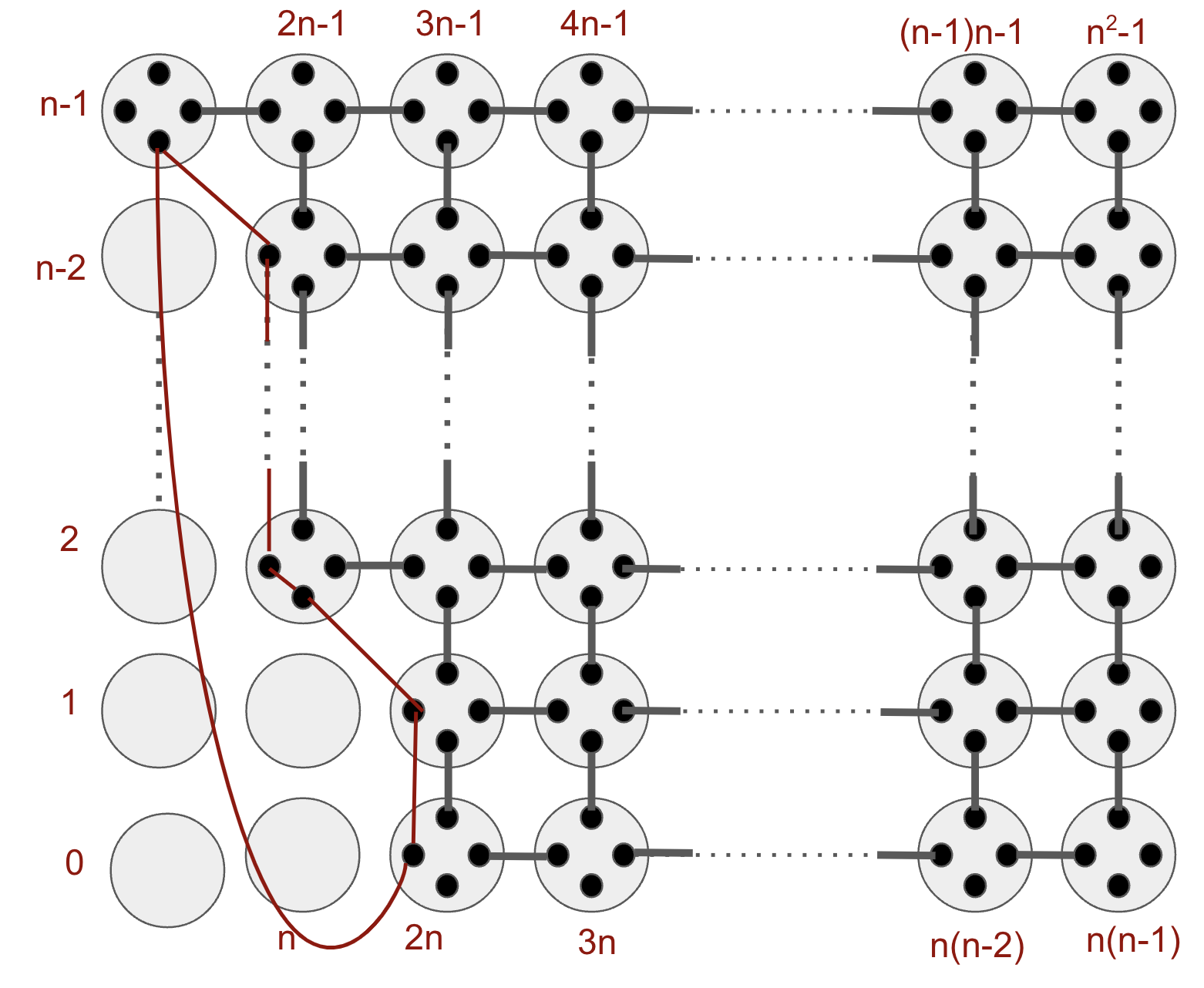}

\par\medskip
\includegraphics[
width=1.62in
]{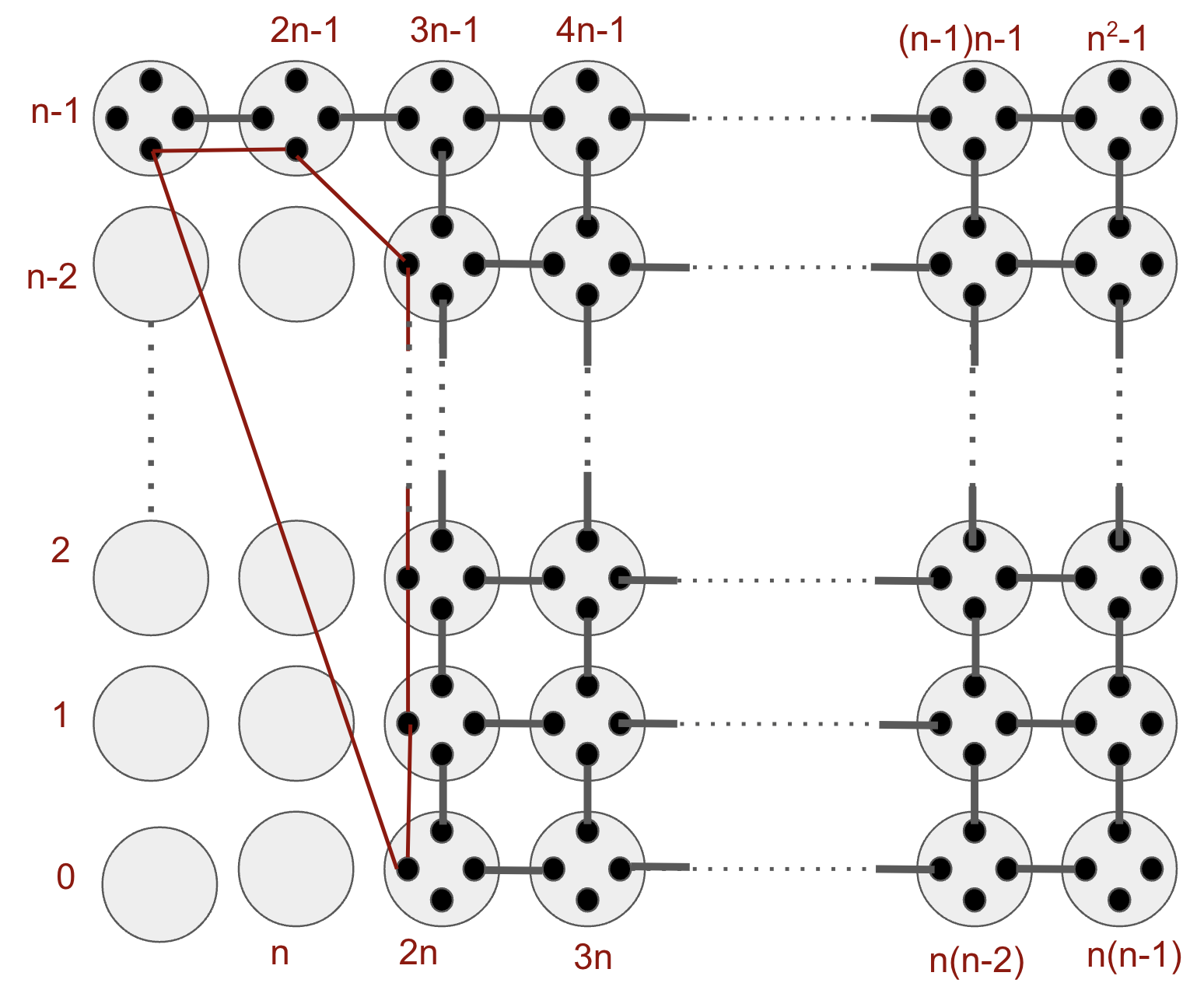}\hfil
\includegraphics[
width=1.72in
]{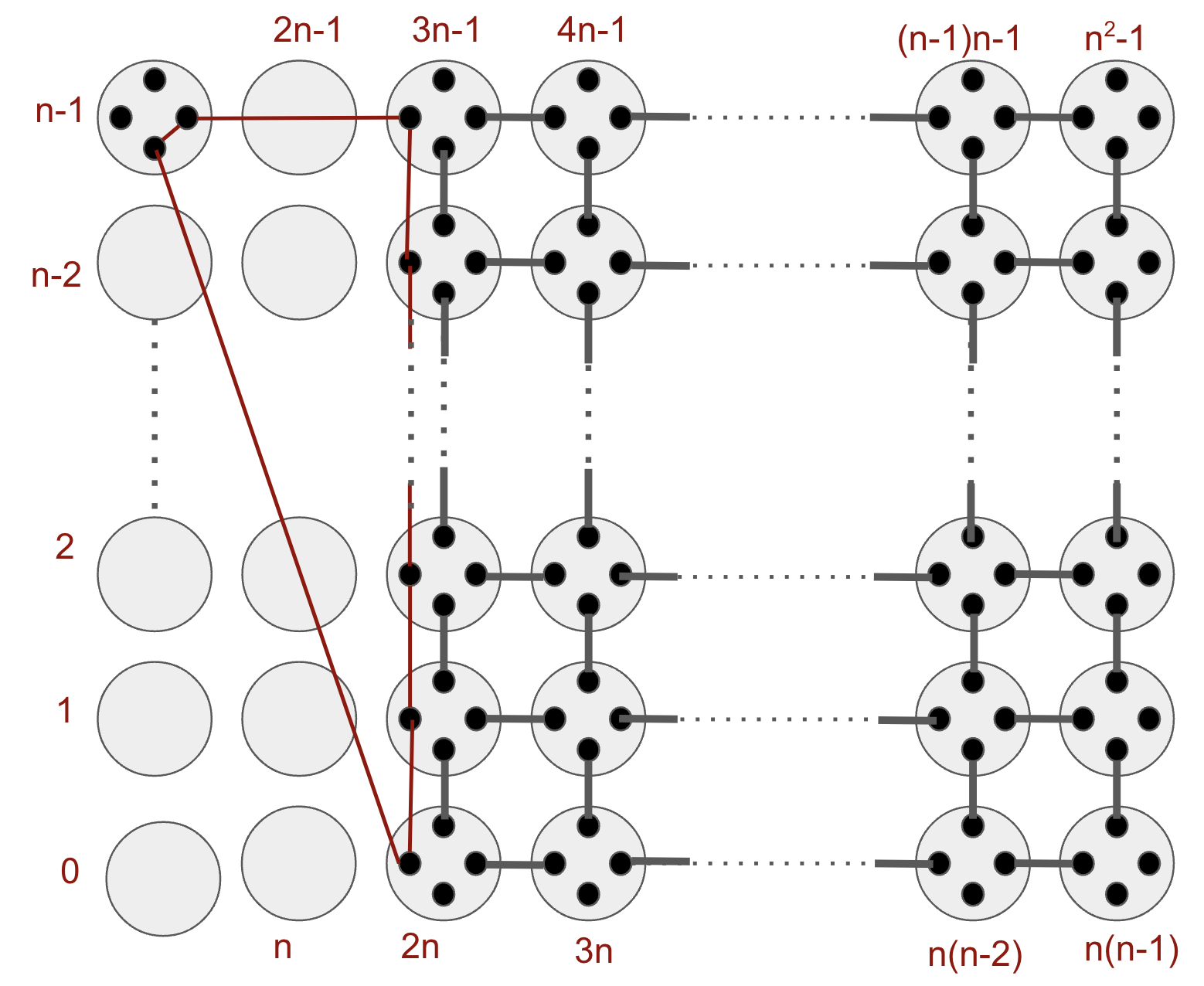}\hfil
\includegraphics[
width=1.72in
]{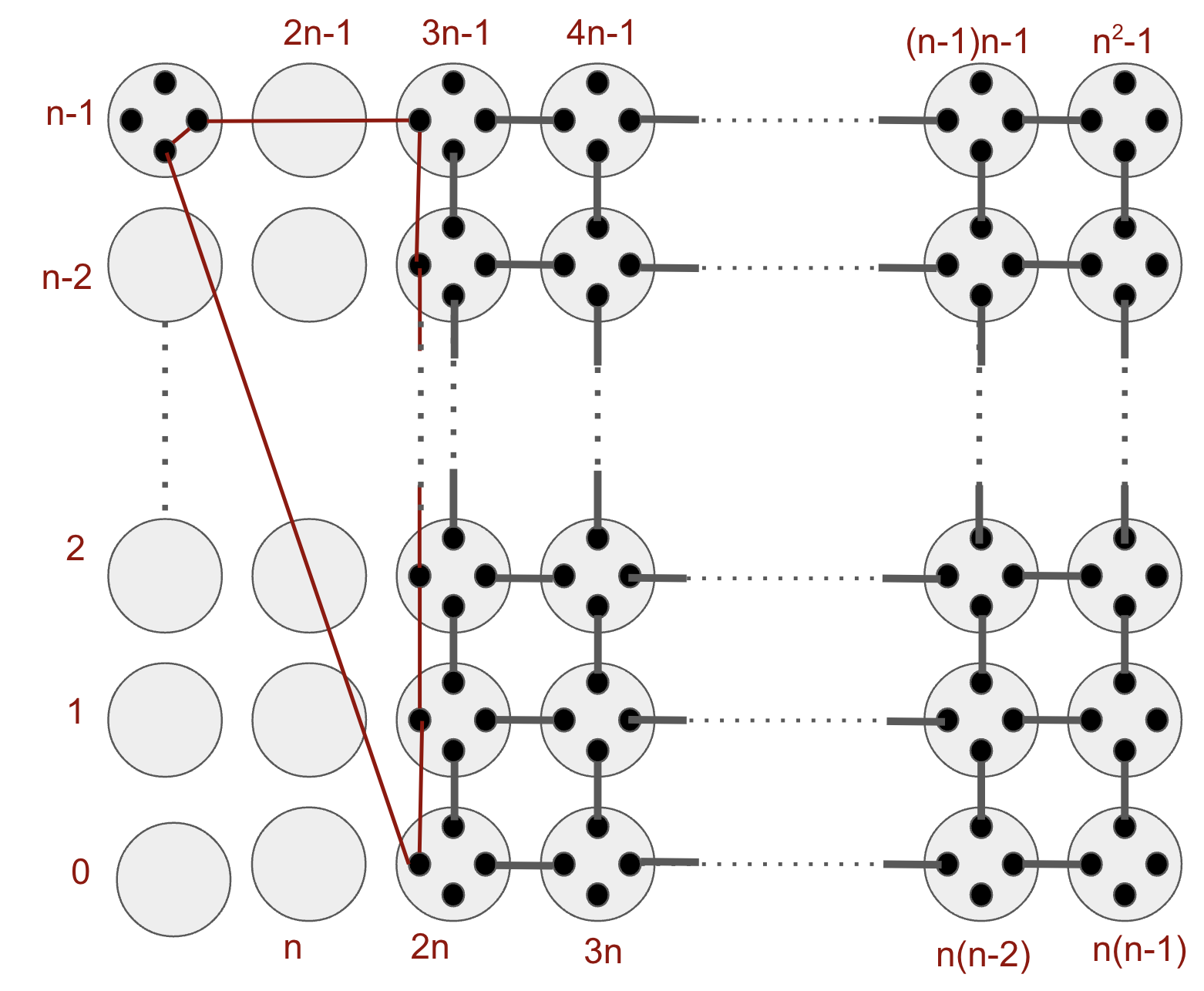}
\par\medskip

\includegraphics[
width=1.72in
]{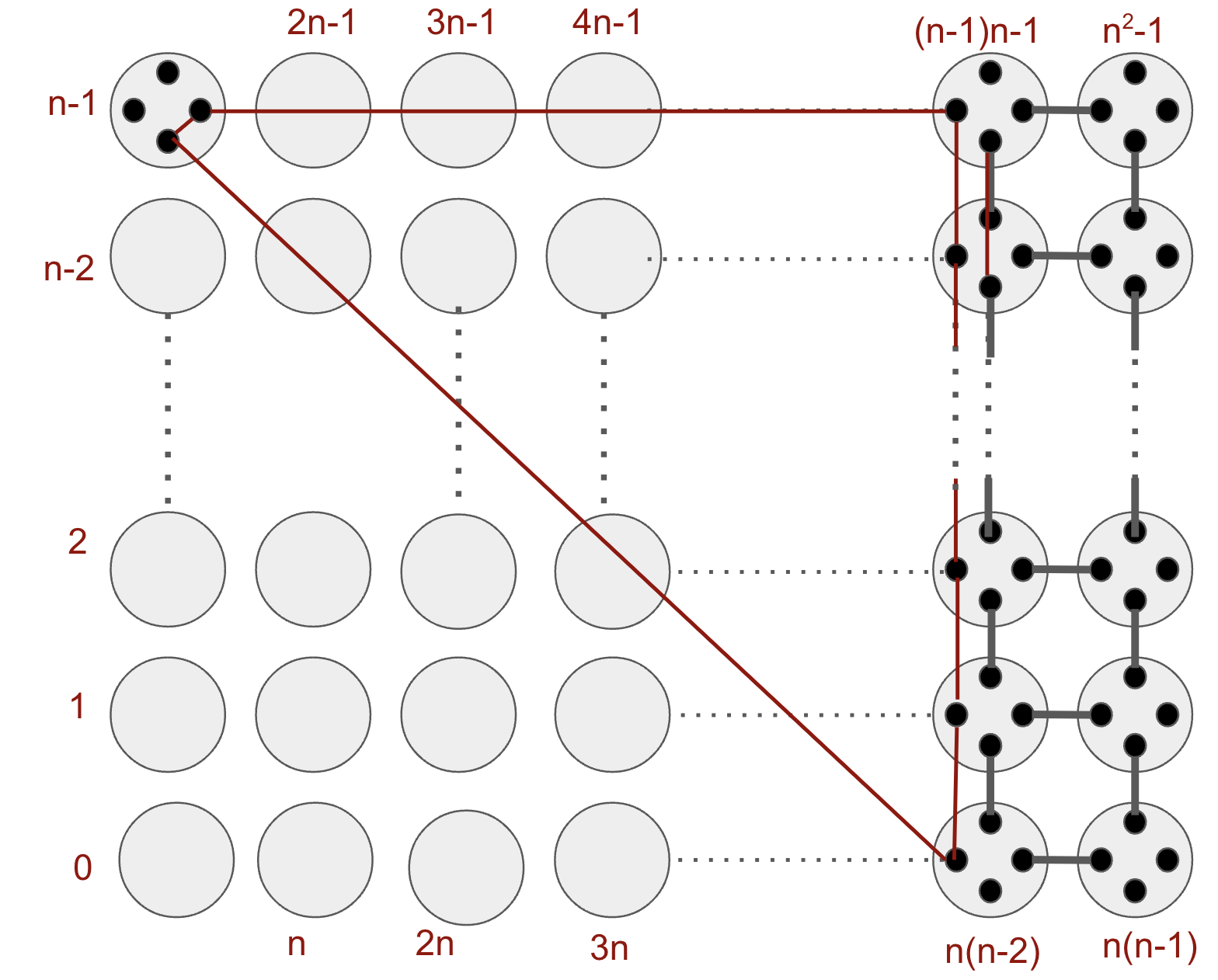}
\hfil 
\includegraphics[
width=1.72in
]{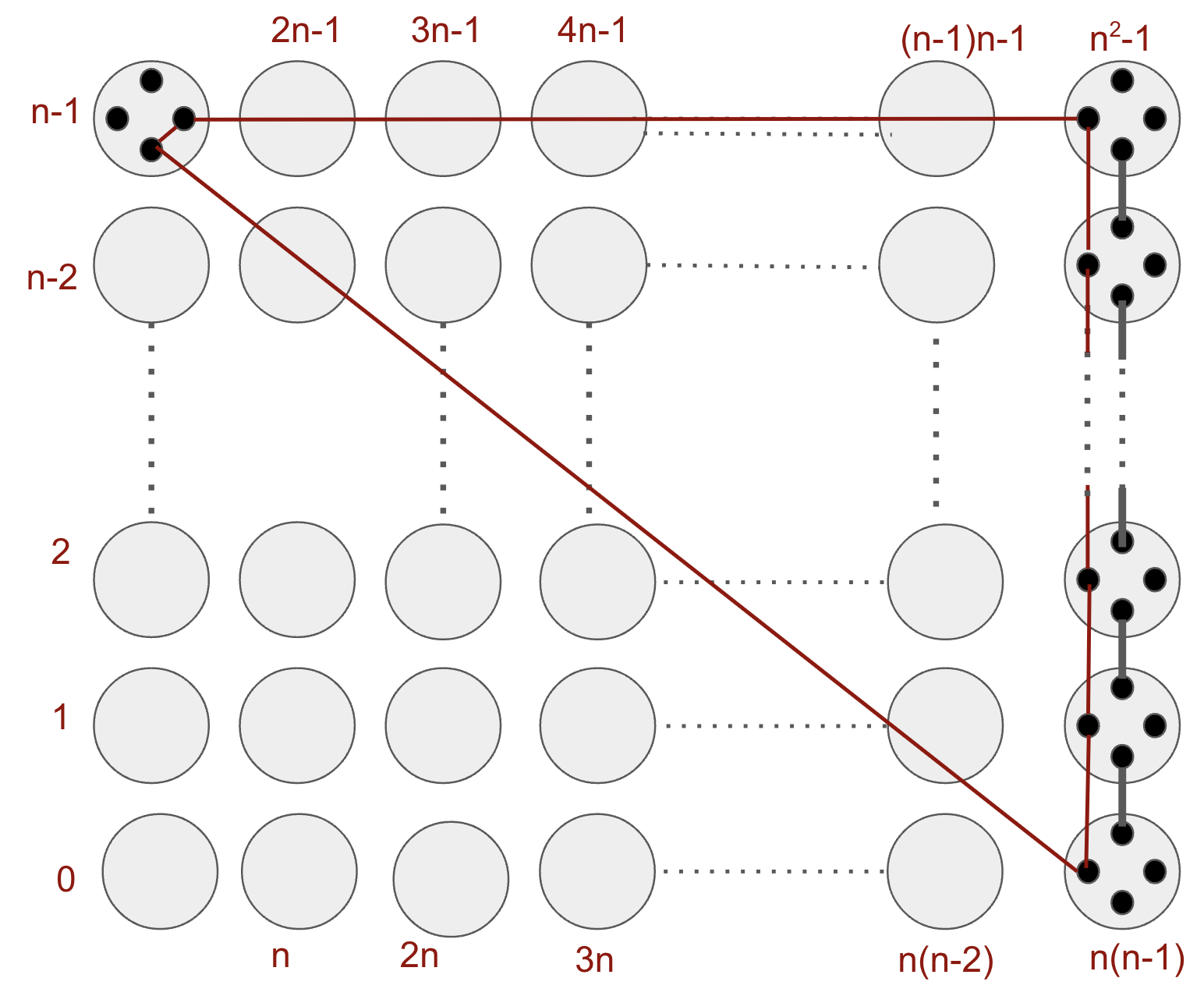}
\hfil
\includegraphics[
width=1.72in
]{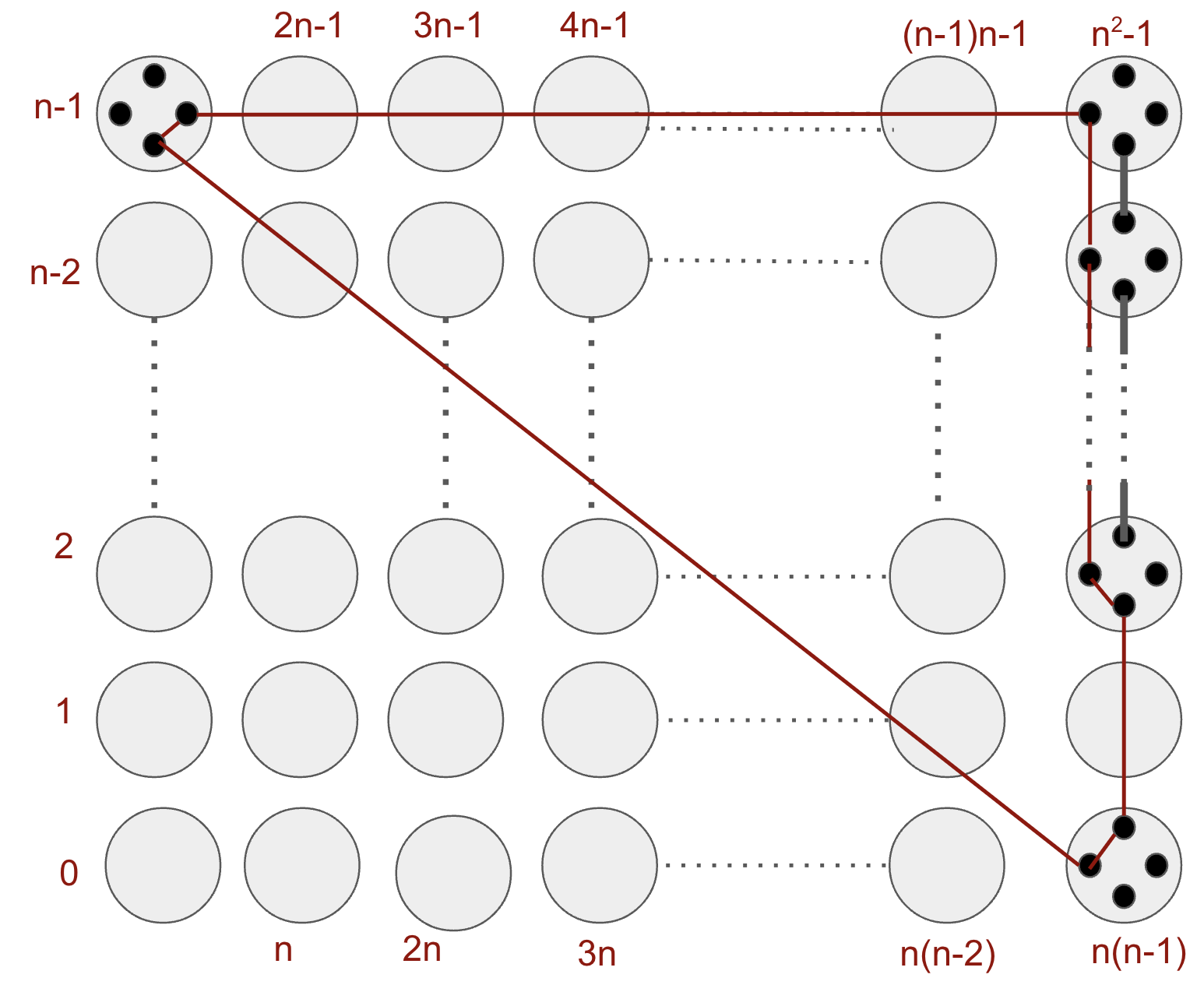}
\par\medskip
\end{center}
\end{figure*}%
\clearpage
\begin{figure*}\ContinuedFloat
\begin{center}
\includegraphics[
width=1.72in
]{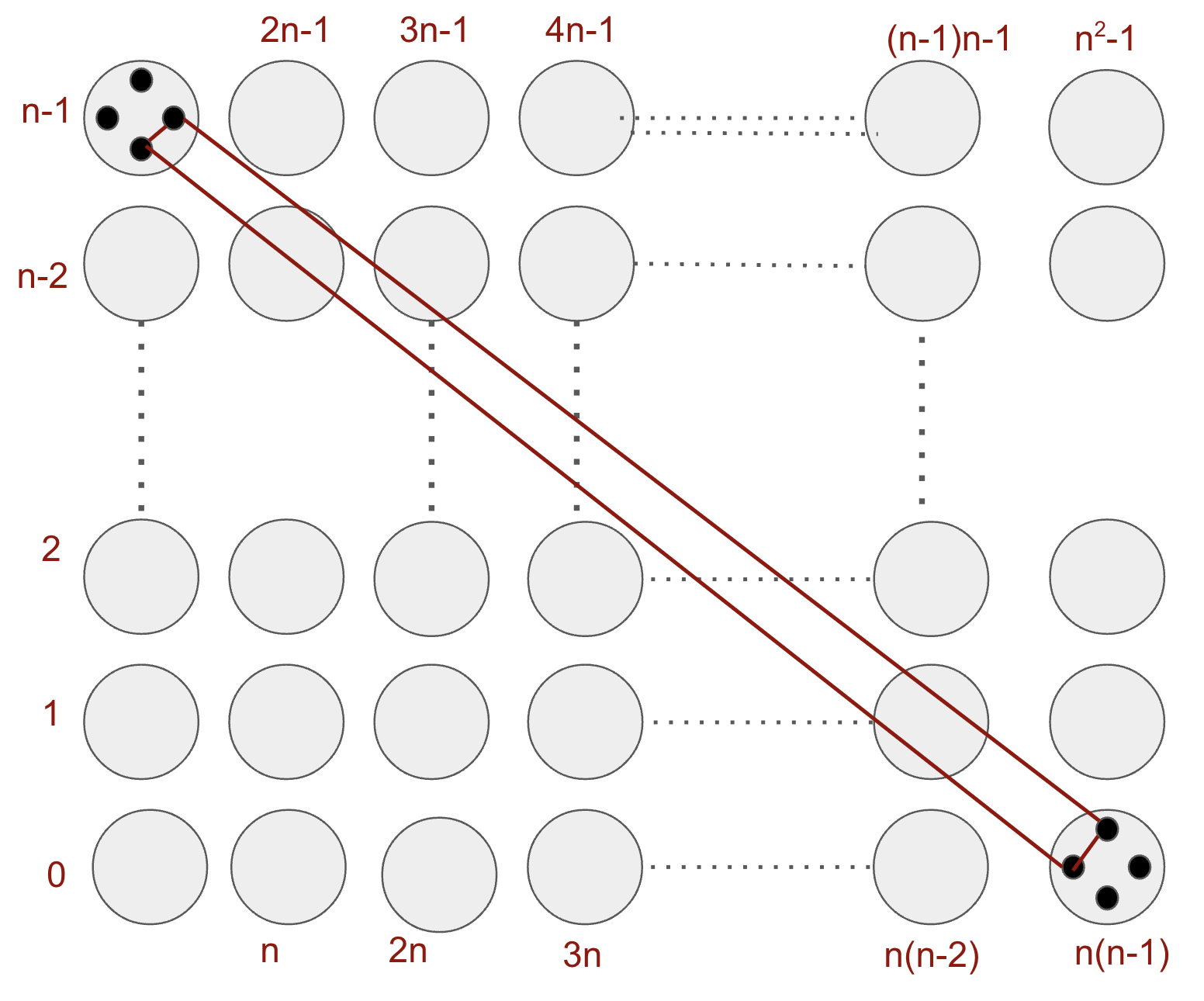}
\end{center}

\label{fig:2d-arch}%
\caption{In this figure, we give a pictorial representation of the proof for linearity of the size of the intermediate entanglement created during the swapping process when Alice and Bob are two-edged node.  (1) In the first figure, we give the state of the network before any swapping occurs. (2) We give the state of the network after the first swap, creating a bipartite entangled state. (3) The state after the network after second swap creating a 3-partite entangled state. (4) The state of network after we have swapped the first $(n-1)$ nodes to create a $n$-entangled state. (5) The state of the network after swapping the $n^{\textrm{th}}$ node, creating an $(n+1)-$entangled state. (6) The state of the network after swapping the $(n+1)^{\textrm{th}}$ node, creating an $(n+1)-$entangled state. The size doesn't change. (7) We again see that the size of the entangled state doesn't change when we swap the $2n-2$ node. (8) The state size increases when swapping the $2n-1$ node to $n+2$. (9) On swapping the $2n$ node, the size increases to $n+3$. We see that the size oscillated between $n+2$ and $n+3$ in the subsequent steps. (10) The state of the network after swapping $n(n-2)-1$. (11) The state of the network after swapping $n(n-1)-1$. We see that the size of the network state starts decreasing after we swap the n(n-1)+1. The size of the entangled state is $n+2$. (12)  The size of the entangled state is $n+1$.  (13) The size of the entangled state is four. }
\end{figure*}
Now, we consider the different node locations of Alice and Bob. 
\begin{itemize}
    \item \textbf{Alice and Bob are on a degree-two node. }
    \begin{itemize}
        \item Alice and Bob are on diagonally opposite nodes. In this case, the maximum size of the created entangled state is $n+3$.
w.l.o.g consider the nodes on position $\left((n-1),n(n-1)\right)$. We start with the swapping on the zeroth node. By repeatedly applying the rules given above and swapping the $1,2,3,\cdots n-2$ nodes, we obtain an $n$- entangled state. We now perform a swapping on $n$ node to obtain an entangled state of size $(n+1)$. We measure on $(n+1,n+2,\cdots 2n-2)$, and the state size doesn't change. On swapping the $2n-1$ node, we obtain an entangled state of size $n+2$. On swapping the $2n$ node, we obtain a $n+3$ state. We then observe that state size oscillates between $n+3,n+2$ depending on the node situation. We then observe that the size decreases after reaching $n(n-1)+1$ node. 
    \item Consider a square grid with Alice and Bob's node on the same edge. W.l.o.g, we assume that the node positions are $n-1$ and $n^2-1$. We can follow the argument above to see that the maximal size of the created entangled state is $n+3$. 
    \end{itemize}
    \item \textbf{Alice is on a degree-two node, and Bob is on a degree-three node. }
    \begin{itemize}
        \item Alice and Bob are on adjacent nodes. w.l.o.g assume that the node position is $n-1$ and $2n-1$. On swapping nodes $\left\{0,1,n-2\right\}$, we obtain an entangled state of size $n+1$. On swapping node $n$, the size of the entangled state is $n+2$. On swapping nodes $\left\{n+1,\cdots 2n-2\right\}$, the size of the entangled state remains $n+2$. On swapping node $2n$, the size of the entangled state changes to $n+3$. On swapping node $3n-1$, the size changes to $n+4$. The size of the entangled state on the subsequent swaps oscillates between $n+4$ and $n+3$. Once we reach node $n(n-1)$, the size of the entangled state starts decreasing. 
        \item Alice and Bob are on adjacent lines but not adjacent nodes. W.l.o.g, assume that the nodes are on $(n-1,n)$. On swapping nodes $\left\{0,1,n-2\right\}$, we obtain an entangled state of size $n$. On swapping the node $n+1$, the size of the resultant state is $n+2$. On swapping the $2n-1$ node, the size of the entangled state changes to $n+3$. On swapping the $2n$ node, the size of the entangled state changed to $n+6$. The size of the entangled state in subsequent steps oscillates between $n+6$ and $n+5$ till we swap the node $n(n-1)$, after which the size of the entangled state decreases further. 
        \item Alice and Bob are on the same edge of the square grid. Let the nodes be $n-1,mn-1$, where $m\leq n$. On swapping nodes $\left\{0,1,n-2\right\}$, we obtain an entangled state of size $n$. On swapping the node $n$, the size of the resultant state is $n+1$. On swapping the $2n-1$ node, the size of the entangled state changes to $n+2$. On swapping the $mn+1$ node, the size changes to $n+3$, and on swapping the $(m+1)n$ node, the size changes to $n+4$. 
        The size of the entangled state in subsequent steps oscillates between $n+5$ and $n+4$ till we swap the node $n(n-1)$, after which the size of the entangled state decreases further.

    \item
       Alice and Bob are not on the same edge of the square grid. Let the nodes be $n-1,mn$, where $m\leq n$.
        On swapping nodes $\left\{0,1,n-2\right\}$, we obtain an entangled state of size $n$. On swapping the node $n$, the size of the resultant state is $n+1$. On swapping the $2n-1$ node, the size of the entangled state changes to $n+2$. On swapping the $2n$ node, the size of the entangled state changes to $n+3$. On swapping the $mn+1$ node, the size changes to $n+4$, and on swapping the $(m+1)n$ node, the size changes to $n+7$. The size of the entangled state in subsequent steps oscillates between $n+6$ and $n+7$ till we swap the node $n(n-1)$, after which the size of the entangled state decreases further. Note that this gives an upper bound on the size. For low enough $n$, we will not encounter all of these situations. 
    \end{itemize}
    \item \textbf{Alice is on a two-edge and Bob is on a four-edge}. 
    \begin{itemize}
        \item Non-adjacent lines: Consider Alice to be in the leftmost corner. We have a state $n+1$ till we encounter Bob. Jumping over the node gives us a $n+3$ state.  Encountering Bob's node on the left gives us a $n+5$ state. The state oscillates between $n+4$ and $n+5$ before the size decreases.

        \item Adjacent lines: Assume Alice's node is $n-1$  We obtain $n+3$ state on jumping node Bob. On hitting the node adjacent to Alice, We obtain a state of size $n+4$.  We then swap on $3n$ node to obtain $n+5$ state. On swapping the node next to Bob, we obtain a node of size $n+7$. After that, the entangled state oscillates between $n+7$ and $n+6$ until the size decreases. 
    \end{itemize}

From here on, we state the results for the size of the intermediate entangled state created during the protocol. The proof for the statement is similar to the ones elucidated above. 
    
    \item \textbf{Alice and Bob are on a degree-three node}.
    \begin{itemize}
        \item Alice and Bob are on the same edge. The maximum size of the entangled state is $n+6$. 
        \item  Alice and Bob are on different edges. The maximum size of the entangled state is $n+6$. 
    \end{itemize}
    \item \textbf{Alice is on a degree-three node, and Bob is on a degree-four node.}
    \begin{itemize}
        \item Alice and Bob are in the same column. The maximum size of the entangled state is $n+7$.  
        \item Alice and Bob are in a different column. The maximum size of the entangled state is $n+8$. 
    \end{itemize}
    \item \textbf{Alice and Bob are on a degree-four node.} 
    \begin{itemize}
        \item Alice and Bob are not on adjacent nodes. The maximum size of the entangled state is $n+9$. 
        \item Alice and Bob are on adjacent nodes. The maximum size of the entangled state is $n+7$.
    \end{itemize}
\end{itemize}

We now observe that the maximum size observed for $p=1$ is $n+c$. If we take $p\leq 1$, we may have to keep track of more than one entangled state in each time step. However, the total size would be less than or equal to the above-mentioned one. Hence, we see that the size of the intermediate state for this protocol is always less than $n+c$. 

\section{Impact of $k$-hop communication on entanglement rates}\label{appen:k-level}

\begin{figure}
\centering
\hspace*{\fill}
\begin{subfigure}{0.45\textwidth}
   \includegraphics[width=\linewidth]{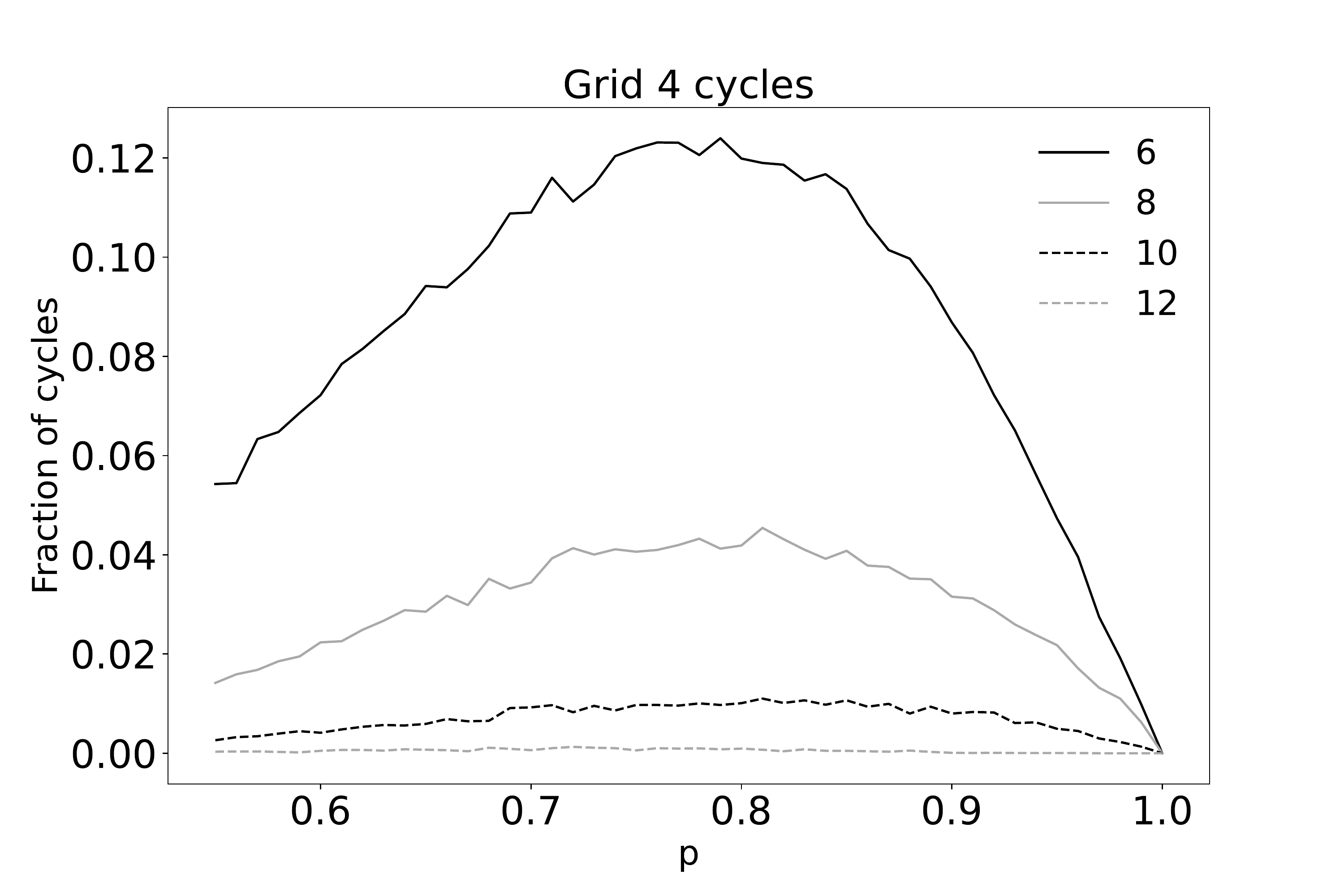}
   \caption{Fraction of $2k+2$ cycles in a graph with $n=4$.} \label{fig:region_d}
\end{subfigure}
\hspace*{\fill}
\begin{subfigure}{0.45\textwidth}
   \includegraphics[width=\linewidth]{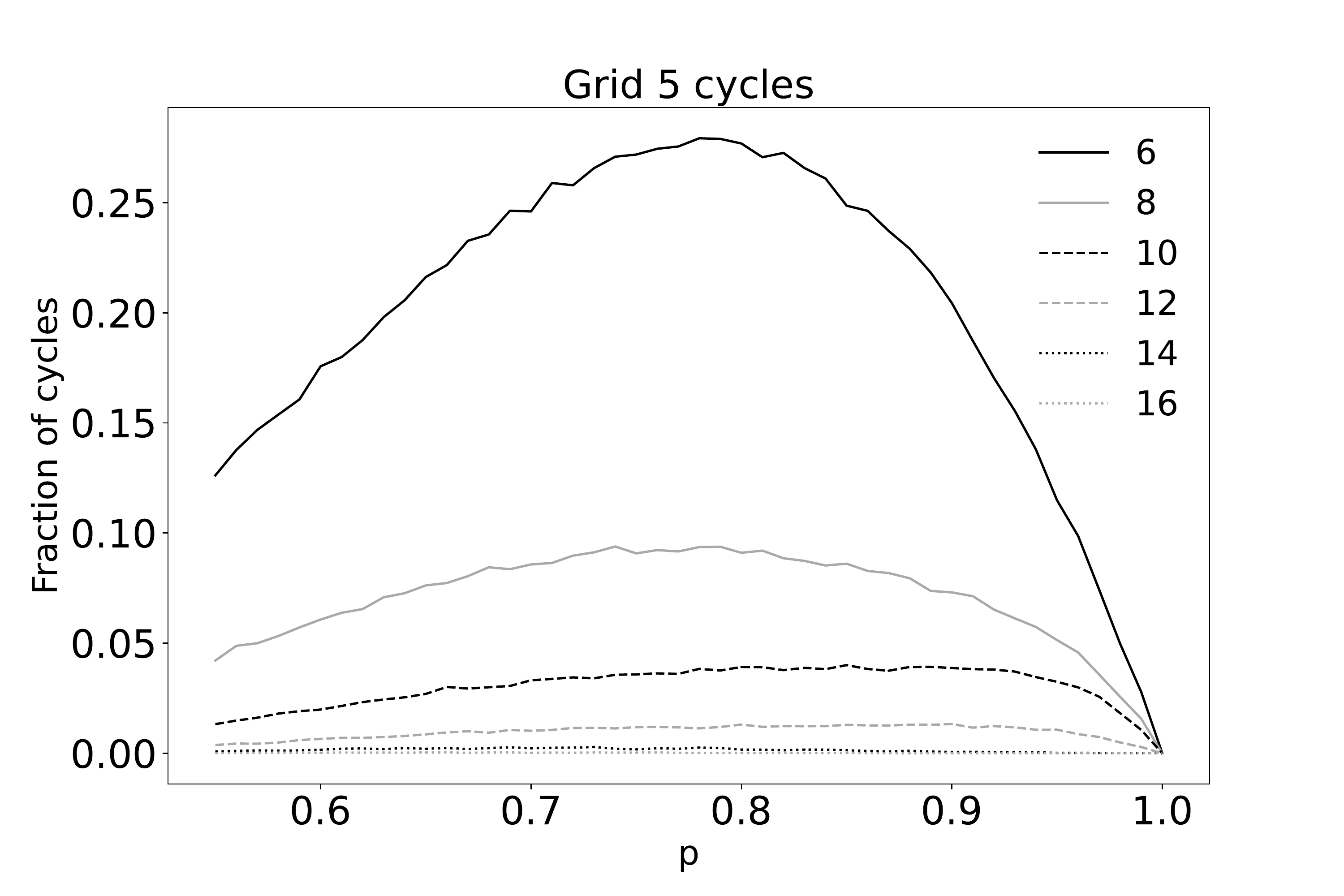}
   \caption{Fraction of $2k+2$ cycles in a graph with $n=5$.} \label{fig:region_e1}
\end{subfigure}

\begin{subfigure}{0.45\textwidth}
\centering
   \includegraphics[width=\linewidth]{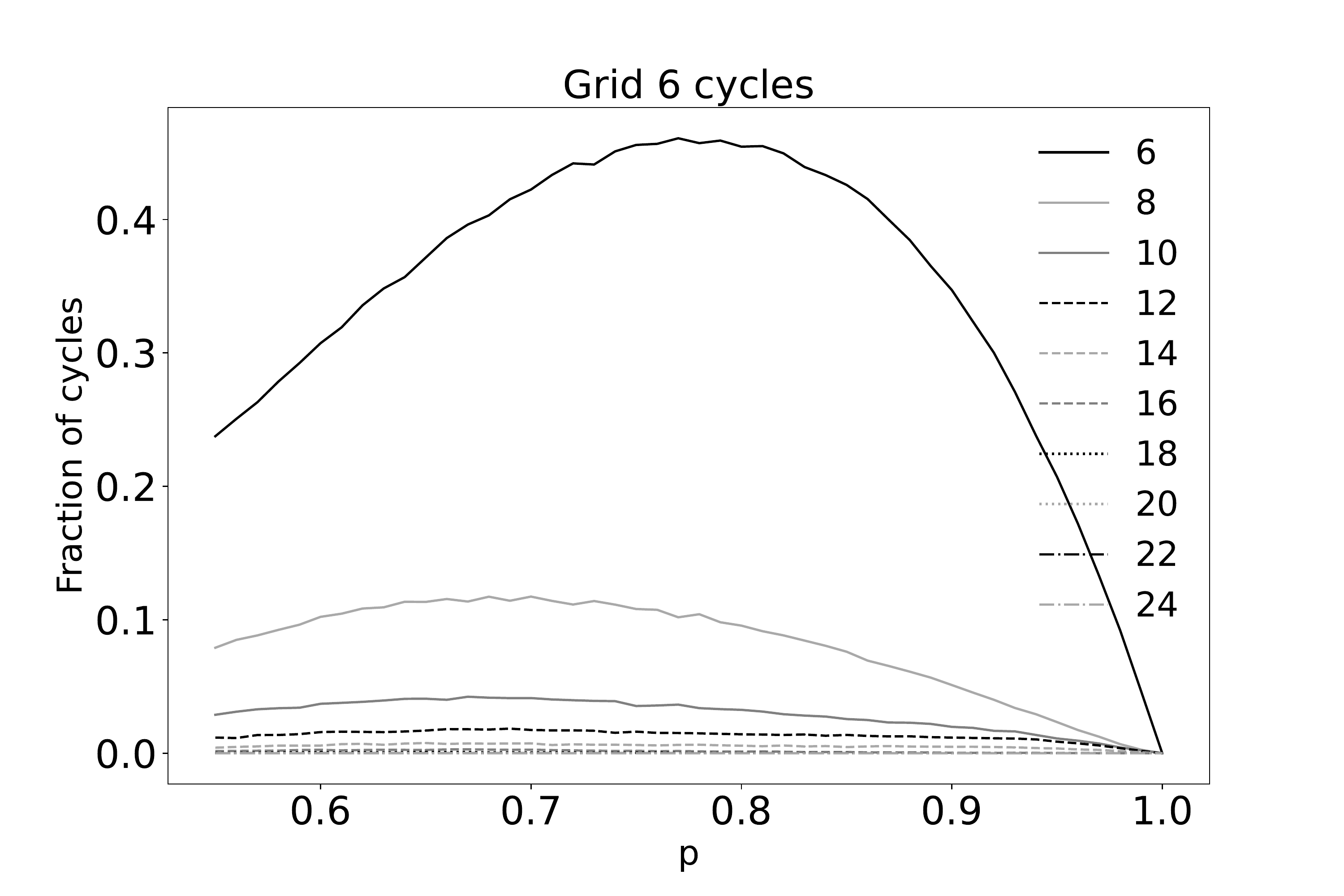}
   \caption{Fraction of $2k+2$ cycles in a graph with $n=6$.} \label{fig:region_f1}
\end{subfigure}

\caption{We plot the fraction of rounds in which we observe $2k+2$-polygons in a graph of size $n$ as a function of link probability $p$. The legend in the plot represents the cycle length}
\label{fig:cycles-fraction}
\end{figure}

The number of $2k+2$-polygons formed in the graph depends primarily on the probability of establishing a link $p$ and the size of the grid $n$. We perform a Monte-Carlo simulation with various values of $p$ and $n$ to obtain the fraction of the $2k+2$-polygons in each round. We find that with lower values and higher values of $p$, the fraction of the polygons is low. For $p=1$, the fraction of polygons of all sizes is zero. The highest fraction is observable around $0.8-0.85$. For lower values of $p$, we see that the probability of the link generation is low, which translates to the probability of cycles or polygons in the graph being low. For higher values of $p$, observe that our protocol with $k=1$ allows for the removal of all the four-edged polygons or four cycles in the network. This translates to the removal of the higher-edged polynomial as well. We plot the fraction of rounds with $2k+2$-polygons as a function of $p$ and $n$ in Figure~\ref{fig:cycles-fraction}

\section{BBPSSW distillation Scheme}\label{appen-distillation}
Here, we use a $2\rightarrow 1$ distillation protocol described by \cite{Benett1996}, also known as BBPSSW protocol. Given two entangled pairs, $\rho_{A_1B_1}$ and $\rho_{A_2B_2}$. The protocol is described as follows: 
\begin{itemize}
    \item Apply $\textrm{CNOT}_{A_1\rightarrow A_2} \otimes \textrm{CNOT}_{B_2\rightarrow B_1}$. 
    \item Measure $A_2$ is $\sigma_z$ (Pauli $Z$) basis and $B_2$ is $\sigma_x$ (Pauli $X$)  basis. Suppose the outcomes of the two measurements are $\zeta_1$ and $\zeta_2$. 
    \item Keep the pair $A_1, B_1$ if $\zeta_1\oplus\zeta_2 = 0$, i.e. if the measurement results coincide.
\end{itemize}
Given two copies of a Werner state with fidelity $f_i$, the resultant state from the distillation procedure has the fidelity $F$, given as
\begin{equation}
    F = \frac{f_i^2 + ((1-f_i)/3)^2}{f_i^2+2f_i(1-f_i)/3 +5((1-f_i)/3)^2}.
\end{equation}
The denominator of the above expression is equivalent to the probability of success of the protocol. That is,
\begin{equation}
    \textrm{prob}(f_i,2) = f_i^2+2f_i(1-f_i)/3 +5((1-f_i)/3)^2.
\end{equation}

We write the probability distribution for the various fidelities observed when there are $t=6$ rounds of link-level entanglement distribution. The probability of obtaining no entangled state is modified to
\begin{align}
    p_{\textrm{nolink}} &= (1-p)^6 + ^6C_2  p^2  (1-p)^{4}(1-\textrm{prob}(f_0,2)) + ^6C_4 p^4 (1-p)^2\ (( 1-\textrm{prob}(f_0,2))^2  +\nonumber\\& \textrm{prob}(f_0,2)^2 (1-\textrm{prob}(f_2,2)) ) + p^6\left((1-\textrm{prob}(f_0,2))^3 + ^3C_2 (1-\textrm{prob}(f_0,2)
    )\textrm{prob}(f_0,2)^2(1-\textrm{prob}(f_2,2))\right).
\end{align}

The probability of obtaining $f_0$ is given as: 
\begin{align}
    p_{f_0} = ^6C_1 \ p (1-p)^5 + ^6C_3 p^3 (1-p)^3(1-\textrm{prob}(f_0,2)) + ^6C_5 p^5(1-p) ((1-\textrm{prob}(f_0,2))^2)
 \end{align}

 The probability of obtaining $f_2$ is given as: 
\begin{multline}
    p_{f_2} =^6C_2 \ p^2 (1-p)^4 \textrm{prob}(f_0,2) +^6C_4 p^4 (1-p)^2 (2\ \textrm{prob}(f_0,2)(1-\textrm{prob}(f_0,2))) +\\p^6 (^3C_1 (1-\textrm{prob}(f_0,2))^2 \textrm{prob}(f_0,2)+ \textrm{prob}(f_0,2)^3(1-\textrm{prob}(f_2,2))) \\+ ^6C_5 p^5(1-p) (2 \ \textrm{prob}(f_0,2)(1-\textrm{prob}(f_0,2))) + ^6C_3 p^3(1-p)^3\textrm{prob}(f_0,2)
 \end{multline}

 The probability of obtaining $f_4$ is given as: 
 \begin{multline}
 p_{f_4} =^6C_4 p^4(1-p)^2 \textrm{prob}(f_0,2)^2 \textrm{prob}(f_2,2) + ^6C_5p^5(1-p)\textrm{prob}(f_0,2)^2\textrm{prob}(f_2,2)+\\p^6(^3C_2 (1-\textrm{prob}(f_0,2))\textrm{prob}(f_0,2)^2\textrm{prob}(f_2,2)+ \textrm{prob}(f_0,2)^3\textrm{prob}(f_2,2))
 \end{multline}

where $\textrm{prob}(f_j,2)$ is the probability of success of the distillation protocol when the initial fidelity is $f_j$.


\end{document}